\documentclass[twoside,12pt]{article}
\usepackage{epsfig}

\newcommand{\be}{\begin{equation}}
\newcommand{\ee}{\end{equation}}
\newcommand{\bea}{\begin{eqnarray}}
\newcommand{\eea}{\end{eqnarray}}

\makeatletter
\def\lesssim{\mathrel{\mathpalette\vereq<}}
\def\vereq#1#2{\lower3pt\vbox{\baselineskip1.5pt \lineskip1.5pt
\ialign{$\m@th#1\hfill##\hfil$\crcr#2\crcr\sim\crcr}}}
\def\gtrsim{\mathrel{\mathpalette\vereq>}}
\makeatother

\begin{document}

\title{\vspace{1cm}
Electromagnetic Dissociation as a Tool for\\ Nuclear Structure 
and Astrophysics}
\author{Gerhard\ Baur,$^{1}$\footnote{E-mail: g.baur@fz-juelich.de}
~Kai\ Hencken,$^2$\footnote{E-mail: k.hencken@unibas.ch}
~Dirk\ Trautmann$^2$\footnote{E-mail: dirk.trautmann@unibas.ch}\\
\\
$^1$Forschungszentrum J\"ulich, D-52425 J\"ulich, Germany\\
$^2$Universit\"at Basel, CH-4056 Basel, Switzerland}
\maketitle
\begin{abstract} 
Coulomb dissociation is an especially simple and important reaction
mechanism. Since the perturbation due to the electric field of the
(target) nucleus is exactly known, firm conclusions can be drawn from
such measurements. Electromagnetic matrixelements and astrophysical
$S$-factors for radiative capture processes can be extracted from
experiments.  We describe the basic elements of the theory of
nonrelativistic and relativistic electromagnetic excitation with heavy
ions.  This is contrasted to electromagnetic excitation with leptons
(electrons), with their small electric charge and the absence of
strong interactions. We discuss various approaches to the study of
higher order electromagnetic effects and how these effects depend on
the basic parameters of the experiment.  The dissociation of neutron
halo nuclei is studied in a zero range model using analytical methods.
We also review ways how to treat nuclear interactions, show their
characteristics and how to avoid them (as far as possible).  We review
the experimental results from a theoretical point of view. Of special
interest for nuclear structure physics is the appearence of low lying
electric dipole strength in neutron rich nuclei.  Applications of
Coulomb dissociation to some selected radiative capture reactions
relevant for nuclear astrophysics are discussed. The Coulomb
dissociation of $^8$B is relevant for the solar neutrino problem.  The
potential of the method especially for future investigations of
(medium) heavy exotic nuclei for nuclear structure and astrophysics is
explored.  We conclude that the Coulomb dissociation mechanism is
theoretically well understood, the potential difficulties are
identified and can be taken care of.  Many interesting experiments
have been done in this field and many more are expected in the future.
\end{abstract}



exotic nuclei, halo nuclei, nuclear astrophysics.

\newpage

\tableofcontents

\newpage

\section{Introduction}
\label{sec:intro}

Nuclear reactions at high energy are very complicated in general,
due to the strong 
interactions between the colliding hadrons. However, in 
the special case of very 
peripheral collisions the nuclei do
not touch each other and the short range strong forces between the 
collision partners are avoided. The reaction mechanism 
becomes simple and unambiguous conclusions can be drawn from
such experiments. In very peripheral collisions the 
nuclei interact with each other 
through the time-dependent electromagnetic field 
caused by the moving nuclei. Especially for heavy nuclei these fields  
are  very strong and  various interesting effects can occur.
It is the purpose of this review to describe them.
We discuss how these collisions are treated theoretically and how they are 
applied to nuclear structure and nuclear astrophysics problems.
Coulomb excitation has been a very powerful tool in the past
to study electromagnetic matrix-elements in nuclei. Classical review 
papers exist, see, e.g., \cite{AlderW75,AlderBHM56}.
For collision energies below the Coulomb barrier
the condition of no nuclear interactions of the nuclei
with each other is very well fulfilled and valuable nuclear structure 
information has been obtained. An important example is the  
investigation of  nuclear rotational and vibrational
collective motion by means of Coulomb excitation \cite{AlderW75}.
Such studies are also of special interest for the investigation of 
exotic nuclei at the radioactive beam facilities
which have become available in the past decades all over the 
world.  Nuclei (far) away from the valley of stability 
(neutron-- or proton--rich) can now be investigated
experimentally. 
They  have widened considerably the landscape of nuclear physics 
and also made possible novel studies in nuclear astrophysics.
New phenomema like halo nuclei have been discovered.
Electromagnetic excitation and dissociation is again a 
very powerful tool in this case. They are 
expected to play an even more important  role in the future
at the new facilities being proposed presently 
around the world
(GSI, Germany \cite{CDR}, RIA, USA \cite{riawhitepaper,riaweb}, RIKEN, Japan
\cite{ribfweb}).

\begin{figure}[tb]
\begin{center}
(a)~~~~~
\resizebox{0.6\textwidth}{!}{%
\includegraphics{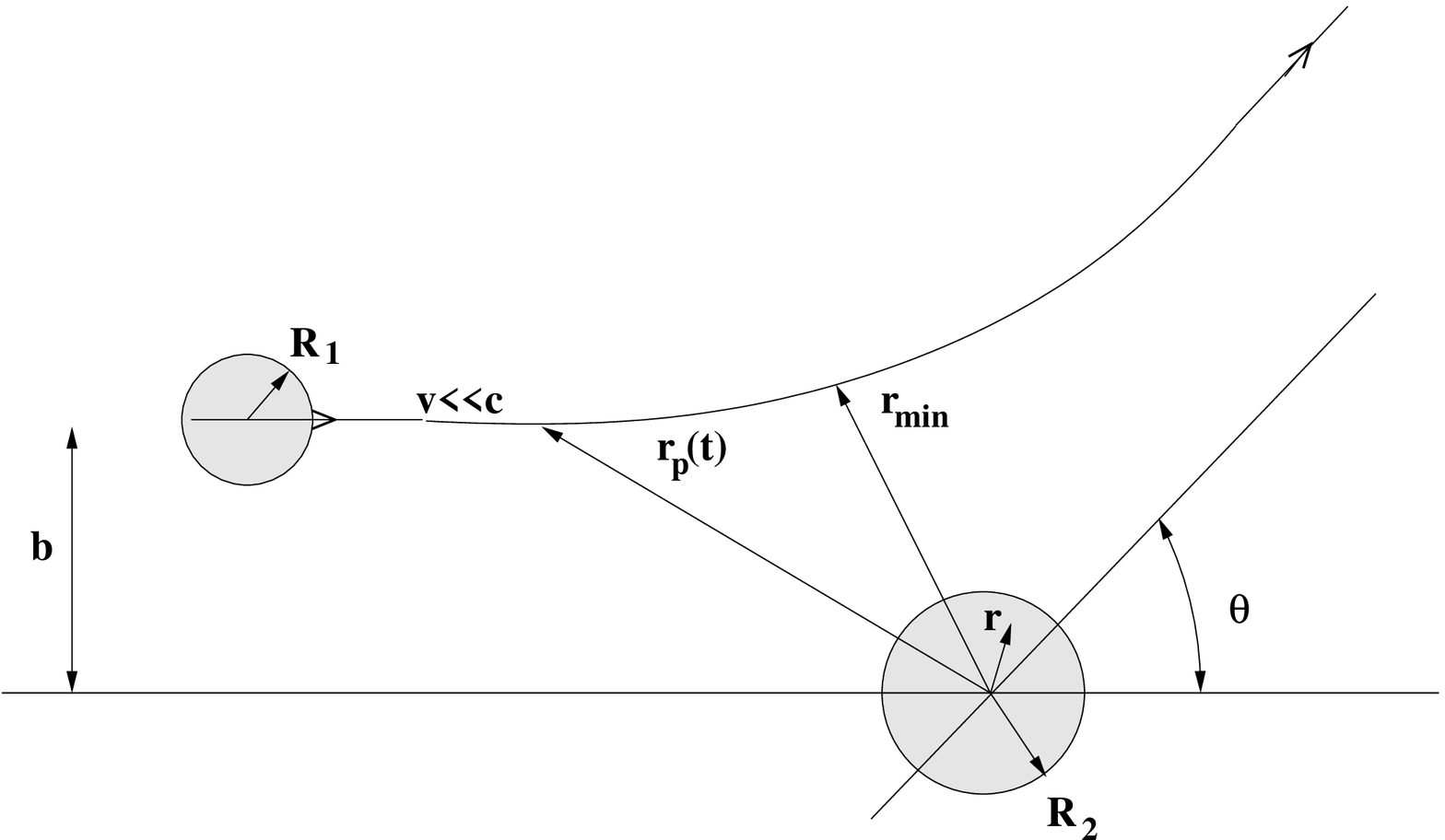}}
\\(b)~~~~~
\resizebox{0.6\textwidth}{!}{%
\includegraphics{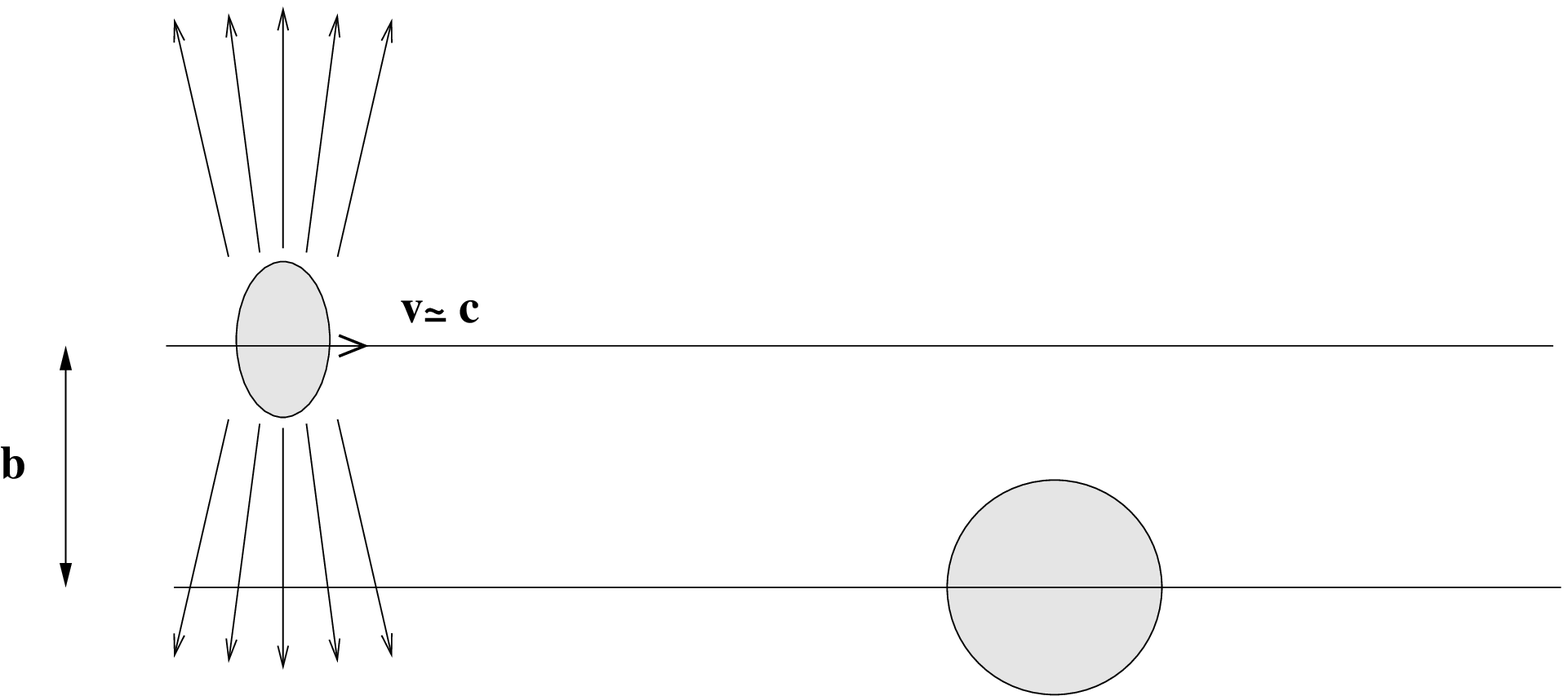}}
\end{center}
\caption{
(a) A projectile with charge $Z_1e$ is scattered on a target nucleus. 
The projectile velocity is nonrelativistic  ($v\ll c$). 
The minimum distance of closest aproach of the Rutherford trajectory 
is $r_{min}> R_1+R_2$, the sum of the nuclear radii.
(b) A projectile with velocity $v \protect\lesssim c$ passes 
by a target nucleus with an impact parameter $b> R_1+R_2$.   
The Rutherford orbit is very close to a straight line.
} 
\label{fig:cxexplain}
\end{figure}

The condition that the hadrons do not touch each other 
can be fulfilled either by using bombarding energies well below the 
Coulomb barrier, or 
by going to very forward scattering angles in {\em intermediate},
($v\lesssim c$, $\gamma \gtrsim 1$)
 and {\em relativistic}\/ ( $v \approx c$,
$\gamma\gg 1$) collisions. These
forward angles (classically) correspond to impact parameters larger 
than the sum of the nuclear radii for high  beam energies.
The two cases are shown schematically in Figs.~\ref{fig:cxexplain}~(a) and~(b).

It is very important to note that with increasing beam energy higher
lying states can be excited with the Coulomb excitation
mechanism. This can lead to Coulomb dissociation, in addition to
Coulomb excitation of particle-bound states. This was reviewed some
time ago in \cite{BertulaniB88}.  It has become more and more clear,
that such investigations are also well suited for secondary
(radioactive) beams.  An (unstable) fast projectile nucleus can
interact with a high $Z$ target nucleus. In this way the interaction
of an unstable particle with a (quasireal or equivalent) photon can be
studied. A similar method is used in particle physics, where it is
known as the {\em Primakoff effect}
\cite{DreitleinP62,PomeranchuckS61}.

Since the electric field of a nucleus with high charge number $Z$ is
much stronger than, e.g., the one of an electron, the nucleus can be a
very suitable electromagnetic probe for certain cases. One can study,
e.g., higher order phenomena, which are inaccessible with conventional
electromagnetic probes like the electron. The excitation of the double
phonon giant dipole resonance observed at GSI
\cite{Emling94,AumannBE98} is an example.

We now give a guided tour through the content of this review: we start
with a general discussion of the theory of Coulomb excitation and
dissociation. Due to the time-dependent electromagnetic field the
projectile is excited to a bound or continuum state, which can
subsequently decay.  Since the electromagnetic interaction is very
well known, rather precise unambiguous information can be obtained
from such experiments.  If first order electromagnetic excitation is
the dominant effect, experiments can directly be interpreted in terms
of electromagnetic matrix elements which also enter, e.g., in
radiative capture cross-sections.  It is therefore very important to
know the conditions of validity of the first order theory and how to
assess quantitatively the higher order effects.

We especially mention two aspects of multiple electromagnetic
excitation: it is a way to excite new nuclear states, like the double
phonon giant dipole resonance \cite{BaurB86}; but it can also be a
correction to the one-photon excitation
\cite{TypelB94a,TypelB94b,TypelWB97}.  The theoretical methods to
treat these problems are discussed in Sec.~\ref{sec:higher}.  They
range from higher order perturbation theory to Glauber calculations,
CDCC (Continuum Discretized Coupled Channel) and solving numerically
the time-dependent Schr\"odinger equation.

Breakup reactions are at least a three-body problem, which is very
complicated in general, and the long range character of the Coulomb
interaction adds to the complication.  We discuss here the Coulomb
dissociation of neutron halo nuclei as an illustrative example. We can
study analytically a rather simple model of the dissociation of a
two-particle system (deuteron, neutron halo nucleus) bound by a
zero-range force in the nuclear Coulomb field.  This sheds light on
the conditions of validity of the usual treatment of the semiclassical
Coulomb excitation theory and of the importance of higher order
effects. This is done in Sec.~\ref{sec:analytic}.

Coulomb excitation is a very useful tool to determine nuclear
electromagnetic matrix-elements.  This is of interest for nuclear
structure and nuclear astrophysics
\cite{BaurBR86,BaurR94,BaurR96,BaurB86}.  In the following we
illustrate the theoretical concepts and discuss examples of
experimental results and their theoretical interpretation. The
emphasis is more on theory.  As theorists we cannot give a detailed
treatment of nor do full justice to the ingenious experimental
developments which make possible the precision studies at these
extreme forward angles.  We refer to the previous reviews, see
e.g. \cite{BaurR94,BaurR96} and to the new proposals for future
facilities like \cite{CDR} or \cite{riawhitepaper}.

In Sec.~\ref{sec:nstruc} we review the results that have been obtained
on nuclear structure from intermediate energy electromagnetic
excitation in the past years.  We briefly mention the Primakoff effect
in particle physics, where the nuclear Coulomb field is used to study
the interaction of fast (unstable) particles with the quasireal
photons due to the Coulomb field of a nucleus. We also mention the
very large effects of electromagnetic excitation in relativistic heavy
ion collisions in fixed target experiments.  These effects become
quite spectacular at the heavy ion colliders RHIC and LHC with the
very high values of the Lorentz parameter. Recently experimental
results from RHIC have become available.

In the next section we discuss applications to radiative capture
reactions (the time reversed process of photodissociation) of
astrophysical interest. This subject has been reviewed before
\cite{BaurR96}. This field is rapidly expanding; this is essentially
due to the new experimental possibilities world-wide.  Future
applications are given in Sec.~\ref{sec:future}.  We discuss new
possibilities, like the investigation of $r$-- and $rp$--process
nuclei, which will be produced with high intensity at the future RIA
facilities, and the experimental study of two-particle capture
processes by means of Coulomb dissociation. We close with conclusions
and an outlook in Sec.~\ref{sec:conclusions}.  This review grew out of
a series of lectures given by one of us in the frame of the
``European Graduate School Basel-T\"ubingen'', for the transparencies see
\cite{eurograd}.

\section{Theory of Electromagnetic Dissociation}
\label{sec:theory}

First we briefly recall the basic points 
of the theory of nonrelativistic (NR) Coulomb excitation. 
Excellent reviews exist, see e.g. \cite{AlderW75}. It is usually a 
very good approximation to treat the relative motion between projectile and 
target classically ({\em semiclassical approximation}). It 
is also well understood how one can obtain the semiclassical 
limit from the quantal theory \cite{AlderBHM56}. 
We  want to concentrate here on intermediate and relativistic energies.
In these cases one can replace the Rutherford 
trajectory by a straight line to a good approximation.
In contrast to low energy Coulomb excitation retardation is now an 
important effect and the long wave-length approximation is not necessarily 
a good approximation. 
Deviations from the straight line approximation
have to be and can be assessed quantitatively. In collisions above the 
Coulomb barrier strong interactions between the nuclei cannot be avoided.
However at very forward angles, corresponding to 
large impact parameters, such nuclear effects 
can become negligible. Even then
there can be some kind of nuclear effects at a certain level 
of accuracy: there is diffraction due to the wave nature of the projectile.
However, essentially due to the small de Broglie wavelength
of the projectile this effect is quite small.
These questions will be addressed in Sec.~\ref{sec:nuclear} below.
  
\subsection{\it Nonrelativistic (NR) Projectile Velocity}
We discuss briefly the basic ideas of (NR) Coulomb excitation.
The condition of {\em no nuclear contact} is ensured 
by using bombarding energies below the Coulomb barrier.
In the semiclassical theory the projectile moves on a classical 
Rutherford trajectory, see Fig.~\ref{fig:cxexplain}(a),
giving rise to a time dependent external electromagnetic field.
Nonrelativistic semiclassical Coulomb excitation theory
is a classical textbook example of the application of  
time-dependent perturbation theory,
see, e.g., \cite{Messiah85} or \cite{EisenbergG88}.

The excitation of the target nucleus is due to the   
static Coulomb interaction, or more generally,
(see \cite{AlderW75}, Eq.~(II.1.2)) the mutual electromagnetic
interaction between the ions. For NR collisions we can {\em neglect retardation}, 
and the electromagnetic interaction can be written as
\begin{equation}
W(1,2)= \int \int d^3r_1 d^3r_2 \frac{\rho_1 (\vec{r_1})\rho_2(\vec{r_2})
-\vec{j}_1(\vec{r_1}) \cdot \vec{j}_2(\vec{r_2})/c^2
}{\left|\vec{r_1}-\vec{r_2}\right|},
\label{eq:theory:W12}
\end{equation}
where $\rho_i(\vec{r_i})$, $\vec{j}_i(\vec{r_i)}$ are the 
charge and current densities for the projectile ($i=1$)
and the target nucleus ($i=2$), respectively.
For non-overlapping charge-densities one can express
$W(1,2)$ in terms of the electromagnetic 
multipole moments of the two nuclei.
The full expression is given in \cite{AlderW75}. We give here only
the most important terms, that is the 
{\em monopole-monopole} and {\em monopole-multipole} interactions
and neglect the less important multipole-multipole interactions.

We are especially interested in target excitation.
In this case we can assume that nucleus 1 is a
point charge, we have $\rho_1(\vec r_1)=Z_1 e \delta(\vec r_p(t))-\vec r_1)$.
We can specialize to the coordinate system $(\vec{r},\vec{r_p})$,
where $\vec r$ is the target coordinate (measured from the center of mass
of the target) and ${\vec r}_p(t)$ is the relative coordinate between 
the center of masses of the two nuclei, see Fig.~\ref{fig:cxexplain}(a).
In the semiclassical theory $\vec{r_p}(t)$ is a classical time-dependent 
parameter.  The charge operator $\rho(\vec{r}\,)$ is given by
\begin{equation}
\rho(\vec{r}\,)=e \sum_i \delta(\vec{r}-\vec{r_i}),
\end{equation}
where the $\vec{r_i}$ denote the proton coordinates.  

Neglecting the multipole-multipole interactions 
we can write the electromagnetic interaction as 
\begin{equation}
W(1,2)=V(1,\vec{r_p}(t))+ V(2,\vec{r_p}(t))+\frac{Z_1 Z_2 e^2}{r_p(t)}.
\end{equation}

The last term is the monopole-monopole 
Coulomb potential between projectile and target which 
is responsible for the Coulomb trajectory of the projectile.
The quantity $V(2,\vec{r_p}(t))$
gives rise to target excitations
and $V(1,\vec {r_p}(t))$ to projectile excitations, respectively.

The monopole--electric-multipole 
interaction can be obtained from the expansion
\begin{equation}
\frac{1}{\left|\vec r - \vec{r_p} \right|} = 4\pi \sum_{\lambda\mu}
\frac{1}{2\lambda+1}
\frac{r^\lambda}{r_p^{\lambda+1}} Y_{\lambda\mu}^*(\hat r) 
Y_{\lambda\mu}(\hat{r_p}),
\label{eq:elmulti}
\end{equation}
where we have made use of the condition of {\em no nuclear contact} so that
always $r<r_p$. In this way, the variables $r$ and $r_p$ are separated.

The total time-dependent interaction relevant for the target excitations
is given by the sum $V(2,\vec r_p(t))$ $=$ $V_E(2,\vec r_p) + V_M(2,\vec r_p)$,
where  
we can write the quantities $V_E$ (electric interaction) and $V_M$ 
(magnetic interaction), see
Eqs.~(II.1.12) and~(II.1.13) in \cite{AlderW75}, as
\begin{eqnarray}
V_E(2,\vec r_p) &=& \sum_{\lambda\ge1,\mu} \frac{4\pi Z_1 e}{2\lambda+1}
M(E\lambda,-\mu) (-1)^\mu r_p^{-\lambda-1} Y_{\lambda\mu}(\hat r_p)
\label{eq_VE2}\\
V_M(2,\vec r_p) &=& \sum_{\lambda\ge1,\mu} \frac{4\pi Z_1 e}{2\lambda+1}
\frac{i}{\lambda} M(M\lambda,-\mu) (-1)^\mu \frac{\dot{\vec r_p}}{c}
r_p^{-\lambda-1} \vec L Y_{\lambda\mu}(\hat r_p).
\label{eq_VM2}
\end{eqnarray}
The electromagnetic interaction can 
be parametrized completely in 
terms of the electromagnetic multipole matrix-elements $M(\pi\lambda\mu)$.
In the long wave length limit they are defined as
\begin{eqnarray}
M(E\lambda\mu) &=& \int d^3r \rho(\vec r) \: r^\lambda 
\:Y_{\lambda\mu}(\hat r)
\label{eq:melambda}
\\
M(M\lambda\mu) &=& \frac{-i}{c(\lambda+1)} \int d^3r \vec{j}(\vec{r}) \cdot
\left[r^\lambda 
\: \vec L \: Y_{\lambda\mu}(\hat r)\right],
\label{eq:mmlambda}
\end{eqnarray}
with $\vec L = -i \vec r \times \vec \nabla$.
Especially we have
$M(E00)=Z_2 e/\sqrt{4\pi}$, where $Z_2$ is the electric 
(monopole) charge of the target nucleus. 

The condition of no nuclear contact is vital to this approach. 
In this case we obtain the complete
separation into the electromagnetic matrixelements of the target.
For $r>r_p$ the strong interaction between the nuclei would come into play
and would 
severely modify the results. This is discussed in detail 
in Sec.~\ref{sec:nuclear} (and also in \cite{EsbensenB02,EsbensenB02c}).

There are a few dimensionless parameters which characterize the  
electromagnetic excitation: we define the {\em adiabaticity parameter}
as the ratio between the {\em collision time}
and the {\em excitation time}
\begin{equation}
\xi= \frac{\tau_{coll}}{\tau_{exc}}.
\end{equation}

We can estimate the collision time to be 
$\tau_{coll}=r_{min}/(\gamma v)$ and 
we have $\tau_{exc} =\hbar/\Delta E$ , where $\Delta E=\hbar \omega$
is the nuclear excitation energy. From this we get
\begin{equation}
\xi=\frac{\omega r_{min}}{\gamma v}.
\label{eq:xidefined}
\end{equation}

For NR collisions we have $\gamma \approx 1$,
whereas in the relativistic case (see below)
the Lorentz parameter $\gamma$ can be 
much larger than one and the collision time can 
become very small due to the Lorentz contraction.
For adiabaticity parameters $\xi \gg 1$ the system can 
follow the adiabatic ground state and no excitation occurs
\cite{Messiah85}
(with probability $\sim \exp(-\xi)$). This means also
that in NR Coulomb excitation one can only excite nuclear states
for which the long-wavelength limit is valid:
Due to the adiabaticity condition $\xi\lesssim 1$ we have
$\omega r_{min} \ll v$. This  leads to 
$k R_2 \ll 1$, where $k=\omega/c$ and
$R_2$ denotes the size of the nucleus (see Fig.~\ref{fig:cxexplain}(a)), 
since $r_{min}>R_2$ and $v<c$.
On the other hand, for relativistic collisions the collision time
can be very small and one is able to excite states 
for which the long wavelength limit 
is no longer valid. 

A second parameter is the {\em strength parameter},
which is defined as the strength of the interaction potential times its
duration (in units of $\hbar$):
\begin{eqnarray} 
\chi&=&\frac{V_{int}\tau_{coll}}{\hbar}.
\end{eqnarray} 
Here $V_{int}$ denotes a typical value of the interaction potential. 
For a multipole interaction of order $\lambda$, this value of the interaction
potential is typically $e\left<f\right\|M(E\lambda)\left\|i\right>/r
_{min}^{\lambda+1}$ (NR case), 
the strength parameter $\chi$ is therefore estimated to be  
\begin{equation}
\chi=\frac{Z_1 e<f||M(E\lambda)||i>}{\hbar v r_{min}^{\lambda}}.
\label{eq:chidefined}
\end{equation}
(The relativistic case will be discussed below: in this case, the collision 
time has a $\gamma^{-1}$-dependence; the interaction $V_{int}$
is proportional to $\gamma$ and $\chi$ becomes independent 
of $\gamma$.)
The strength parameter for the monopole-monopole case is the 
Coulomb parameter, which is given by 
\begin{equation}
\eta=\frac{Z_1 Z_2 e^2}{\hbar v}.
\label{eq:etadefined}
\end{equation}

In order to obtain the equation describing the time evolution
of the system under the influence of the  electromagnetic field, 
we expand the wave function of the target nucleus 
into  the complete set of eigenstates of the target Hamiltonian 
$\Phi_n$ with energy $\hbar\omega_n$.
\begin{equation}
\Psi(t)=\sum_n a_n(t)\exp(-i\omega_n t) \Phi_n,
\end{equation}
One finds the following system of coupled equations
(see,  e.g., Eq.~(5.29) of \cite{AlderW75})
for the excitation amplitudes $a_n(t)$ 
\begin{equation}
i\hbar \dot a_n=\sum_m \left<n|V(t)|m\right>\exp[i(\omega_n-\omega_m)t] \:
a_m(t),
\label{eq:adotcc}
\end{equation}
where the time-dependent electromagnetic interaction is 
denoted by $V(t)$. A formal solution  can be given
using the time ordering operator ${\cal T}$ as
(see Eq.~(II.3.13) of \cite{AlderW75})
\begin{equation}
a_n(\infty) = \left< n \right| {\cal T} \exp\left(-\frac{i}{\hbar}
\int_{-\infty}^{+\infty} dt \tilde V(t) \right) \left| 0 \right>,
\label{eq_anformalsol}
\end{equation}
where the interaction $\tilde V$ is given by
\begin{equation}
\tilde V(t) = \exp(i H_0 t/\hbar) V(t) \exp(-i H_0 t/\hbar)
\label{eq:Vtilde}
\end{equation}
and $H_0$ is the Hamiltonian of the nuclear system.

We mention two important limits: if the electromagnetic 
interaction $V(t)$ is weak enough (strength parameter $\chi \ll 1$), 
it  can be treated in first order ({\em one photon exchange}) and
one obtains an expression which factorizes into the orbital integrals 
and the electromagnetic matrixelements. One has \cite{AlderW75}
\begin{equation}
a_{n}=\frac{4\pi Z_1e}{i\hbar} \sum_{\lambda\mu}
\frac{(-1)^{\mu}}{2\lambda+1} <n|M(E\lambda,-\mu)|0>S_{E\lambda\mu}.
\label{eq:anrelativistic}
\end{equation}
The orbital integrals $S_{\pi\lambda\mu}$ can be calculated from 
the kinematics of the process \cite{AlderW75}
(We have given here only the electric part of the excitation,
$\pi=E$, the magnetic part $\pi=M$ can be found in \cite{AlderW75}).
Thus the experimentally 
determined excitation probabilities (or cross-sections)
are directly proportional to the 
reduced transition probabilities $B(\pi\lambda)$ (see Eq.~(4), p.~93 of
\cite{AlderW75}). 

Another important limit is the {\em sudden approximation}, where the 
collision time $\tau_{coll}=r_{min}/\gamma v$ is 
much smaller than the nuclear excitation time
$\tau_{exc}=1/\omega$, i.e., the adiabaticity parameter $\xi$ 
is much smaller than one. 
But the strength parameter $\chi$ can have arbitrary values.
In this case, we have $\tilde V \approx V$ and 
we can neglect the time ordering in Eq.~(\ref{eq_anformalsol})
and the electromagnetic interaction can be summed to all orders.
For large values of $\xi$ and $\chi$ one may use 
an adiabatic representation of the Schr\"odinger equation, 
see \cite{AlderW75}. It is also often 
useful to solve Eq.~(\ref{eq:adotcc}) numerically in these cases. 
(The case of the excitation of a 
harmonic oscillator can be treated exactly analytically, see below.)
In contrast to the NR case considered in this subsection the velocity 
for intermediate energy and relativistic collisions $v$ is close to $c$,
and the strength parameter $\chi$, see Eq.~(\ref{eq:chidefined}),
cannot be too large in these cases and 
higher order effects tend to be small, see also Sec.~\ref{ssec:equiphot}.
This is in contrast to Coulomb excitation below the barrier,
where $v\ll c$ and  $\chi$ can be much larger than one.
An example is the excitation of a rotational band in 
deformed nuclei \cite{AlderW75}, 
more details will be given in Sec.~\ref{sec:higher}. 

Although we have shown above that we can treat the electromagnetic
interaction in the limit of no retardation for NR collisions,
one may also treat the electromagnetic interaction
in its full relativistic form, taking retardation 
effects into account. This is done, e.g., in \cite{AlderBHM56}
using the Coulomb gauge.
In addition to the contribution from the instantaneous 
Coulomb interaction ({\em longitudinal photons\/})
one gets the one photon exchange contribution due to the 
$H_{int}=\vec{j} \cdot \vec{A}$ interaction
({\em transverse photons\/}).
The multipole expansion of the vector potential
$\vec{A}$ is performed, and it is found \cite{AlderBHM56}
that the instantaneous Coulomb interaction is canceled by 
a certain term (see Eq.~(\ref{eq:app:int3}) in the appendix) and only
the contribution from transverse photon remains. Again there is 
a factorization into an orbital integral and 
an electromagnetic matrixelement; this time it is the full electromagnetic
multipole matrix element (see Eqs.~(\ref{eq:appME}) and~(\ref{eq:appMM}))
in the appendix), where the wave number is equal to $k=\omega/c$,
i.e.,  {\em at the photon point}
({\em please note that the  long wavelength approximation is 
not made in this case}).
One may say that (NR) Coulomb excitation is due to longitudinal 
(Coulomb) photons, yet it is true that the nuclear structure 
information is entirely contained in electromagnetic 
matrixelements {\em at the photon point} $k=\omega/c$.
(For more details see the appendix.)
 
There is an important difference of Coulomb excitation
to electron inelastic scattering. 
In inelastic electron scattering the photon is {\em virtual}
(spacelike, i.e., $|\vec{k}| > \omega$),
and one can study electromagnetic matrixelements as 
a function of momentum transfer. In this way, e.g., 
the spatial distribution of the nuclear charge can be investigated.
Also monopole transitions can occur.

The matrixelements occurring in Coulomb excitation 
determine the interaction of 
real ({\em transverse}, $q^2=0$) photons with the nuclear system.
The equivalence of the current and charge matrix-elements
(see Eqs.~(\ref{eq:melambda}) and~(\ref{eq:appME})) 
in the long wavelength limit is sometimes called ``Siegert's
theorem'', see e.g. p.~89ff of \cite{EisenbergG88}. From the above it 
is clear that the condition of no nuclear contact is
the more basic condition than the long wave length limit.
In the following discussion of relativistic electromagnetic excitation,
the long wavelength limit does not generally apply. Yet, due to 
the condition of no nuclear contact,
the information about nuclear structure is again entirely contained in 
the electromagnetic matrixelements at the photon point.

Before going to intermediate energy and relativistic
energy electromagnetic excitation let us mention the straight 
line limit of the NR Coulomb excitation theory: 
From the standard formalism of Coulomb excitation 
using Rutherford trajectories one can also obtain this limit
by going to small scattering angles $\theta$. There remains a 
characteristic difference between the straight line theory and 
the straight line limit of the Rutherford trajectory: a factor 
$\exp(-\pi\xi/2)$, see Eq.~(II.E.79) of \cite{AlderBHM56}.
This factor can be traced back to the influence of 
Coulomb repulsion; the impact parameter $b$ should be replaced by an 
effective impact parameter.
This is explained in detail in \cite{AlderBHM56}and \cite{WintherA79}.
We come back to this in the next section, where the relativistic case 
is discussed.

\subsection{\it Equivalent Photon Method, Theory of Relativistic Coulomb 
Excitation}
\label{ssec:equiphot}

{\bf 1. Basic Idea}\\
The equivalent photon method was introduced by 
Fermi in a classic paper in 1924 \cite{Fermi24,Fermi25},
see also the translation by Gallinaro and White \cite{FermiGW02}.
He considered only the NR case, the relativistic generalisation
was subsequently given by Weizs\"acker and Williams 
\cite{Weizsaecker34,Williams34}. This method is lucidly described in the
textbook of Jackson \cite{JacksonED},
so let us only summarize very briefly the idea:

We consider the classical straight line motion 
of a charged particle with charge $Ze$ and impact parameter $b$
(see Fig.~\ref{fig:cxexplain}(b)). This motion 
gives rise to a time-dependent electromagnetic field at 
a given point (the target nucleus). 
The Fourier-transform of the electric field is {\it ``compared to 
the corresponding electric field of a suitable spectrum of light''}
\cite{Fermi24}. In this way, Fermi introduced the idea of 
{\it ``\"aquivalente Strahlung''} (equivalent photons). 
In this method only the $E1$-multipole
is treated correctly, since the spatial variation of the electric
field strength over the nuclear volume is neglected.
Higher multipoles ($E2$, $E3$, \dots,
$M1$, $M2$, \dots) show a different behaviour (see, e.g., 
\cite{BertulaniB88} and below). Higher multipole contributions 
are given correctly only in the limit $\gamma\rightarrow\infty$
in this approach.

In the equivalent photon approximation the cross section for an
electromagnetic process is written as
\begin{equation}
\sigma = \int \frac{d\omega}{\omega} \: n(\omega) \sigma_{\gamma}(\omega).
\end{equation}

where $\sigma_{\gamma}(\omega)$ denotes the appropriate cross section
for the photo-induced process and $n(\omega)$ is  the equivalent
photon number. 
The equivalent photon number in the case of a point particle 
is given in terms of the modified Bessel functions $K_n$ as:
\begin{equation}
n(\omega) = \frac{2 Z_1^2 \alpha}{\pi} \frac{c^2}{v^2}
\left[\xi K_0(\xi)K_1(\xi)-\frac{v^2\xi^2}{2c^2}(K_1^2(\xi)-K_0^2(\xi))\right],
\label{eq:nomegaexact}
\end{equation}
where $\xi=\omega b_{min}/\gamma v$.
The minimum impact parameter is denoted by $b_{min}=R_1+R_2$. 
Eq.~(\ref{eq:nomegaexact}) is well approximated by 
\begin{equation}
n(\omega) = \frac{2}{\pi} Z_1^{2} \alpha \ln \frac{\gamma v}{\omega b_{min}}
\label{eq:nomegaln}
\end{equation}
for $v \sim c$ and $\xi\ll1$.

For $\xi\gg1$ one can use the asymptotic 
expression for the modified Bessel functions and
obtain a spectrum which decreases exponentially with~$\xi$:
\begin{equation}
n(\omega )\approx \frac{Z_1^2 \alpha}{2} \exp(-2\xi).
\end{equation}
Already from this simple approach the major properties 
can be read off: the equivalent photon spectrum is very soft, it 
shows essentially an $1/\omega$ dependence. There is an 
adiabatic cutoff at $\omega=\gamma v/b_{min}$. There is a $(c/v)^2$
dependence, which gives a large photon flux for small beam velocities.
However, this is restricted to small photon energies,
$\omega< \gamma v /b_{min}$.

{\bf 2. Semiclassical straight line approximation and
exact multipole decomposition}\\ 
In the above method equivalent photon numbers of higher 
($E2$, $E3$, \dots, $M1$, $M2$,\dots) 
electromagnetic multipoles are in general not treated properly. 
A beautiful analytic expression which treats all multipoles
exactly was given in \cite{WintherA79},    
see also \cite{BertulaniB88}. We discuss it briefly now.

In the relativistic case, retardation can no longer be neglected.
One can  use, e.g., the Li\'enard-Wiechert potential 
(see Eq.~(\ref{eq:lienardwiechert})) in order to describe the 
(retarded) electromagnetic interaction \cite{WintherA79},
corresponding to the 
Lorentz gauge. In the multipole expansion the condition
$r<r_p$ is used and again (see appendix for more details) one can 
separate the first order amplitude into an {\em orbital part} and a 
{\em structure part}, which consists of the electromagnetic matrix elements 
at the photon point ({\em no long wave length limit is made}).
We give only the final formula here:
\begin{equation}
a_{fi}=-i \frac{Z_1e}{\gamma\hbar v} \sum_{\pi l m}
(-1)^m \sqrt{2l+1} \left(\frac{\omega}{c}\right)^l G_{\pi l m} \left(\frac{c}{v}\right)
K_m\left(\frac{\omega b}{\gamma v}\right) \left<I_fM_f\right|M(\pi l -m\left|
I_i M_i\right>,
\label{eq:afirelat}
\end{equation}
where the functions $G_{\pi lm}$ can be expressed in terms of 
the associated Legendre polynomials, see \cite{WintherA79}.

This formula is the basis for the analysis of electromagnetic excitation.  
The dependence on the 
electromagnetic multipolarity $\pi l$, the impact parameter $b$ and 
the Lorentz factor $\gamma$ is clearly exposed. We note that 
the above equation is a first order result, i.e. 
the strength parameter $\chi$ should be much smaller than one. 
In the straight line relativistic case
we can write $\chi$ as (see  Eq.~(1.6) of \cite{WintherA79})
\begin{equation}
\chi \sim \frac{Z_1 e <f|M(E \lambda )|i>}{\hbar v b^{\lambda}}.
\label{eq:chidefined2}
\end{equation}
We may define an {\em intermediate energy region}
where $\gamma v/c =\gamma\beta =\sqrt{\gamma^2-1}=
\lesssim$ 1--2. In this region there are relativistic effects,
see \cite{AguiarAB90}, but also effects from the Coulomb repulsion.
This region  is of special interest to us: 
at GSI we have $\gamma\beta \sim$ 1--2, 
at MSU, RIKEN, and  GANIL
there is $\beta\gamma \lesssim 0.5$.

While in the nonrelativistic theory the Coulomb repulsion 
of the ions is included in the Rutherford trajectory, the 
problem for the relativistic case (where we have assumed a 
straight line trajectory) is more delicate.
A very interesting approach to treat the classical trajectory in a
relativistic Coulomb problem can be found in \cite{AguiarAB90,AleixoB89}.

It was shown in \cite{WintherA79} that a simple redefinition of the
impact parameter accounts well for Coulomb repulsion:
\begin{equation}
b'=b+ a_0 \frac{\pi}{\gamma}.
\end{equation}
An alternative procedure is suggested in \cite{CharagiG90,BertulaniCG02} where
the following substitution is made:
\begin{equation}
b'=a_0+\sqrt{a_0^2+b^2}.
\end{equation}
This amounts to replacing the asymptotic impact parameter $b$
by the distance of closest approach in small angle  scattering.  
The quantity $a_0$ is given in the NR, as well as, in the  relativistic case
by $a_0=\hbar \eta/(kv)$ with $\hbar k=\gamma m_a v$ the momentum of the
projectile nucleus. Recently the effects of retardation on the electromagnetic
excitation in intermediate energy collisions were carefully
studied in \cite{BertulaniSMD03}

We note some further properties:
The high energy $\gamma \rightarrow \infty$ limit is investigated
in \cite{WintherA79}. The question of which value of the magnetic quantum
number $m$ gives the largest
contribution to a given transition depends on the value of $k$. 
For small values of $k$ the 
amplitudes for $m=\pm l$ are largest. The excitation amplitude 
is proportional to $1/b^l$.  Integrating over $b$ this leads to the  
logarithmic rise of the  cross section  with $\gamma$
only for $l=1$, i.e., for dipole transitions ($E1$, $M1$).
Higher multipoles on the other hand tend 
to a constant in this approximation (i.e. neglecting the 
$m=\pm1$ contribution). For large impact parameters
$kb>l$ the largest contribution is due to $m=\pm1$ . This 
contribution corresponds to the Weizs\"acker-Williams approximation,
where the excitation corresponds to the absorption of a photon 
traveling in the beam direction with helicity $m=\pm 1$ (see the 
Eqs.~(2.20)--(2.23) in \cite{WintherA79}). 
Now there is the logarithmic rise of the cross section
for all multipolarities, albeit only for extremely high
values of $\gamma$.

{\bf 3. Validity of the Semiclassical Method:}\\
It is well known that for Coulomb parameters 
$\eta\gg 1$ the semiclassical approximation can be used.
This can be seen from the following argument: two different transverse 
momenta $q_T$ are associated with an impact parameter $b$: 
The  Coulomb repulsion gives a (classical) momentum transfer
to the nucleus perpendicular to the beam direction. It is given by 
$q_{T,Coul}=2\eta\hbar/b$. Due to quantal diffraction we have
a transverse momentum 
$q_{T,diff}= \hbar/b$ associated with the impact parameter $b$.
The semiclassical approximation is appropriate 
as long as the Coulomb push $q_{T,Coul}$ is much larger than 
the uncertainties associated with the wave nature of the particle, i.e.,
$q_{T,Coul} \gg q_{T,diff}$, or $\eta \gg 1$.
A more refined derivation of this condition can be found in \cite{Bohr48}
and also in App.~B of 
\cite{AlderW75}. Another condition for the validity of the 
semiclassical approximation is $\Delta E/E \ll1$, i.e., the
energy loss due to the excitation is small. This is well fulfilled
due to the adiabaticity condition Eq.~(\ref{eq:xidefined}).

From the criterion for the validity of the semiclassical approximation
(i.e. $\eta \gg 1$) we find:
for highly charged ions with $Z_1 Z_2> 137$ (like the Pb--Pb system) 
the semiclassical approximation is valid
for all velocities $v$, whereas in general we should satisfy the condition
$Z_1 Z_2>137 v/c$ or $v/c< Z_1 Z_2/137$; for sufficiently low 
velocities the semiclassical approximation becomes always applicable.

\subsection{\it Quantum Mechanical Treatment of Projectile Motion: PWBA
and Glauber Method}
For cases where the Coulomb parameter $\eta$ is smaller than one
the semiclassical approximation is not valid. Well-known prominent examples 
are electron or proton scattering on light target nuclei.
In these cases one uses fully quantal approaches like PWBA 
or Glauber theory \cite{BertulaniB88,TypelWB97}. The nuclear interaction
with the nucleus can often be taken into account
using the black disk
approximation. Characteristic diffraction effects arise, 
see also \cite{MuendelB96,BaurR96} and \cite{TypelWB97}. 

For $\eta \gg 1$ one can obtain the semiclassical limit
starting from a quantum-mechanical approach \cite{AlderW75}.
In Glauber theory this can be done as follows, e.g., see
\cite{BertulaniN93}
(see also Eqs.~(73)-(80) of \cite{BertulaniCG02}).
The excitation amplitude has the form
(see Eq.~(73) of \cite{BertulaniCG02})
\begin{equation}
f^{\mu}_{inel}(\theta)=ik\int_0^\infty db \: b \: J_\mu(qb)\exp [i\chi(b)]
\: a_\mu(b).
\label{eq:eikonalf}
\end{equation}
The eikonal phase for a Coulomb potential
can be written as $\chi(b)=2\eta ln(kb)$. The inelastic 
excitation amplitude is denoted by $a_\mu$ (cf. Eq.~(\ref{eq_anformalsol})
or~(\ref{eq:anrelativistic}))
where $\mu$ denotes the $z$-component of the angular 
momentum transfer to the nucleus.
The integrand in Eq.~(\ref{eq:eikonalf}) oscillates rapidly as a function 
of $b$.
In the semiclassical approximation we obtain the main contribution
for those values of $b$ where the phase is stationary.

We can approximate the Bessel function for large values of $qb$ as
\begin{equation}
J_\mu(qb)\approx \frac{1}{\sqrt{2\pi qb}}
\left[\exp\left(iqb-i\pi\frac{\mu+1/2}{2}\right)
+ \exp\left(-iqb+i\pi\frac{\mu+1/2}{2}\right)\right].
\end{equation}

We use the stationary phase (saddle point) approximation
\begin{equation}
\int db G(b) \exp\left[ i\Phi (b)\right]
\approx \sqrt{\frac{2\pi i}{\left|\Phi^{''}(b_0)\right|}}
G(b_0) \exp(-i\pi/4) \exp\left[ i\Phi(b_0) \right],
\end{equation}
where a phase $\exp(-i\pi/4)$ needs to be chosen for our case, where 
$\Phi^{\prime\prime}(b_0)=- q^2/(2\eta) < 0$
and which is valid for slowly varying functions $G(b)$.
The point of stationary phase is denoted by $b_0$, it is determined from
the condition 
\begin{equation}
\Phi^{'}(b_0)=0. 
\label{eq:statphasecond}
\end{equation}
With
\begin{equation}
\Phi(b)=-qb+2\eta ln(kb).
\end{equation}
the condition Eq.~(\ref{eq:statphasecond}) leads to 
\begin{equation}
b_0=\frac{2\eta}{q},  
\end{equation}
i.e., the classical
connection between impact parameter and scattering angle, or 
momentum transfer. 
We obtain
\begin{equation}
f^{\mu}_{inel}(\theta)\approx \frac{ik}{\sqrt{q}} 
\sqrt{\frac{i b_0}{\left|\Phi''(x_0)\right|}}
\exp\left[i\chi(b_0)+i\pi \mu/2 \right] a_\mu(b_0),
\label{eq:eikonalscl}
\end{equation}
Finally we obtain
\begin{equation}
\frac{d\sigma_{inel}}{d\Omega}=\frac{d\sigma_{Rutherford}}{d\Omega} P(b_0),
\end{equation}
where $P(b_0)=\frac{1}{2I_i+1} \sum_{m_i,m_f}|a_{\mu}|^2$, analogous
to the expression in the nonrelativistic case \cite{AlderBHM56,AlderW75}.

The main difference of electromagnetic excitation in hadron-hadron 
scattering, as opposed to electron- (lepton-) hadron scattering,
is due to the strong absorption at impact parameters less 
than the sum of the two nuclear radii. Thus, for $\eta \gg 1$ the 
semiclassical method is very suitable to take these effects into account,
because the impact parameter is the relevant variable and not
the momentum transfer.
Neglecting strong absorption effects it can be shown that 
(first order) SCA and PWBA give identical 
results for total cross sections (where the limit of the
momentum transfer $q_{max}$ must be extended to 
$\infty$; this is in general a good approximation
as the contribution to the cross section for large $q$ is rather small
). For the nonrelativistic case this is shown, e.g., in 
\cite{BetheJ68},
the relativistic generalization is given by \cite{BertulaniB88}.
Likewise the equivalence of the coupled channel semiclassical theory 
and Glauber theory (i.e., with the electromagnetic interaction treated to 
all orders) is given in \cite{Baur91c}.
It should be noted however that {\em differential cross sections} are 
different in the PWBA and SCA approaches, and thus the appropriate
theory should be used in each individual case. 

\begin{figure}
\begin{center}
\resizebox{0.25\textwidth}{!}{%
\includegraphics{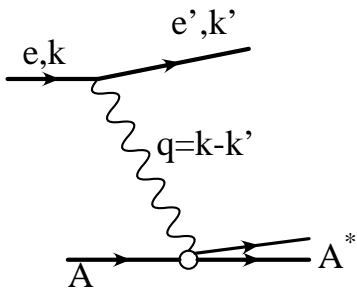}}
\end{center}
\caption{One-photon excitation e.g. in the  $(e,e')$ reaction. In  PWBA
there is a definite four-momentum transfer $q_{\mu}=
k_{\mu}-k'_{\mu}$ to the nucleus. In contrast to this there is no definite
value of $q$ in the Coulomb excitation
due to the large number of additionally exchanged elastic photons.}
\label{fig:eeprime}
\end{figure}
In the PWBA there is a definite four-momentum transfer $q_{\mu}=
k_{\mu}-k'_{\mu}$, that is, a definite energy transfer $\omega=q_0$ and
a definite three-momentum transfer 
$\vec{q}=(\vec{q_T},q_l)$. 
To a good approximation $q_l$ is given by the  minimum momentum transfer  
$q_l=q_{min}= \omega/v$
for small angle scattering and small energy loss. The invariant mass
of the photon is $Q^2=-q^2=q_T^2+(\omega/(\gamma v))^2$.
We always have $Q^2>0$: the exchanged photon is {\em virtual} (spacelike).

In contrast to this, the photon, which is exchanged in Coulomb
excitation, contains a sum over virtual photon momenta; they conspire 
in such a way that only an electromagnetic matrix-element survives,
which corresponds to the interaction with a real photon $Q^2=0$
(see appendix for details).   

In principle, heavy ion electromagnetic excitation can also be treated 
within field theory. This is however rather complicated and has to our 
knowledge never fully been done in practice. Due to the validity of the 
semiclassical approximation it is also not 
necessary to do so. Let us indicate how one would proceed
\cite{AsteBHT03,BaurHAT03}:
There are infinitely many graphs which have elastic photon exchanges 
between the two ions. For small angle scattering and high energies one 
can make use of an relativistic eikonal approach \cite{LevyS69,HayotI72}.
In this approach infinitely many Feynman graphs are summed.
The result can again be expressed in terms of an expression in impact parameter
space 
with a relativistic eikonal $\chi$. 
The semiclassical
approach can be recovered in exactly the same way as already explained above.
From the saddle point (stationary phase) approximation one obtains again the
relation $b=2\eta/q$.
For $b<b_{min}=R_1+R_2$ there is nuclear contact and and graphs which involve 
the strong interaction have to be added, see \cite{HayotI72}.
In the case of nucleus-nucleus collisions the black disk model can be 
applied. This means that the eikonal $\chi$ is a very large 
imaginary number for $b<b_{min}=R_1+R_2$.

\section{Higher Order Effects}
\label{sec:higher}

The importance of higher order effects is governed by the strength
parameter $\chi$. This parameter is proportional to $1/v$, see
Eq.~(\ref{eq:chidefined}) and Eq.~(\ref{eq:chidefined2}) above.  For
slow collisions the Coulomb field acts for a long time and this
parameter can be quite large. E.g., rotational bands in deformed
nuclei can be very well excited in this way and this has been a
fruitful field of research \cite{AlderW75}.  For collisions with $v
\lesssim c$ higher order effects tend to be small.  The strength
parameter $\chi$ will soon reach its lower limit, which we have for
$v=c$ in Eq.~(\ref{eq:chidefined}) and Eq.~(\ref{eq:chidefined2}).

Let us deal in the first subsection with a very interesting higher
order electromagnetic excitation process at intermediate and
relativistic energies: the excitation of the Multiphonon Giant Dipole
Resonance, especially the Double Phonon Giant Dipole resonance
(DGDR). In the following subsection we will discuss a more disturbing
effect: corrections of one-photon exchange effects due to multiphoton
exchange (a similar problem exists for $(e,e')$ scattering named {\em
dispersion corrections} see, e.g. \cite{EisenbergG88}).

\subsection{\it Multiphonon Giant Dipole Resonances}
\label{sec:higher:dgdr}

The Giant Dipole Resonance (GDR) is a collective mode par
excellence. This can directly be seen from the fact that the
energy-weighted TRK-sum rule (Eq.~(\ref{eq:TRKsum}) below) is
exhausted by this mode.  With its large $B(E1)$-value this transition
is very strongly excited electromagnetically in intermediate energy
heavy ion collisions. For heavy systems the excitation probability in
grazing collisions is of the order of 30~percent. From this one can
conclude directly that higher order effects must play a role.  This
was the situation in the middle of the eighties, and the question was
to which states does the GDR couple most strongly. There is the
Axel-Brink hypothesis that a GDR is built on any nuclear (excited)
state. In this spirit a multiphonon harmonic oscillator model was
adopted in \cite{BaurB86,BraunMunzinger85} and higher order
electromagnetic excitation was calculated within this model. In the
harmonic oscillator model the interaction can be included to all
orders and a Poisson distribution of the multiphonon states is found,
as is well known, see, e.g., \cite{AlderW75}.  We give some key steps
in the derivation of this result, see also \cite{Merzbacher70}.

In terms of the corresponding creation and destruction operators 
$a^{\dagger}$ and $a$ the Hamiltonian of a harmonic oscillator
is given by 
\begin{equation}
H=\hbar\omega(a^{\dagger}a+\frac{1}{2}),
\label{eq:hamil}
\end{equation}
where $\omega$ denotes the energy of the oscillator.  We have the
boson commutation rules $[a,a^{\dagger}]=1$ and
$[a,a]=[a^{\dagger},a^{\dagger}]=0$, respectively.  Only one mode is
shown explicitly, in general one has to sum (integrate) over all the
possible modes.

We assume that the interaction $V$ is linear in the creation and
destruction operators:
\begin{equation}
V(t)=f(t) (a+a^{\dagger}).  
\end{equation}
In this case we can calculate $\tilde V$ (Eq.~(\ref{eq:Vtilde}))
explicitly using the boson commutation rules given above and the
expansion
\begin{equation}
\tilde V(t)= V(t)+ it [H_0,V(t)]+\frac{(it)^2}{2!}[H_0,[H_0,V(t)]]+ \cdots .
\end{equation}
One finds
\begin{equation}
\tilde V(t)=f(t) \left(a^{\dagger} e^{- i\omega t} +a e^{i\omega t}\right).
\end{equation}
Now we can convince ourselves that the commutator of $\tilde V$ at
different times $t$ and $t'$ is a pure $c$-number. In this case we can
disregard the time ordering operator in Eq.~(\ref{eq_anformalsol}) and
obtain an exact analytical answer, up to an unimportant overall phase
factor (for details see, e.g., \cite{AlderW75}).

This leads to the excitation of a so-called
coherent state, see \cite{Glauber63,KlauderS85}.  For the excitation
of multiphonon states this is explicitly shown, e.g., in
\cite{BertulaniB88c}. One has
\begin{equation}
a_n=<n|e^{-i(u*a^{\dagger}+ua)}|0>=
\frac{(-iu^*)^n}{\sqrt{n!}}e^{-\frac{1}{2}u u^*},
\label{eq:sudden}
\end{equation}
where $u$ is the  $c$-number 
\begin{equation}
u=\int_{-\infty}^{\infty}dt f(t)\exp(- i\omega t)
\end{equation}
In order to show this the operator identity $e^{A+B}=e^A e^B
e^{-\frac{1}{2}[A,B]}$ was used, which is valid for two operators $A$
($=-iu^*a^{\dagger}$) and $B$ ($=-iua$) for which the commutator is a
$c$-number.

\begin{figure}[tb]
\begin{center}
\resizebox{0.3\textwidth}{!}{%
\includegraphics{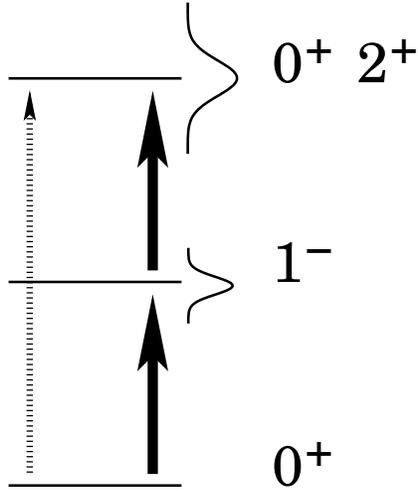}
}
\end{center}
\caption{
One- and two-photon excitation of the DGDR.
The one-photon excitation to the DGDR is forbidden 
($0^+ \to 0^+$) or strongly hindered ($0^+ \to 2^+$). 
}
\label{fig:DGDR1}
\end{figure}
The excitation probability $P_N(b)$ of an $N$-phonon state is given by
\begin{equation}
P_N(b)=\frac{\left[P^{(1)}(b)\right]^N \exp\left[-P^{(1)}(b)\right]}{N!},
\end{equation}
where $P^{(1)}(b)$ is the excitation probability of the GDR for an
impact parameter $b$ calculated in first order perturbation theory. 
It is given by \cite{BertulaniB88}
\begin{equation}
P^{(1)}(b) =\frac{2\alpha^2Z_1^2N_2Z_2}{A_2m_N\omega}\frac{1}{b^2},
\label{eq:phi}
\end{equation}
where the energy $\omega$ of the resonance is given by $\omega=80
A^{-1/3}\mbox{MeV}$ and where $N_2$ and $A_2$ are the neutron and mass
number of the nucleus, which is excited.

Certainly the most difficult question at that time was the width of
the multi-phonon states (It still is today.)

\begin{figure}[tb]
\begin{center}
\resizebox{0.5\textwidth}{!}{%
\includegraphics{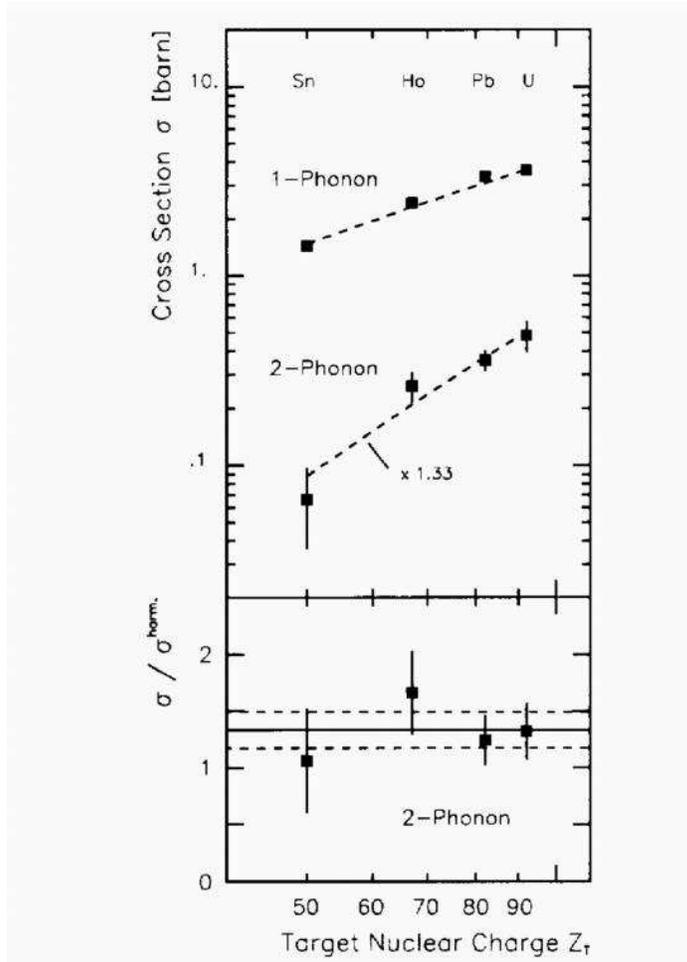}
}
\end{center}
\caption{(upper panel) Integrated experimental 
cross sections for the one-phonon GDR and the DGDR
in $^{208}$Pb obtained from different targets 
of nuclear charge $Z$. In case of the DGDR the calculated 
values are multiplied by a factor of 1.33 to fit to the experimental data.
(lower panel): ratio of the experimental cross sections
for the DGDR in $^{208}$Pb to the harmonic approximation.
The mean value and its error 
are indicated by solid and dashed lines , respectively.
Taken from \protect\cite{Boretzki96}, Fig.~2 there,
where further details can be found.}
\label{fig:DGDR2}
\end{figure}
Subsequently detailed experiments \cite{Ritman93,Schmidt93} 
were carried out at GSI and the
two-photon excitation of the DGDR was clearly observed in a two step excitation
process, see Fig.~\ref{fig:DGDR1}.  Generally, the experiments also confirmed the
simple theoretical approach. In Fig.~\ref{fig:DGDR2} we show the
dependence of GDR and DGDR excitation on the charge number $Z$ of the
target. For the GDR we have a one-photon exchange, i.e., there is a
$Z^2$ dependence. For the DGDR we have a two-photon exchange, i.e., a
$Z^4$ dependence, essentially in agreement with the experimental
result \cite{Boretzki96}.

However, interesting differences to the harmonic model showed up,
which motivated various groups to do more detailed theoretical
calculations.  Most conspicuously, the width of the DGDR tends to be
somewhat less than twice the width of the one-phonon GDR. The DGDR is
a remarkable example of a large amplitude collective motion with
damping.  We refer in this respect to the reviews \cite{AumannBE98}
and \cite{Bertulani02}, where also further references can be found.
The influence of the damping on the GDR- and DGDR-excitation cross
sections was investigated in \cite{BaurBD00}, a more sophisticated
approach was adopted in \cite{GuW01}. This work also puts the approach of
Carlson et al. \cite{Carlson99a,Carlson99b} on a firmer theoretical basis.
A schematic model which treats the coherent (collective) excitation modes 
and their coupling to more complicated states was considered in
\cite{BaurB86e}.

It would also be very interesting to study the DGDR in deformed nuclei
like $^{238}$U. What will the splitting of the DGDR be due to the
nuclear deformation? What is the influence of the nuclear shape on the
minimum impact parameters? Clearly, a closer look at these problems
would be very interesting in the future.

Due to the very collective nature of the GDR, the electromagnetic
couplings in this case are very strong. As we have seen in this
section, interesting effects follow from this.  Even at the
relativistic heavy ion colliders RHIC and LHC the GDR, and possibly
also DGDR, play an interesting role, which is briefly discussed in
Sec.~\ref{ssec:rhic}.

Finally we mention that multiphonon states were also reviewed in
\cite{ChomazF95}, where emphasis is also put on the nuclear excitation
mechanism (which becomes important for lower beam energies) and
quadrupole phonons. Multiphonon states were first investigated in
pion-charge exchange reactions \cite{MordechaiM91}.

\subsection{\it Corrections to One-Photon Exchange} 
\label{sec:higher:corr}

Now we turn to cases where the higher order effects are less
pronounced. But still they are there and they are essentially a
correction to the most interesting dominant first order effect.  There
are various methods to take higher order electromagnetic effects into
account theoretically.  If the coupling is strong, a coupled channels
approach is appropriate (see Eq.~(\ref{eq:adotcc})), sometimes higher
(especially second) order perturbation theory is sufficient. There is
a sum over all intermediate states $n$, which are (considered to be)
important \cite{AlderW75}.  Another approach is to integrate the
time-dependent Schr\"odinger equation directly for a given model
Hamiltonian \cite{MelezhikB99,EsbensenBB95,Utsunomia99,TypelW99}.
In this way all orders are taken into account.

If the collision is sudden ($\xi \ll 1$), one can neglect the time
ordering ${\cal T}$ in the usual perturbation approach, see
Eq.~(\ref{eq_anformalsol}) ({\em sudden approximation}). The
interaction is again summed to infinite order.  In order to obtain the
excitation cross section one has to calculate the matrix element of
this operator between the initial and final state. Intermediate states
$n$ do not appear explicitly.

In order to see the typical effects and how they depend on the
relevant parameters we now study higher order effects in the simple
model of \cite{TypelB94a}.  Let us briefly describe this model. It
will be discussed also in the following section, where this model is
discussed in the so-called post-form DWBA ({\em distorted wave Born
approximation}).  (The method used here is more closely related to the
{\em prior-form DWBA.})

We assume a target nucleus with charge $Z$ (to avoid unnecessary
complications we assume an infinite mass for a pedagogical discussion,
this is however not essential). We want to describe the breakup of a
bound state $a=(c+n)$ with binding energy $E_0$ in the Coulomb field
of this target. There is the Coulomb interaction $V_c=Z_c Z e^2/r_c$
between the target and the core $c$.  We assume a zero range
interaction between the core and the neutron. This can be viewed as a
deep square well potential with depth $V$ extending up to a radius $a$
with $a^2 V$ held constant when $a \rightarrow 0$.  In this potential
an $s$-wave bound state can exist.

In order to calculate the breakup process $Z+(c+n) \rightarrow Z+c+n$
we assume that the $(c+n)$-system moves on a straight line trajectory
with velocity $v$ and impact parameter $b$ in a semiclassical model.
We restrict ourselves to electric dipole transitions. The $E1$ and
$E2$ effective charges are given by
$Z^{(l)}_{eff}=Z_c \left[\frac{m_n}{m_n+m_c}\right]^l$,
$l=1,2,\cdots$.  Due to the smallness of $Z^{(2)}_{eff}$ it is a good
approximation to neglect the $E2$ transitions for the neutron-core
case (This is in contrast to the proton-core case to be discussed
below).  Analytical results were obtained for $1^{st}$ and $2^{nd}$
order electromagnetic excitation for small values of the adiabaticity
parameter $\xi$.  We are especially interested in collisions with
small impact parameters, where higher order effects tend to be larger
than for the very distant ones. In this case the adiabaticity
parameter $\xi$ is small. For $\xi=0$ ({\em sudden approximation}) we
have a closed form solution.
(This is a special example of the general formula, see
Eq.~(\ref{eq_anformalsol}) but without the time-ordering operator). 
In Eq.~(37) of \cite{TypelB94a} the angle
integrated breakup probability is given.  We expand this expression in
a scaling parameter
\begin{equation}
y=\frac{2 Z Z_c e^2 m_n}{\hbar v (m_n+m_c) b \kappa}
= \frac{m_n \eta}{(m_n+m_c) b \kappa},
\label{eq:ydefined}
\end{equation}
where the parameter $\kappa$ is related to the binding energy $E_0$ by
\begin{equation}
E_0=\frac{\hbar^2 \kappa^2}{2\mu}.
\label{eq:kappadefined}
\end{equation}
and the reduced mass $\mu$ is given by
\begin{equation}
\mu = \frac{m_n m_c}{m_n+m_c}.
\end{equation}

This parameter $y$ is directly related to the strength parameter
$\chi$.  We define another scaling parameter
\begin{equation}
x=\frac{q_{rel}}{\kappa},
\label{eq:xdefined}
\end{equation}
where the wave number $q_{rel}$ is related to the energy $E_{rel}$
of the continuum final state by
\begin{equation}
E_{\rm rel}= \frac{\hbar^2 q_{rel}^2}{2\mu}.  
\end{equation}
In leading order (LO) we obtain
\begin{equation}
 \frac{dP_{LO}}{dq_{rel}}=\frac{16}{3\pi \kappa} y^2 \frac{x^4}{(1+x^2)^4}.
\label{eq:PLO}
\end{equation}
The next to leading order (NLO) expression is proportional to $y^4$
and contains a piece from the 2$^{nd}$ order $E1$ amplitude and a
piece from the interference of 1$^{st}$ and 3$^{rd}$ order. We find
\cite{TypelB01}
\begin{equation}
 \frac{dP_{NLO}}{dq_{rel}}=\frac{16}{3\pi\kappa} y^4 
 \frac{x^2 (5-55x^2+28 x^4)}{15 (1+x^2)^6}.
\label{eq:PNLO}
\end{equation}
These are very simple and transparent results for the Coulomb
dissociation of neutron-halo nuclei. There are two scaling parameters
$x$ and $y$. The parameter $x$ controls the shape of the relative
energy distribution, whereas the parameter $y$ plays the role of the
strength parameter. We have given these formulae explicitly, since
they show most of the important features in a simple analytical
formula: as is to be expected, $y$ is proportional to the target
charge $Z$ and the effective $E1$-charge $Z_c m_n/(m_n+m_c)$.  It is
proportional to $1/v$, again this is the special case of a general result.
We also have the $1/b$ dependence characteristic of $E1$
transitions.

\begin{figure}[tb]
\begin{center}
\resizebox{0.6\textwidth}{!}{%
\includegraphics{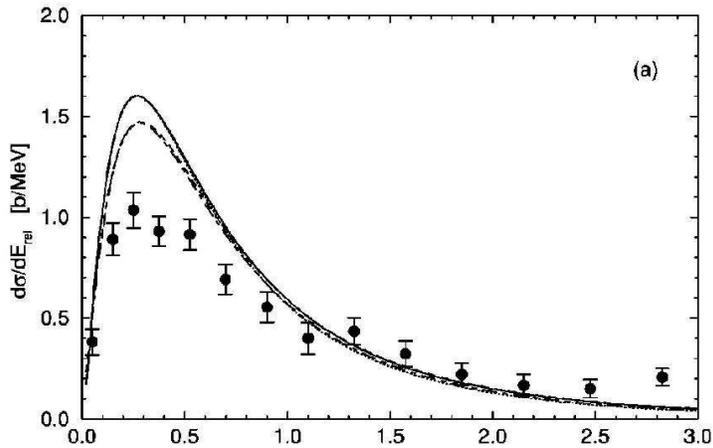}
}
\end{center}
\caption{Differential cross sections integrated over scattering
angle from $0^0$ to $3^0$ for the Coulomb dissociation of 67~$A$MeV
$^{19}$C scattered on $^{208}$Pb as a function of the relative energy.
The solid line shows the results of the LO expression
(Eq.~(\protect\ref{eq:PLO})), the dashed one includes also the NLO
contribution (Eq.~(\protect\ref{eq:PNLO})). The dotted line shows
the results of a semiclassical calculation for $E1$ only in first order,
the dashed-dotted line (which almost coincides with the (dashed) 
LO+ NLO results) for the numerical solution of the Sch\"odinger 
equation with $E1$. Both 1$^{st}$ order calculations (solid and dotted line)
are also in close agreement with each other.
The points are the experimental results from 
\protect\cite{Nakamura99}. 
Reproduced from Fig.~3 of \protect\cite{TypelB01}. 
Copyright (2001) by the American Physical Society.}
\label{fig:typelb01}
\end{figure}
The integration over $x$ and the impact parameter $b$ can also be
performed analytically in good approximation, for details see
\cite{TypelB01}. We can insert the corresponding values for the
Coulomb dissociation experiments on $^{11}$Be and $^{19}$C
\cite{Nakamura94,Nakamura99} in the present formulae.  We find that
the ratio of the NLO contribution to the LO contribution in the case
of Coulomb dissociation on $^{19}$C \cite{Nakamura99} is given by
$-2\%$.  This is to be compared to the results of \cite{Tostevin99},
where a value of about $-35\%$ was found (A further discussion of this
is found in \cite{TypelB01}). A version of the theoretical model used
in \cite{Tostevin99}, see also \cite{TostevinRJ98}, is discussed below
in Sec.~\ref{sec:analytic}.

In Fig.~\ref{fig:typelb01} we show a comparison of theoretical and
experimental differential cross sections for Coulomb dissociation of
$^{19}$C , see Fig.~3 of \cite{TypelB01}. In this figure, LO and NLO
results (Eqs.~(\ref{eq:PLO}) and~(\ref{eq:PNLO})) are also compared to
full dynamical calculations (solution of the time-dependent
Schr\"odinger equation).

Let us now deal with the breakup of a proton-core system as another
typical case.  Examples are $^8\mbox{B}={}^7\mbox{Be}+p$ or
$^{17}\mbox{F}={}^{16}\mbox{O}+p$. In this case, the $E1$-effective
charge is small and the $E2$-effective charge is large, as compared to
the neutron-core case.  The effective charges are given by
$Z^{(l)}_{eff}=Z_c(\frac{-m_p}{m_p+m_c})^l
+Z_p(\frac{m_c}{m_p+m_c})^l$ $(l=1,2,...)$ where $m_p$ and $m_c$ are
the masses of the proton and the core, $Z_c$ is the charge number of
the core and $Z_p=1$.  Also in view of the importance of the $^8$B
Coulomb dissociation for the solar neutrino problem (more see below in
Sec.~\ref{sec:astro}) this case has been studied extensively with all
the methods available, which we now briefly survey. In most models,
some nuclear structure quantities enter and it is a pity that somewhat
different assumptions (for example about $E1$ or $E2$ matrix elements)
are used in the different calculations by the various groups.  This
makes the comparison of the different reaction calculations somewhat
more difficult.  How does one separate the effects from the nuclear
structure input and the effects of the different approximations in the
reaction theory? Yet, quite a conclusive picture has emerged.  Using
perturbation theory, higher order effects in the Coulomb dissociation
of $^8$B were studied in \cite{TypelB94b,TypelWB97}. This approach
works best for the high beam energies. Results are given in
Fig.~\ref{fig:typel97}, see Sec.~\ref{sec:astro} below.

The CDCC ({\em Coupled Discretized Continuum Channels}) method was
also applied, for recent references see
\cite{MortimerTT02,MatsumotoKOI03}.  The effects of $E2$ and higher
order excitations are included.  Since the relative motion of the
projectile and target, with its huge number of partial waves is
treated in a quantum mechanical way, this kind of method becomes
numerically more and more involved for high energies.  In
\cite{Marta02} it was seen that (not unexpectedly, since the
corresponding $\eta$-value is much larger than 1) the semiclassical
approximation compares well to corresponding fully quantal
calculations at sub-barrier energies.  This is a great simplification
in the numerical evaluation.  At higher energies, where also nuclear
(grazing) collisions can become important, a Glauber approach is very
useful \cite{BertulaniCG02}.

Instead of expanding the wave function into the nuclear basis states
one may also study the time-development of the nuclear wave function
directly.  With modern powerful computational methods it is now
possible to solve the time-dependent Schr\"odinger equation very
efficiently 
\cite{TokimotoEA01,EsbensenB02,EsbensenBB95,EsbensenB96,EsbensenB95}.  
It may be
considered to be a drawback of these approaches that they rely on
choosing a specific (simple) nuclear Hamiltonian, like a particle-core
model with a phenomenological interaction between them. However, it
will cover the essential aspects of the problem in most cases.  In
practice such approaches turn out to be very useful and also give us
some general insight into, e.g., the importance of higher order
effects.
\begin{figure}[tb]
\begin{center}
\resizebox{0.6\textwidth}{!}{%
\includegraphics{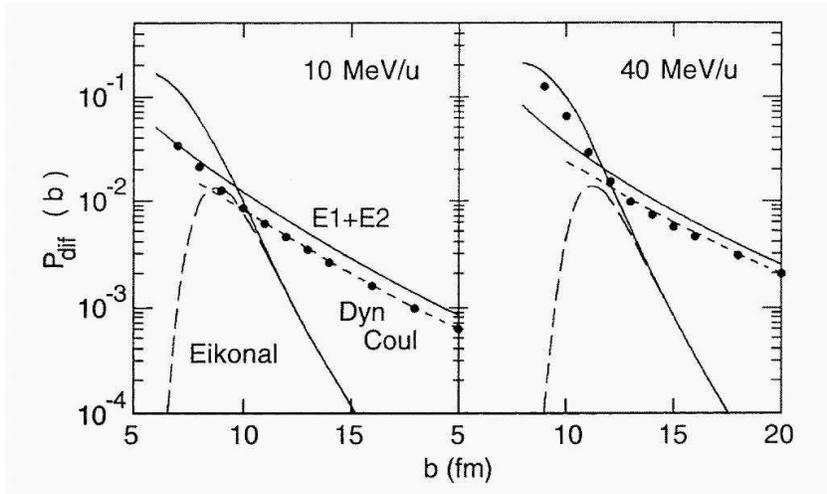}
}
\end{center}
\caption{
Diffraction dissociation probabilities of the $^{17}$F ground state at
10 and 40~$A$MeV on a $^{208}$Pb target. The points are the results of
the full dynamic calculation. The long-dashed and steep solid line is
the result of an eikonal calculation with and without core-absorption
effects. The flatter solid line is the result of a first order
calculation including $E1$ and $E2$ contribution. The short dashed
line is the result of the dynamic calculation including only the
Coulomb interaction. Taken from \protect\cite{EsbensenB02}, Fig.~5 there.}
\label{fig:dynf17}
\end{figure}

Esbensen and Bertsch \cite{EsbensenB02} investigated the two-body
($p+^{16}$O)breakup of $^{17}$F on Ni and Pb targets at
10-40 $A$MeV. Higher order, as well as, nuclear effects are important
and we report on some of their results now. A {\it ``dynamic
polarization (Barkas) effect''} is found. It is essentially due to the
interference of a (first order) $E1$ amplitude and a (second order)
$E1$-$E2$ amplitude.  The effect is proportional to $Z^3$.  In
Fig.~\ref{fig:dynf17} we show the breakup probabilities of $^{17}$F on
a Pb target at 10 and~40 $A$MeV as a function of the impact parameter.

In \cite{AlderW75} one may find various approaches to study deviations from
the sudden approach, see, e.g., the discussion in Sec.~II.3. It would be
of interest to further explore this in the future.

In summary we may say that higher order effects are well under
control.  After all we have a fair knowledge about nuclear structure
and a very good knowledge of how to treat the electromagnetic
interaction.  Due to present days' computing power many quite involved
calculations have become possible and they are essential for the
analysis of the experimental data. Simple models and analytical
solutions are also useful: they show the dependence on the various
parameters.

\section{Analytically Solvable Model for the Coulomb Breakup of Neutron
Halo Nuclei}
\label{sec:analytic}

Breakup processes in nucleus-nucleus collisions are complicated, in
whatever way they are treated. They constitute at least a three-body
problem, which is further complicated due to the long range Coulomb
force.  Exact treatments (like the Faddeev-approach) are therefore
prohibitively cumbersome.  On the other hand, many approximate schemes
have been developed in the field of direct nuclear reactions, and
these approaches have been used with considerable success
\cite{Austern70}. In this context we wish to investigate a realistic
model for the Coulomb breakup of a neutron halo nucleus.  This model
has already been mentioned in the previous section.  With the
operation of exotic beam facilities all over the world, these
reactions (previously restricted essentially to deuteron induced
reactions) have come into focus again. The Coulomb breakup of these
nuclei is of interest also for nuclear astrophysics, since the breakup
cross section can be related to the photo-dissociation cross section
and to radiative capture reactions relevant for nuclear astrophysics
\cite{BaurR96}, see Sec.~\ref{sec:astro} below.

In this section we study the Coulomb breakup of a neutron-halo nucleus
with a zero-range neutron-core interaction.  A full quantum mechanical
treatment can be done with two different treatments of the final
state. In the so-called {\em prior-form} the Hamiltonian $H$ is split
in the initial and final channel into
\begin{equation}
H=H^{prior}_f+V^{prior}_f ,
\label{eq:prior1}
\end{equation}
where 
\begin{equation}
H^{prior}_f=T+ V_a + V_{cn}
\label{eq:prior2}
\end{equation}
and 
\begin{equation}
V^{prior}_f= V_c + V_n -V_a,
\label{eq:prior3}
\end{equation}
where $V_i$ denotes the interaction of $i=c$, $n$ with the target
nucleus and $V_{cn}$ is the relative core-neutron interaction.  The
interaction $V_a$ is an optical model interaction of the $a=(c+n)$
bound state with the target nucleus.  The final state is a product of
the wave function of the center of mass (with the coordinate vector
$\vec R$) and the relative wave function of the unbound $(c+ n)$
system \cite{Austern70}.

In the so-called {\em post-form} the Hamiltonian $H$ is split in the
final channel into
\begin{equation}
H=H^{post}_f+V^{post}_f,
\label{eq:postform1}
\end{equation}
where 
\begin{equation}
H^{post}_f=T+ V_n + V_c
\label{eq:postform2}
\end{equation}
and 
\begin{equation}
V^{post}_f=V_{nc}. 
\label{eq:postform3}
\end{equation}
The interactions $V_n$ and $V_c$ are treated to all orders in this
way, whereas the relative interaction between $c$ and $n$ is only
treated in first order.  In the previous section we have studied the
{\em prior-form} DWBA (in its semiclassical limit, i.e. for large
values of the Coulomb parameter $\eta$). Accordingly we have used a
decomposition into a c.m. motion of the projectile system and its
relative wave function in the initial and final channel.

An important benefit of the present model is that it can be solved
analytically in the CWBA (neglecting the nuclear interaction of the
projectile with the target we can call the DWBA the {\em Coulomb Wave
Born Approximation} CWBA), see Sec.~\ref{ssec:CWBA}.  Analytic
expressions also exist for the PWBA limits of both prior- and
post-form, which we can compare with each other. Thus this model
constitutes an ideal {\em theoretical laboratory} to investigate the
physics of breakup reactions. We can study certain limiting cases and
its relation to the semiclassical approximation, which is mainly used
in the analysis of experiments. Especially the effect of {\em
postacceleration} (to be explained in more detail below) can be
studied in a unique way in this approach. We will first study the
plane wave limits of both the prior- and post-form and compare them
with the semiclassical limit.  The question of higher-order effects
will then be discussed in the full CWBA approach.

\subsection{\it Theoretical Model, Plane Wave Approximations 
and Qualitative Discussion}
We consider the breakup of a particle $a=(c+n)$ (deuteron,
neutron-halo nucleus) consisting of a loosely bound neutral particle
$n$ and the core $c$ (with charge $Z_a=Z_c$) in the Coulomb field of a
target nucleus with charge $Z$
\begin{equation}
a+Z \rightarrow c+n+Z.
\label{eq:1}
\end{equation}
To simplify the kinematical relations we assume in this section that
the target is infinitely heavy (this is not necessary, see below).  We
assume that the $a=(c+n)$ system is bound by a zero range force, see
Sec.~\ref{sec:higher}.  The potential $V_{cn}$ is adjusted to give one
$s$-wave bound state with a binding energy $E_0$.  Neglecting the
nuclear interaction of $c$ and $n$ with the target ({\em pure Coulomb}
case) the Hamiltonian of the system is given by
\begin{equation}
H=T_n + T_c + \frac{Z Z_c e^2}{r_c} + V_{cn}(r).
\end{equation}
The bound-state wave function of the system is given by 
\begin{equation}
\phi_0 = \sqrt{\frac{\kappa}{2\pi}} \frac{\exp(-\kappa r)}{r},
\label{eq:phigs}
\end{equation}
where the quantity $\kappa$ is related to the binding energy $E_0$,
see Eq.~(\ref{eq:kappadefined}) above.

The incoming particle $a$ has a momentum $\vec q_a$ and the momenta of
the outgoing particles $c$ and $n$ are denoted by $\vec q_c$ and $\vec
q_n$, respectively; see also Fig.~\ref{fig:2} below.  It is useful to
introduce also the relative and c.m. momenta $\vec q_{rel}$ and $\vec
q_{cm}$ which are related to $\vec q_c$ and $\vec q_n$ in the
following way:
\begin{eqnarray}
\vec q_{rel}&=&\frac{m_c \vec q_n - m_n \vec q_c}{m_n + m_c}\\
\label{eq:qrel}
\vec q_{cm}&=&\vec q_c + \vec q_n.
\label{eq:qcm}
\end{eqnarray}  
These momenta are further constrained by energy conservation
\begin{equation}
\frac{q_a^2}{2 m_a}-\frac{\kappa^2}{2 \mu}=\frac{q_c^2}{2 m_c}
+\frac{q_n^2}{2 m_n} =\frac{q_{cm}^2}{2(m_c+m_n)}+ \frac{q_{rel}^2}{2 \mu}.
\label{eq:econserved}
\end{equation}

From the experimental point of view it is interesting to mention at
this point a {\em magnifying glass effect}: for a given scattering
angles of the particle $c$ and $n$ there is a minimum relative energy
$E_{rel}^{min}$.  The relative energy $E_{rel}=q^2_{rel}/(2\mu)$ is a
small quantity, close to $E_{rel}^{min}$.
It can be determined accurately from the measurement of the momenta
$\vec q_c$ and $\vec q_n$ (which are both large), see, e.g.,
\cite{BaurBR86} and Fig.~4 of \cite{BaurR94}.

It is interesting to write down the amplitude for the breakup reaction
in Born (plane wave) approximation.  This approximation is valid for
small values of the Coulomb parameter $\eta_a$.  This parameter
characterizes the strength of the Coulomb interaction.  In the
applications considered here $\eta_a$ is usually much larger than one,
but still, this example will expose many characteristic features of
the breakup process.  In the prior-form we split the Hamiltonian, see
above, into $H=H_0 + V_{c}$ with $H_0=T+ V_{cn}$ (i.e. we we set
$V_a=0$ ).  Accordingly we have in the initial state a plane wave for
the c.m. motion and a bound state wave function; in the final state we
have again a plane wave of the c.m. motion and the relative continuum
wave function, which in the zero-range approximation is given by
\begin{equation}
\left|\vec q_{rel} \right>= \exp(i\vec q_{rel} \cdot \vec r)- \frac{\exp(iq_{rel} r)}{(\kappa+iq_{rel})r},
\label{eq:phicontinuum}
\end{equation}
i.e., only the $s$-wave is modified by the short range potential
$V_{cn}$.  Using the Bethe integral
\begin{equation}
\frac{1}{r_c}=\frac{1}{\left|\vec R -\frac{m_n}{m_a} \vec r\right|}
=\frac{1}{2\pi^2} \int 
d^3q \frac{\exp\left[i\vec q \cdot (\vec R- \frac{m_n}{m_a}\vec r\/)\right]}
{q^2},
\end{equation}
we can separate the matrix element into the relative and c.m.\/
coordinates.  The $T$-matrix in the Born approximation (PWBA) is found
to be
\begin{equation}
T^{Born}=4 \pi \frac{Z Z_c e^2}{q^2} a_{fi}(\vec \Delta p),
\label{eq:tprior}
\end{equation}
where $\Delta p$ is related to the momentum transfer (or {\em Coulomb
push}) to the target nucleus
\begin{equation}
\vec q = \vec q_a-\vec q_{cm}=\vec q_a-\vec q_c-\vec q_n
\label{eq:qcoul}
\end{equation}
by $ \vec \Delta p=(m_n/m_a) \vec q$.  This {\em Coulomb push} has a
perpendicular component $q_{\perp}$ and a component in the beam
direction $q_{\|}$. For high energies we have $q_{\|}=\omega/v$
(corresponding to the {\em minimum momentum transfer}).  The amplitude
$a_{fi}$ can be calculated analytically, see, e.g., Eqs.~(33), (34) of
\cite{TypelB94a}.  It is found to be
\begin{equation}
a_{fi}=\sqrt{8 \pi \kappa} (a_{FT} + a_S).
\end{equation}
The quantity $a_{FT}$ is essentially the Fourier transform of the
Yukawa wave function, given by
\begin{equation}
a_{FT}= \frac{1}{(\vec q_{rel}-\vec \Delta p)^2+ \kappa^2}
\end{equation} 
and $a_S$ takes the $s$-wave scattering part of the continuum wave
into account:
\begin{equation}
a_S=\frac{i(\kappa+iq_{rel})}
{2|\Delta p|(\kappa^2+q_{rel}^2)} \ln{\frac{\kappa +i(q_{rel}+|\Delta p|}
{\kappa+i(q_{rel}-|\Delta p|)}}.
\end{equation}
We are interested in forward angle scattering at high energies. The
value of $q_{\|}$ is usually small and we have $q \approx q_a \theta$
where $\theta (\ll1)$ is the scattering angle. We can introduce the
modulus of the elastic scattering Coulomb amplitude
\begin{equation}
f_{coul}= \frac{2\eta_a q_a}{q^2} \approx \frac{2\eta_a}{q_a \theta ^2}.
\label{eq:fcoul}
\end{equation}
We find $T=(2\pi/m_a) f_{coul} a_{fi}$.  For very small values of $q$,
or equivalently $\Delta p$, the terms $a_{FT}$ and $a_S$ nearly cancel
each other (the bound and the continuum wave functions are orthogonal)
and the dipole term will be dominant. We find
\begin{equation}
a_{fi}(\vec q_{rel}, \vec \Delta p) \approx
\sqrt{8 \pi \kappa} \frac{2 \vec q_{rel} \cdot \vec \Delta p}
{( q_{rel}^2+ \kappa^2)^2}.
\label{eq:afiprior}
\end{equation}
We can compare this expression with two others: the semiclassical
straight line expression \cite{TypelB94a} in Sec.~\ref{sec:higher}
above and the Born limit of the post form CWBA.

First, we recall the results of \cite{TypelB94a} and
Sec.~\ref{sec:higher} above. In this approach the semiclassical
straight line approximation is used. It is valid for high beam
energies and for $\eta_a\gg 1$. In the {\em semiclassical approach}
(corresponding to the {\em prior form} decomposition of $H$), one now
calculates ({\em impact parameter dependent}) breakup amplitudes
$a_{fi}(b)$; the breakup amplitude is given by
\begin{equation}
f_{breakup}=f_{coul} a_{fi}
\end{equation}
and the cross section is 
\begin{equation}
d\sigma/d\Omega= d\sigma_{ruth}/d\Omega\ |a_{fi}(b)|^2,
\end{equation}
where
\begin{equation}
\frac{d\sigma_{ruth}}{d\Omega}=|f_{coul}|^2.
\end{equation}
The impact parameter $b$ is related to the {\em Coulomb push} by the
semiclassical relation $b= 2\eta_a/q$.  We see that the formula for
the Born approximation is the same as the one derived for the
semiclassical sudden limit.  Please keep in mind however that the
ranges of validity of the two approaches, do not overlap: for the Born
approximation we have $\eta_a\ll 1$ while the semiclassical
approximation requires $\eta_a \gg 1$.  In the sudden limit we have
$\omega=0$ and there is only a transverse momentum transfer
$q_{\perp}$.  The breakup amplitude in the sudden limit is given by,
see Eq.~(33) of \cite{TypelB94a}
\begin{equation}
a_{fi}=\left<q\right| \exp(i \vec \Delta p \cdot \vec r) \left| 0\right>.
\end{equation}
It should be kept in mind that in \cite{TypelB94a} only the electric
dipole component is considered.  The above equation leads to
\begin{equation}
a_{fi}=i{\vec \Delta p} \cdot\left<q\right| \vec r\left|0\right>
\end{equation}
in lowest order of $\Delta p$ or $q$. This expression is the same as
Eq.~(\ref{eq:afiprior}). In this case there is no interaction between
the $c$ and $n$ in the final state $p-$wave. In higher orders of $q$,
however, this interaction will show up.

Now we compare these {\em prior form} expressions to the {\em
post-form} DWBA scheme.  In the prior form we have a c.m.-motion
(either {\em classical} or {\em quantal}) of the center of mass (with
coordinate vector $\vec R$) of the $(c+n)$-system (either in the
ground- or excited state) in the Coulomb field $Z Z_c e^2/R$. If
this potential is neglected because of the high projectile energies,
one has a plane wave or a straight line trajectory. The relative wave
function of the $(c-n)$-system is governed by the internal Hamiltonian
$H_{cn}=T_{cn}+V_{cn}$.

In the post-form the Hamiltonian in the final state is split as in
Eqs.~(\ref{eq:postform1})--(\ref{eq:postform3}).
I.e., we take the Coulomb interaction between $Z$ and $c$ fully into
account in the final state (i.e., postacceleration effects are fully
considered) and treat $V_{cn}$ to lowest order only.  In our model
there is no resonance structure in the $c+n$ continuum.  We can expect
that $V_{cn}$ is not so important.  This is clearly a good assumption
for the deuteron and will also hold for other neutron halo systems.
Similarly a c.m. Coulomb wave function is used for the initial state.
In the final state the broken up system is treated in a different way,
more appropriate if there are strong postacceleration effects. This
is, e.g., found to be the case for low energy deuteron breakup, see
\cite{BaurT72,Jarczyk72}.  For this see the next subsection.

We first give explicitly the (Born) plane wave limit \cite{BaurHTT01}
of the post-form DWBA (or CWBA). The Born approximation in this case
is obtained by substituting for the full Coulomb wave functions in
Eq.~(\ref{eq:tintegral}) below their 1$^{st}$ order expansions in
$V_{c}$.  We find (see Eq.~(3) of \cite{BaurHTT01})
\begin{equation}
T=D_0 f_{coul} 4 \pi \left[\frac{1}{q_a^2-(\vec q_n+\vec q_c)^2}+
\frac{m_c}{m_a}\frac{1}{(q_c^2-(\vec q_a-\vec q_n)^2)}\right],
\label{eq:tpost}
\end{equation}
where the {\em zero-range constant} is given by 
\begin{equation}
D_0 = \frac{\hbar^2}{2 \mu} \sqrt{8 \pi \kappa}.  
\label{eq:D0defined}
\end{equation}
As above, $f_{coul}$ is the modulus of the Coulomb scattering
amplitude Eq.~(\ref{eq:fcoul}).

This expression shows a similarity to the Bethe-Heitler formula for
brems\-strahlung $ Z+e^- \rightarrow Z+e^- + \gamma$. The
Bethe-Heitler formula has two terms which correspond to a Coulomb
interaction of the electron and the target followed by the photon
emission and another one, where the photon is emitted first and then
the electron scatters from the nucleus. Now we have a Coulomb
scattering of the incoming particle followed by breakup $a=(c+n)
\rightarrow c+n$ and another term, where the projectile a breaks up
into $c+n$, and subsequently, $c$ is scattered on the target $Z$.

\begin{figure}[tb]
\begin{center}
\resizebox{0.35\textwidth}{!}{%
\includegraphics{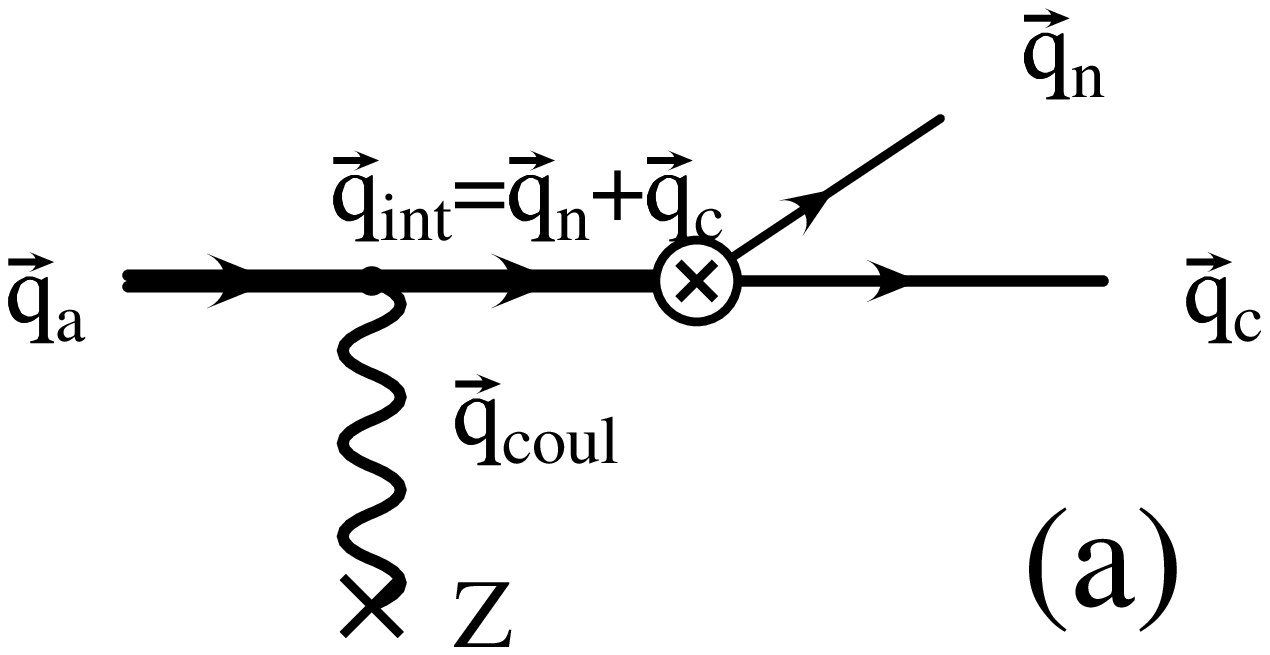}}
~~~
\resizebox{0.35\textwidth}{!}{%
\includegraphics{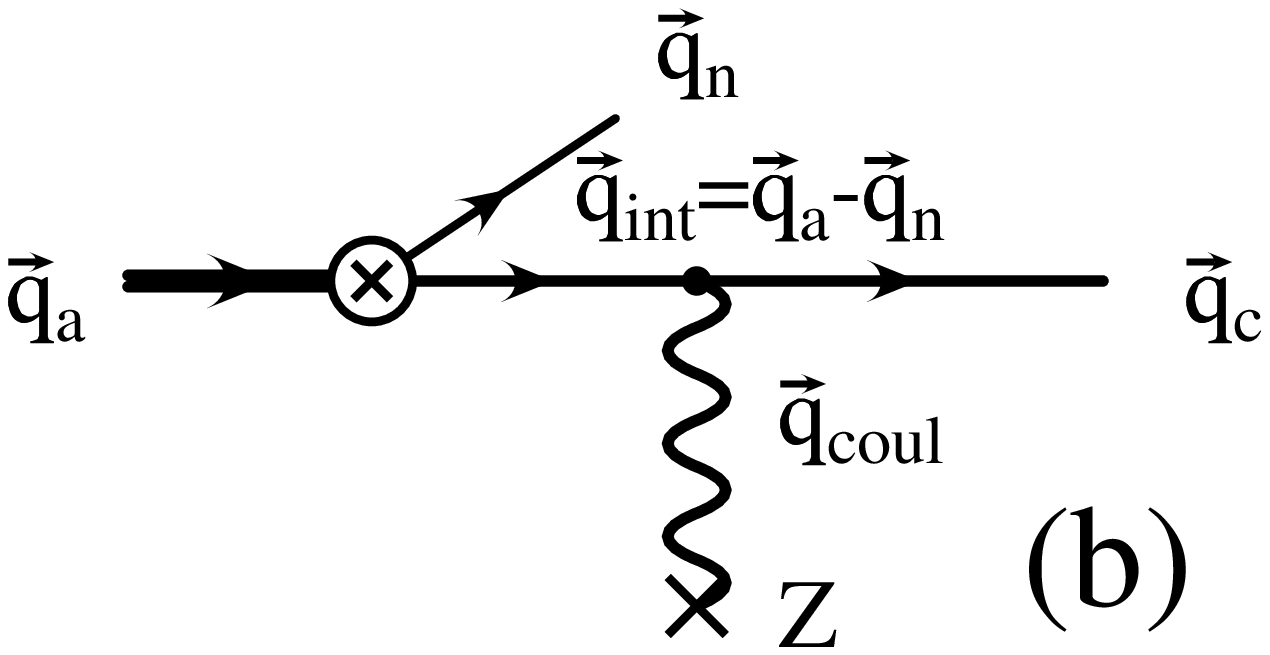}}
\end{center}
\caption{The two bremsstrahlung type of graphs, which describe the Coulomb
breakup in the post-form 
Born approximation. Three-momentum conservation at each vertex
determines the intermediate momenta $\vec q_{\textrm{int}}$.}
\label{fig:2}
\end{figure}
The two terms in the parenthesis correspond to the two graphs shown in
Fig.~\ref{fig:2}.

It is useful to rewrite the denominators of Eq.~(\ref{eq:tpost}) in
terms of the relative and c.m. variables $q_{rel}$ and $\Delta p$
(Eqs.~(\ref{eq:qrel}) and~(\ref{eq:qcoul})) as
\begin{equation}
q_a ^2- q_{cm}^2=\frac{m_a^2}{m_n m_c}(q_{rel}^2
+\kappa^2)
\end{equation}
and 
\begin{equation}
q_c^2-(\vec q_a-\vec q_n)^2=- \frac{m_c+m_n}{m_n}(\kappa^2+(
\vec q_{rel}-\vec \Delta p)^2)),
\end{equation}
where we have used the energy conservation Eq.~(\ref{eq:econserved})
to obtain this result.  For the $T$-matrix we obtain
\begin{equation}
T=\sqrt{8 \pi \kappa} \frac{2 \pi f_{coul}}{m_a}
(\frac{1}{q_{rel}^2+\kappa^2} -\frac{1}
{(\vec q_{rel}-\vec \Delta p)^2+\kappa^2}).
\label{eq:tpost2}
\end{equation}
For small values of $\Delta p$, or equivalently $q$, the two terms
almost cancel.

We can expand this equation in the small parameter $\Delta p$.  In
lowest order we obtain
\begin{equation}
T=\sqrt{8 \pi \kappa} f_{coul}\frac{2\pi }{ m_a}
\frac{2 \vec \Delta p \cdot \vec q_{rel}}{(\kappa^2+ q_{rel}^2)^2}.
\end{equation}
We can compare this to the limit of small $\vec \Delta p$ in the prior
form PWBA, Eqs~(\ref{eq:tprior}) and~(\ref{eq:afiprior}).  We see that
these limits agree with each other.

For {\em larger} values of $q$ characteristic differences will appear:
in Eq.~(\ref{eq:tprior}) the $n$-$p$ final state interaction in the
$l=0$ channel is treated properly. This is not the case for
Eq.~(\ref{eq:tpost}) or~(\ref{eq:tpost2}).  In this equation, on the
other hand the Coulomb force on the core $c$ in the final state is
treated properly.

In the Born approximation it is only the momentum transfer $\vec q=
\vec q_a - \vec q_c - \vec q_n$ and not the momentum $q_a$ itself
which enters into the matrix element.  In the semiclassical
approximation, on the other hand, the quantity $v_a=q_a/m_a$ appears
explicitly.  This quantity enters in the adiabaticity parameter $\xi$,
a concept not present in the plane wave approach.  The
(nonrelativistic) adiabaticity parameter 
(see Eq.~(\ref{eq:xidefined})) is given by
\begin{equation}
\xi= \omega b/v_a = \frac{2 \eta_a \omega/v_a}{q}
=\frac{2 \eta_a q_{\|}}{q_{\perp}}.
\end{equation}
We have $q_{\|}=\omega/v_a$ and, for not too large impact parameters,
$q_{\perp} \sim q$.  As we know, for large values of the adiabaticity
parameter, i.e. when $2\eta_a q_{\|} \gg q_{\perp}$, the breakup
probability is suppressed exponentially with $\xi$.  In the next
subsection we will also encounter a simple limit of the full CWBA
expression, very much related to the plane wave (Born) treatment of
the post form CWBA given above. It is valid for small values of $\xi$
and {\em arbitrary} values of $\eta_a$. We will also treat the case of
large values of $\eta_a$ and $\xi$ there.

It is an interesting and difficult task to specify more clearly the
ranges of validity of the post and prior formulations. In the simplest
high energy limit we have just found that there are regions where both
approaches agree with each other. For low energy deuteron breakup the
post-form CWBA is very reasonable, whereas the prior form is very poor
\cite{RybickiA71}. What about high energies?  In second order of
$\Delta \vec p$ differences show up. Which of the two approximations
is reasonable, and in which region of parameter space?  The relevant
parameters are $x$, $y$, and $\xi$, what is the range of validity of
$x$- and $y$-scaling?  In this context we mention a recent paper
\cite{AltIM03}, where final state three-body Coulomb effects are
discussed for the
$^{208}\mbox{Pb}(^8\mbox{B},^7\mbox{Be}+p)^{208}\mbox{Pb}$ Coulomb
breakup reaction.

\subsection{\it CWBA}
\label{ssec:CWBA}

In the post-form CWBA the Hamiltonian is split in two ways
corresponding to the initial and final states.  We have $H=H_i+V_i$
where
\begin{equation}
H_i = T + V_{nc}(r) + V_a(R) = T_R+T_r+ V_{nc}(r)+\frac{Z Z_c e^2}{R},
\end{equation}
i.e., the initial state separates in the $\vec r$ and $\vec R$
coordinates.  In the final state we have $H=H_f+V_f$ with
\begin{equation}
H_f = T + V_c(r_c) = T_{r_c}+T_{R_{n-(Ac)}}+\frac{Z Z_c e^2}{r_c}.
\end{equation}
The $T$-matrix for the reaction Eq.~(\ref{eq:1}) can be written as
\cite{BaurT72b}
\begin{equation}
T = \left< \chi^{(-)}_{\vec q_c} \psi_{\vec q_n} \right| V_{nc}
\left| \chi_{\vec q_a}^{(+)} \phi_0 \right>
= D_0 \int d^3R \chi^{(-)}_{\vec q_c}(\vec R) e^{-i \vec q_n\cdot\vec R}
\chi_{\vec q_a}^{(+)}(\vec R),
\label{eq:tintegral}
\end{equation}
with the {\em zero range constant} $D_0$ as given in
Eq.~(\ref{eq:D0defined}) above.

In order to compare with experimental data, one has to take the finite
target mass $m_A$ into account in a standard way.  For this one replaces the
momenta of the previous section (where the target mass was assumed to
be infinite) by the ones with the correct kinematics and the masses
$m_a$, $m_c$, and $m_n$ would have to be replaced by the reduced
masses $(m_a m_A)/(m_a+m_A)$, $(m_c m_A)(m_A+m_c)$ and
$(m_n(m_A+m_c))(m_n+m_c+m_A)$, respectively.  Note that if we would
have included also an interaction $V_{n}$ such a separation would not
have been possible.

The initial state is given by the incoming Coulomb wave function
$\chi^{(+)}_{\vec q_a}$ with momentum $\vec q_a$ and the halo wave
function $\phi_0$. The final state is given by the independent motion
of the core described by the outgoing Coulomb wave function
$\chi^{(-)}_{\vec q_c}$ in the Coulomb field of the target nucleus $Z$
with asymptotic momentum $\vec q_c$ and the free neutron with momentum
$\vec q_n$, described by a plane wave.

\begin{figure}[tb]
\begin{center}
\resizebox{0.35\textwidth}{!}{%
\includegraphics{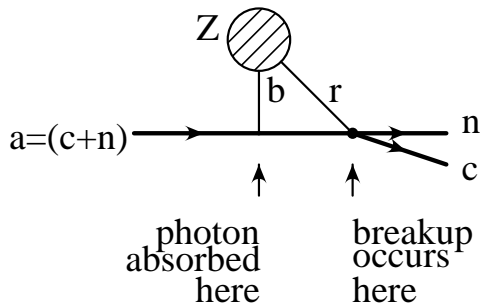}}
\end{center}
\caption{A schematic classical view of Coulomb-breakup, as adapted from 
\protect\cite{Galonsky94}. The distance from the target nucleus to the breakup 
point is denoted by $r$.}
\label{fig:1}
\end{figure}
\begin{figure}[tb]
\begin{center}
\resizebox{0.5\textwidth}{!}{%
\includegraphics{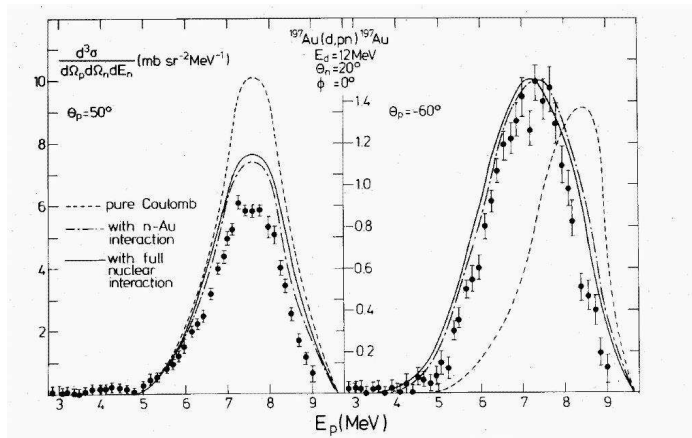}}
\end{center}
\caption{Comparison of calculations and measurement for the deuteron breakup
coincidence cross section on $^{197}$Au at $E_d=12$~MeV (Fig.~4 of
\protect\cite{BaurRT84}). The postacceleration effect can clearly be
seen, as the maximum of the proton energy ($\sim 7.5$~MeV) is larger
than the one of the neutron ($\sim 2.5$~MeV).  The experimental data
are taken from \protect\cite{Jarczyk72}.  }
\label{fig:post}
\end{figure}

In these wave functions the Coulomb interaction is taken into account
correctly to all orders.  The present {\em post-form} description,
Eqs.~(\ref{eq:tintegral}), includes therefore the
effects of {\em postacceleration}.  {\em Postacceleration} can be viewed as
a higher order electromagnetic effect, see Sec.~\ref{sec:higher} and refers to
the fact that (at low beam energies) the core $c$ has a larger final
state energy than what one would get from sharing the kinetic energy among
the fragments according to their mass ratio.  {\em Postacceleration}
arises in a purely classical picture of the breakup process. This is
nicely discussed in \cite{Galonsky94} (We show their Fig.~5 here as
our Fig.~\ref{fig:1}). The nucleus $a=(c+n)$ moves up the Coulomb
potential, loosing the appropriate amount of kinetic energy. At the
{\em breakup point} (marked as {\it ``breakup occurs here''}, see
Fig~\ref{fig:1}), this kinetic energy (minus the binding energy) is
supposed to be shared among the fragments according to their mass
ratio (assuming that the velocities of $c$ and $n$ are equal). Running
down the Coulomb barrier, the charged particle $c$ alone (and not the
neutron) gains back the Coulomb energy, resulting in its {\em
postacceleration}. Of course this picture is based on the purely
classical interpretation of this process, and will be modified in a
quantal treatment, where such a {\em breakup point} does not
exist. The correct semiclassical limit of the theory in this case can
be found, e.g., in \cite{BaurPT74}. A purely classical formula for
this postacceleration, where the {\em breakup point} corresponds to
the distance of closest approach --- i.e., $b=r$ in Fig.~\ref{fig:1}
--- is given in \cite{BaurBK95}. Postacceleration is clearly observed
in low energy deuteron breakup, in the theoretical calculations, as
well as, in the corresponding experiments, see Fig.~\ref{fig:post} and
also, e.g., \cite{BaurT76,BaurRT84}.

The formula Eq.~(\ref{eq:tintegral}) is also useful for the
description of the Coulomb dissociation of halo nuclei at high beam
energies, see \cite{ShyamBB92}. Within this theory postacceleration
effects become negligibly small in the high energy region. This is
seen in the numerical calculations \cite{ShyamBB92} and in the
analytical investigations to be described below. It can, e.g., be
applied to $^{11}$Be and $^{19}$C Coulomb dissociation experiments
\cite{Nakamura99,Nakamura94} (We disregard here the importance of
finite range effects).

On the other hand the 1$^{st}$ order semiclassical Coulomb excitation
theory was widely applied in the past years to the Coulomb
dissociation of high energy neutron halo nuclei, see, e.g.,
\cite{BaurHT01}.  The theory corresponds to the {\em prior form},
mentioned above.  The question of higher order electromagnetic effects
was studied recently in \cite{TypelB01} within this framework. These
effects were found to be small, for zero range, as well as, finite
range wave functions of the $a=(c+n)$-system.  Higher order effects
were recently studied in a post-form DWBA approach
\cite{Tostevin99,TostevinRJ98}, which was already discussed in
Sec.~\ref{sec:higher}.

We now show that the full CWBA amplitude is directly related to the
Born approximation, Eq.~(\ref{eq:tpost}) or the semiclassical result,
see Eq.~(\ref{eq:afiprior}).  This is true (with modifications to be
discussed below) not only for $\eta_a \ll 1$ but also for arbitrary
values of $\eta_a$ and $\eta_c$. For this to be the case, the beam
energy must be high (compared to the binding energy $E_0$) and the two
fragment must be scattered to forward angles.

As was discussed above, the plane-wave approximation to the post-form
DWBA has some similarity to the bremsstrahlung process.  Replacing the
neutron by a photon the diagrams of Fig.~\ref{fig:2} are analogous to
the bremsstrahlung diagrams in lowest order. The electron-photon
vertex corresponds to the $(c+n) \rightarrow c+n$ breakup amplitude,
the neutral massless photon to the massive neutron.  In the case of
bremsstrahlung it is well known \cite{LandauLQED} that {\em for
arbitrary values of $\eta$}, even for $\eta \gg 1 $, one obtains the
Born approximation result, as long as the scattering is into a narrow
cone in the forward direction.  This leads one to suspect that higher
order effects are not very large in the case of high energy Coulomb
dissociation also, where the fragments are emitted into the forward
direction.

The $T$-matrix can be evaluated analytically in this model due to the
well known Nordsieck formula \cite{Nordsieck54}.  Using this formula
one obtains the $T$-matrix Eq.~(\ref{eq:tintegral}) in terms of a
hypergeometric function $F$ as well as its derivative $F'$. The
argument of the hypergeometric function $F$ (and $F'$) is given by
\cite{BaurT72,ShyamBB92}:
\begin{equation}
\zeta_0 = 
\frac{2 q^2 (q_a q_c + \vec q_a \cdot \vec q_c)
-4(\vec q \cdot \vec q_a )(\vec q \cdot\vec q_c)}
{(q^2-2\vec q \cdot \vec q_a)
(q^2+2\vec q \cdot \vec q_c)}.
\end{equation}
We observe that the parameter $\zeta_0$ is found to be negative and
$-\zeta_0 \gg 1$ for perpendicular momentum transfers $q_\perp \gg 2
\eta_a q_\|$ (nonadiabatic case) and beam energy large compared to the
binding energy.

This was already noticed in the numerical evaluation of the process:
the hypergeometric series therefore does not converge and an analytic
continuation had to be used. Here we use this fact to our advantage
and make a linear transformation to get the argument of the
hypergeometric function close to zero. The transformation we are using
leads to the argument of the hypergeometric function $z=
1/(1-\zeta_0)$ (Eq.~(15.3.7) of \cite{AbramowitzS64}). In this respect
our approach differs from the one used in the bremsstrahlung case,
where a transformation giving an argument close to one is used. Using
only the lowest order term in the hypergeometric series one obtains
after some algebra (up to an overall phase)
\begin{equation}
 T \approx  4\pi \ D_0 \ f_{coul} \ e^{-\frac{\pi}{2}\xi} 
\biggl[
e^{-i\phi} \frac{1}{q_a^2-(\vec q_n + \vec q_c)^2}
+ e^{+i\phi} \frac{m_c}{m_a}\ \frac{1}{q_c^2-(\vec q_n - \vec q_a)^2}
\biggr].
\label{eq:tdwba}
\end{equation}
Hereby, the relative phase is $\phi = \sigma_0(\eta_c) -
\sigma_0(\eta_a) - \sigma_0(\xi) - \xi/2 \log |\zeta_0|$. The
$\sigma_0(\eta)=\textrm{arg }\Gamma(1+ i \eta)$ are the usual Coulomb
phase shifts, and $\xi=\eta_c-\eta_a$\footnote{The parameter $\xi$ is
the one used commonly in NR Coulomb excitation. The parameter $\xi$ as
introduced above (see Eq.~(\ref{eq:xidefined}) in
Sec.~\ref{sec:theory}) corresponds to the parameter $\xi(\theta)$,
see, e.g., p. 6 of \cite{AlderW75} in the literature on NR Coulomb
excitation. Instead of the scattering angle $\theta$ we use the
corresponding impact parameter $b$.}.

The correspondence to the Born result Eq.~(\ref{eq:tpost}) is clearly
seen. One only has an additional prefactor $e^{-\frac{\pi}{2}\xi}$ and
a relative phase $e^{\pm i\phi}$ between the two terms. The phase
$\phi$ is $O(\xi)$.  Since $v_c\sim v_a$ the quantity $\xi$ is usually
very small and so is $\phi$ for the cases of, e.g.,
\cite{Nakamura99,Nakamura94}. The prefactor is also well known in the
semiclassical theory, where it accounts for the replacement of the
{\em Coulomb bended} trajectories with the straight line trajectories.
Both corrections vanish in the limit $\xi\rightarrow 0 $ and the
result coincides with the usual Born approximation ({\em even if
$\eta_a$ and $\eta_c$ are not small}).

Above we have established the correspondence of the full CWBA result
with the (Born) plane wave result and with the semiclassical result in
the sudden limit.  One may expect also to find a connection between
the two theories in the adiabatic limit, corresponding to $q_\perp < 2
\eta_a q_\|$.

For $\eta_a,\eta_c\gg1$ one can define a classical path for both $a$
and $c$ in the initial and final state. (For high beam energies, the
straight line approximation can be applied and the impact parameter
$b$ is given by $b=2\eta_a/q_\perp$. The adiabatic case corresponds to
the case of large impact parameters $b > 1/q_\|$.) For the adiabatic
regime ($q_\perp < 2 \eta_a q_\|$) the well known exponential decrease
with the adiabaticity parameter is observed in the numerical
calculations. In this case, the inequality $-\zeta_0\gg1$ is no longer
generally satisfied.

The semiclassical limit can be obtained with analytical methods in
this case also.  Such a method is valid in both the adiabatic and
nonadiabatic case as long as $\eta_a,\eta_c\gg1$.

We use two methods to show this result: in the first approach we use
the confluence
\begin{equation}
_1F_1(\alpha,\gamma,x)= \lim_{\beta \rightarrow \infty} 
\ {_2F_1}(\alpha,\beta,\gamma,\frac{x}{\beta})  
\end{equation}
together with some relations of \cite{AbramowitzS64} to relate
hypergeometric functions to (modified) Bessel functions, which appears
also in the semiclassical limit.

A second approach makes use of the partial wave expansion of the
neutron plane wave. The semiclassical limit is made by approximating
the radial wave function with the WKB approximation. In the case of
the plane wave expansion the resulting sum over $l$ can then by done
and an analytic expression can be found for the absolute value of the
matrixelement $T$ \cite{TrautmannHB03}.

A similar situation is encountered in the theory of bremsstrahlung and
Coulomb excitation, see Sec.~II~E of \cite{AlderBHM56}.  There a fully
quantal expression for the differential cross-section for dipole
Coulomb excitation is given in II~E.62. It looks similar to the
corresponding expression for Coulomb breakup (see \cite{BaurT72}). The
semiclassical version of this formula is found in II~E.57. It is noted
there that it can be obtained from the quantal expression by letting
both $\eta_a$ and $\eta_c$ go to infinity and perform a confluence in
the hypergeometric functions.

Postacceleration effects are also important for Coulomb dissociation
studies of radiative capture reactions of astrophysical interest. We
expect that our present investigations will shed light on questions of
postacceleration and higher order effects in these cases
also. Postacceleration was studied numerically in
\cite{BanerjeeBHST02}; as expected these effects turned out to be small at
high energies.

We also note the following important physics point: in the post-form
DWBA we can introduce in a straightforward way neutron-target
interactions \cite{BaurRT84}. One may say that breakup is a kind of
{\it ``transfer into the continuum''}, see also
\cite{BonaccorsoBB03,Bonaccorso99}.  On the other hand, in the prior
form DWBA the final state wave function is the product of a
c.m. motion and an internal $n$-core wave function, governed by the
interaction $V_{nc}$. In this case the breakup process may be viewed
as an inelastic excitation of the bound ($c+n$)-system into the
continuum.

\section{Nuclear Effects, Coulomb-Nuclear Interference}
\label{sec:nuclear}

A common problem for the Coulomb dissociation method is the influence
of the strong interaction between projectile and target. For heavy
target nuclei Coulomb excitation often dominates due to the coherence
factor $Z^2$, whereas the total nuclear cross section increases with
$R^2 \sim A^{2/3}$ , direct reactions (elastic nuclear breakup is a
special case) scale with $R \sim A^{1/3}$, see, e.g.,
\cite{HenckenBE96}, as they take place in a ring area around the
nucleus.

A good method to avoid the nuclear interaction altogether is to use
beam energies below the Coulomb barrier, which suppresses nuclear
effects very effectively.  As was discussed above in
Sec.~\ref{sec:intro}, one can access only low energy nuclear states in
this way. We are mainly interested in higher lying states and higher
beam energies are necessary. Again, we can suppress nuclear effects,
this time by going to forward scattering angles. This corresponds, in
a semiclassical picture, to trajectories where the nuclei do not touch
each other.  The (total) electromagnetic excitation probability falls
off at least with $1/b^2$; it is maximal for grazing impact
parameters, where also the nuclear effects set in. For too large
impact parameters, the adiabatic cutoff sets in and the excitation
probability drops exponentially to zero.  I.e. we have the same
problem as Wilhelm Tell \cite{Schiller05}, who had the task to hit a
rather well defined spot, with grave consequences in the case of
failure.  By the wave nature of the projectile nuclear effects cannot
be totally avoided, they will enter somehow, but often this effect is
negligible.
Even for small scattering angles there can be some effect
coming from the interference of that part of the wave function that is bent 
by the Coulomb interaction with that part of the wave function, which is 
diffracted by the ``black disc'' of the target. This will lead to 
diffraction patterns, see the discussion in \cite{Frahn85}.

It should be kept in mind that one does understand nuclear effects, at
least in a semiquantitative way. It is certainly best to choose the
experimental conditions (beam energy, scattering angle,\dots) in such
a way that these effects are virtually negligible. Such
considerations can be based on the good knowledge of nuclear
effects. As a second best approach they can be taken into account
with confidence by using the results of some kind of nuclear reaction
theory.  Direct (nuclear) reactions have been extensively studied in
the past decades and it can safely be said that one understands
rather well their main characteristic features, but not as
well as Coulomb excitation itself. This section is devoted to these
nuclear processes, Coulomb-nuclear interference and the theoretical
methods used to describe them. These methods sometimes involve heavy
computing.  Entire conferences are devoted to this subject
\cite{intweb}, which we only briefly summarize here.

\subsection{\it Various Kinds of Nuclear Reaction Mechanisms}

Nuclear breakup is dominant for light target nuclei and is used by
itself as a tool to explore the structure of exotic nuclei, especially
nuclear halo systems.  In a nuclear projectile-target interaction all
of the fragments (typically two) can emerge while the nucleus remains
in the ground state.  This is called {\em diffractive breakup} and it
is a reaction mechanism which leads to the same final state as Coulomb
dissociation. In addition to {\em diffractive breakup} ({\em diffractive 
dissociation}) there is
also {\em stripping} and {\em absorption}.  In {\em stripping reactions},
only part of the fragments comes out of the reaction in the very forward
direction, while the rest is absorbed; also all fragments can interact 
violently with the target and are absorbed, see
\cite{Serber47,Glauber55,Sitenko90} for details.

Stripping reactions have been the ``workhorse'' of the study of halo
systems in nuclear reactions, mainly due to the fact that they can be
interpreted in a simple (sometimes too simple) way: in the so-called
{\em transparent limit} the longitudinal momentum distribution of
the core after stripping is related to the momentum space probability 
distribution of the halo nucleon. This picture was more refined in recent 
years \cite{EsbensenB95,HenckenBE96,Hansen96} in order to incorporate the
so-called {\em core-shadowing effects}. The longitudinal momentum
distribution is related to the {\em momentum-distribution at the
surface,}, that is, to the Wigner transform at the surface
\cite{HuefnerN81}.

Diffractive breakup leads to the same final state as Coulomb
excitation.  As all fragments are generally measured in order to
reconstruct, e.g., their relative excitation energy, it is only this
process which needs to be considered as a
background to Coulomb excitation. We concentrate in the
following on this process. We refer here to a number of reviews
\cite{HansenS01,Tanihata96,HansenJJ95,Bonaccorso03a,Bonaccorso03b} 
of direct nuclear reactions with exotic nuclei.

\subsection{\it DWBA}

Nuclear effects can be treated in complete analogy to the DWBA
approach explored for Coulomb excitation above. The pure-Coulomb wave
function $\Psi$ is replaced by the one where an optical potential
between projectile and target acts in addition to the Coulomb
potential. In addition one has to add the nuclear excitation, treated
in first order here, for higher order effects see also the discussion
below in Sec.~\ref{ssec:nuclhigher}.

\begin{figure}[tb]
\begin{center}
\resizebox{0.18\textwidth}{!}{%
\includegraphics{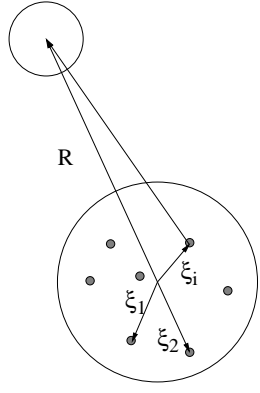}}
\end{center}
\caption{The potential $V(\vec R,{\bf \xi})$ is
given by the sum of the (Coulomb and nuclear) interactions between the
individual components of the target with the projectile $V_i(\vec R -
\xi_i)$.  }
\label{fig:nuclearpot}
\end{figure}

The $T$-matrix for inelastic scattering in the DWBA is given by, see,
e.g., Eq.~(7) of \cite{BertulaniBH91}:
\begin{equation}
T = \int d^3R \Psi_f^{(-)}{}^*(\vec R)\; F_{fi}(\vec R) \;\Psi_i^{(+)}(\vec R),
\end{equation}
The c.m. distorted waves functions are denoted by $\Psi^{(\pm)}_{f,i}$.  The
nuclear form-factor $F_{fi}(R)$ is given by
\begin{equation}
F_{fi}(\vec R) = \int d\xi \; \phi_f^*({\bf \xi}) \; V(\vec R,{\bf \xi}) 
\; \phi_i({\bf\xi}),
\end{equation}
where $\bf \xi$ denotes the internal coordinates, see
Fig.~\ref{fig:nuclearpot}. The initial and final internal wave
functions are $\phi_{i,f}({\bf \xi})$. The projectile-target interaction
is given by
\begin{equation}
V(\vec R,{\bf \xi}) = \sum_i V_i(\vec R - \vec \xi_i).
\end{equation}

We see that nuclear effects enter in two ways: first there is an
influence on the distorted waves of the c.m. motion $\Psi^{\pm}_{f,i}$
due to the nuclear interaction.  Second there is an additional breakup
component due to the nuclear form-factor $F$.  This form-factor can be
expanded in terms of multipoles and added to the Coulomb multipoles.

At lower energies the interactions of protons, neutrons, and also
light ions with different target nuclei have been analysed and
tabulated \cite{PerreyP76}, so that they are usually well known.  At
higher energies (few 100 MeV) the knowledge of the phenomenological
optical potentials is rather sparse. Often global parametrisations
(fitting optical potentials for protons and neutrons) are used 
for a whole range of energies and targets, see, e.g., \cite{VarnerTAL91}, or
for individual target nuclei, see \cite{KoningD03} for a recent
reference.  The interaction between the projectile nucleus and
the target (or some fragments with the target) are then calculated
within a folding approach of the projectile density with the
proton-target and neutron-target optical potentials.
\begin{equation}
V(\vec r)= \int d^3r_1 \rho_1(r_1) 
V_{NA}(|\vec r-\vec r_1|),
\label{eq:singlefoldpot}
\end{equation}
where $V_{NA}$ is the nucleon-target potential.

Another way is to use a double folding approach over {\em both} 
densities (projectile and target). The nuclear interaction is then given by
\begin{equation}
V(r)= \int d^3r_1 d^3r_2 \rho_1(r_1) \rho_2(r_2) 
V_{NN}(|\vec r-\vec r_1 + \vec r_2|),
\label{eq:foldingpot}
\end{equation}
where $V_{NN}$ is the nucleon-nucleon potential.

Only the nucleon-nucleon interaction $V_{NN}$ is needed in addition
to the nuclear densities $\rho_{1,2}$.
A typical example of this approach is the DDM3Y interactions
\cite{Kobos82,Brandan88}, based on a fit of the nucleon-nucleon
interaction in nuclear matter \cite{BertschBML77} together with an
assumed density dependence. Only the real (elastic) part of the
potential can be found in this way and the imaginary part needs to be
adjusted.  In many cases it is chosen for simplicity to be
proportional to the real part, with a proportionality factor in the range
of 0.5--0.9. This proportionality is in general only a first approximation;
especially at lower energies there are corrections, which can be described
by a polarisation potential, see e.g., \cite{KhoaSO95}.

\subsection{\it Eikonal DWBA}
\label{ssec:edwba}

At high energies the partial wave decomposition of the wave function
of projectile and target $\Psi^{\pm}_{f,i}$
is rather cumbersome. High energy methods,
like the {\em eikonal approach} (``Glauber method''), are more useful and at
the same time accurate. The wave nature of the projectile is still
taken fully into account and leads to characteristic diffraction
effects. The use of eikonal wave functions for the initial and final
states in the DWBA matrixelement is discussed in \cite{BertulaniN93},
where also its relation to the semiclassical case is explored. Higher
order effects can be treated within a coupled channels method
\cite{BertulaniCG02}, a corresponding computer program is described
there.  In this approach the differential cross section can be written
as (see also Eq.~(\ref{eq:eikonalf})):
\begin{equation}
f^\mu_{inel} = i k \int_0^\infty b db J_\mu(qb) \exp(i\chi(b)) a_\mu(b).
\label{eq:nucleiko}
\end{equation}
The index $\mu$ denotes the angular momentum transfer to the target in the 
beam direction.
The {\em excitation amplitudes} $a_\mu(b)$ are calculated in the
semiclassical straight line approximation for impact parameter $b$.
The eikonal phase $\chi(b)=\chi_C(b)+\chi_N(b)$ takes Coulomb, as well
as, nuclear effects into account. It involves the Coulomb
phase given by
\begin{equation}
\chi_C(b)= - 2 \eta K_0(kb) \approx 2\eta ln(kb)
\end{equation} 
and a nuclear phase, which can be calculated from an optical model potential
$U_{opt}(r)$ as
\begin{equation}
\chi_N(b)=-\frac{1}{\hbar v} \int^{\infty}_{-\infty} dz 
U_{opt}(r=\sqrt{b^2+z^2}).
\label{eq:chin}
\end{equation}

The simplest way to take nuclear effects into account is the {\it
``black disk model''}, where one sets $\exp i\chi=0$ for $b<R_1+R_2$
and $\chi =\chi_C$ for $b>R_1+R_2$. This is justified due to the strong
imaginary part of the nuclear interaction at these energies.

Diffraction effects in this black
disk model are studied numerically in \cite{BertulaniN93} and
\cite{MuendelB96}.  In Figs.~1 and~2 of \cite{MuendelB96} the
importance of nuclear diffraction can be seen.  A reduced integral is
defined there in Eq.~(8); it depends on the characteristic parameters
$\eta$ and $\xi$.  Plots are given for a reduced scattering angle
$\theta/\theta_{gr}$ where $\theta_{gr}=2\eta \theta_{diff}$ and the
diffraction angle is given by $\theta = 1/(kR)$.  It can be seen in
theses figures that for $\eta\gg1$ the semiclassical approximation is
very good, especially for small values of $\xi$.

A model which takes into account the smoothness of the surface is
discussed in \cite{CharagiG90}, where surface normalized Gaussians are
used for the density of the nuclei. In this way analytic expressions
for the eikonal phase can be found of the form
\begin{equation}
\chi(b) = \chi_0 \exp\left[-b^2/(a_1^2+a_2^2)\right],
\label{eq:charagi}
\end{equation}
with
\begin{equation}
\chi_0=\frac{\pi^2 \bar \sigma_{NN} \rho_1(0) \rho_2(0) a_1^3 a_2^3}
{10 (a_1^2+a_2^2)}  
\end{equation}
and $a_i$ and $\rho_i(0)$ are adjusted to reproduce the experimental
nuclear density at the surface of the nucleus; for a tabulation see
\cite{CharagiG90}. $\bar\sigma_{NN}$ denotes the average
nucleon-nucleon cross section.

An advantage of the (Glauber) eikonal approach is the fact, that one
does not need a nuclear potential in order to derive the eikonal
phase. Scattering at high energies is concentrated in the forward direction.
A knowledge of the elastic 
scattering amplitude $f_{el}(q)$ is sufficient, as it is related via 
a Fourier transform to the eikonal phase
\begin{equation}
\exp\left[i\chi(b)\right] -1 = \frac{i}{2\pi k} \int d^2q
f_{el}(q) \exp(-i \vec q \vec b).
\end{equation}
In order to determine the eikonal phase in nucleus-nucleus 
scattering $\chi_{AA}$ the so called $t\rho\rho$
formalism can be used, where the interaction is described through
a folding of the densities together with the nucleon-nucleon
scattering amplitude \cite{HusseinRB91}, cf. to Eq.~(\ref{eq:foldingpot}).
In terms of the nucleon-nucleon profile function 
$\Gamma_{NN}(b)=1-S_{NN}(b)=1-\exp\left[i\chi_{NN}(b)\right]$
it can be written as
\begin{equation}
\chi_{AA}(\vec b) = i \int d^3r_1 d^3r_2 \rho(r_1) \rho(r_2) 
\Gamma_{NN}(\vec r_{\perp1}-\vec r_{\perp2}-\vec b),
\label{eq:chiAA}
\end{equation}
This is related to the double-folding approach, see
Eq.~(\ref{eq:foldingpot}). In the limiting case of a zero-range 
interaction one gets
\begin{equation}
\Gamma_{NN}(\vec x)= \delta^{(2)}(\vec x) \frac{1}{2}(1 - i\alpha_{NN}) \sigma_{NN},
\end{equation}
where $\sigma_{NN}$ is the total 
nuclear cross section and $\alpha_{NN}$ the ratio
of the real to imaginary part of the forward scattering amplitude.
Parameterizations of this forward amplitude can be found in the
literature \cite{Ray79}. 

The $t\rho\rho$ formalism, which is often also called the {\em optical limit 
approximation} can be derived from the multiple particle Glauber
formalism by taking the first term of the {\em cumulant expansion}.
One wants to find an eikonal $\chi_{AA}$, which reproduces the elastic 
scattering amplitude of the microscopic multiple scattering model, 
see Eqs.~(\ref{eq:fm0eikonal}) and~(\ref{eq:profileS}) below,
\begin{equation}
\exp\left[ i \chi_{AA}(\vec b)\right] 
= \left< 0 \right| \prod_i S_i(b_i) \left| 0 \right>.
\end{equation}
By taking the logarithm of both sides and expanding
to first order in powers of $\Gamma_{NN}$,
see also \cite{Glauber59}, one obtains 
Eq.~(\ref{eq:chiAA}) above.
In this way the interaction is approximately included in all orders.
A discussion of this approach can be found also in 
\cite{JoachainQ74,EsbensenB99,HuefnerN81}.

\subsection{\it Higher Order Effects}
\label{ssec:nuclhigher}

Whereas the Coulomb interaction is long ranged and often the major
contribution comes from larger impact parameters, where the interaction
is not strong and first order theories are adequate, the nuclear
interaction is short ranged and strong. Therefore higher order effects
can become important and should be taken into account.

Within the eikonal DWBA higher order effects in $F(R)$ can be taken
into account as discussed above, e.g., in the (semiclassical) coupled
channel approach as given in \cite{BertulaniCG02}. In the case of
the deuteron, $^6$Li and other halo systems, quite elaborate
calculations have also been made in terms of the CDCC ({\em continuum
discretized coupled channel}) see \cite{SakuragiYK86,MoroCNT02}, or
\cite{YahiroNIK82,YahiroIKK86,AusternIKK87}.
In both approaches one needs to identify the dominant structures especially
in the continuum in order to restrict oneself to a manageable number of
channels to be incorporated in the calculation.

A theory which is well suited for intermediate and high energy nuclear
breakup of halo nuclei is the Glauber multiple scattering theory
\cite{Glauber59,JoachainQ74}. In this approach one uses the eikonal 
approximation together with the sudden limit to describe the elastic, as
well as, inelastic scattering from the ground state to a state $m$ as
\begin{equation}
f_{m0}(q) = \frac{k}{2\pi i} \int d^2b \exp(i \vec q \vec b)
\int d{\bf \xi} \phi_m^*({\bf \xi}) \phi_0({\bf \xi}) 
\left(S(b,{\bf\xi}_{\perp}) - 1\right),
\label{eq:fm0eikonal}
\end{equation}
where $k$ is the momentum of the projectile and $\vec q=\vec k_i - \vec k_f$.
The profile function $S$ depends on the individual impact parameters
$b_i=b-\xi_{i\perp}$. It is given by 
the product of the individual profile functions
\begin{equation}
S(b,{\bf\xi}_{\perp}) = \prod_i S_i(b_i) = \prod_i \exp(i \chi_i(b_i)).
\label{eq:profileS}
\end{equation}
It is possible to evaluate this expression numerically with
fully microscopic wave function $\phi_0$ and $\phi_m$,
see e.g. \cite{VargaPSW02}.

An intermediate model, which is well suited for halo nuclei, is the 
{\em Serber model} \cite{Serber47}, improved by Glauber \cite{Glauber55}, 
see also \cite{Sitenko90}; in this model only
the relevant clusters are taken into account in the Glauber multiple scattering
model.

In the eikonal DWBA, see Eq.~(\ref{eq:nucleiko}), a global projectile-target
profile function is used to describe the c.m. motion
of the projectile. In the improved Serber model, 
see Eq.~(\ref{eq:mpglauber}) below, 
one uses individual profile functions $S_i(b_i)$ for each fragment $i$, e.g.,
the core $c$ and the halo neutron $n$ 
\cite{BertulaniM92,BanerjeeS93,SagawaT94,BarrancoVB96,HenckenBE96}.
The individual profile functions can then be calculated in the same way as
in Eqs.~(\ref{eq:chin}), (\ref{eq:charagi}) or (\ref{eq:chiAA}).
The differential cross section for the
elastic breakup ({\em diffraction}) is then found to be, see also 
Eq.~(\ref{eq:fm0eikonal}),
\begin{equation}
f_{fi}(q) = \frac{k}{2\pi i} \int d^2b \exp(i \vec q \vec b) 
\left< f\right| \exp\left(i \chi_c(b_c)\right) \exp\left(i \chi_n(b_n)\right) 
-1 \left|i\right>,
\label{eq:mpglauber}
\end{equation}
where $b_c=b+m_n/(m_n+m_c) r_\perp$ and $b_n=b-m_c/(m_n+m_c) r_\perp$ 
are the individual impact parameter of the core and the
neutron with the target, respectively.  In this way the most important parts
of the internal structure of the projectile nucleus are taken into account. 
This approach can easily be extended also to the case of more than two
clusters \cite{OgawaYS92,OgawaSY94}. Only initial and final state wave
functions but no intermediate states are needed in this way.

\begin{figure}[tb]
\begin{center}
\resizebox{0.65\textwidth}{!}{%
\includegraphics{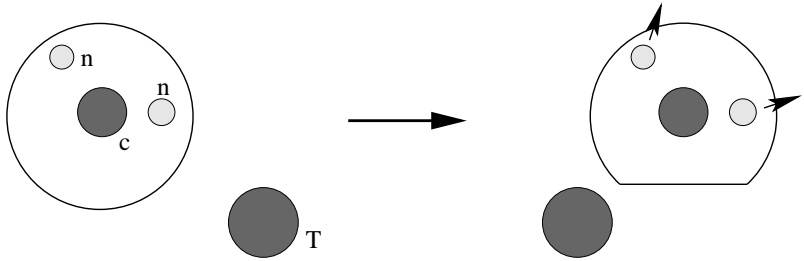}
~~~~~~~~~~~~~~~~
}
\end{center}
\caption{The ``wounded nucleus'' picture of nuclear elastic breakup ({\em
diffractive breakup}) of a halo nucleus. Due to the strong absorptive 
(imaginary) part of the nuclear interaction, those parts of the wave function,
which are touched by the target nucleus are set to zero. This means that
the nucleus is no longer in its ground state but also in excited states.
The projection of this excited state to the continuum states gives
the amplitude for diffractive dissociation, see also Eq.~(\ref{eq:mpglauber}).}
\label{fig:woundednucleus}
\end{figure}

A simple picture of the diffractive breakup can be found in the limit of the
``black disc'' model, where the individual profile functions are approximated
by black discs, see Fig.~\ref{fig:woundednucleus}.  

\subsection{\it Coulomb-Nuclear Interference}

Nuclear effects enter most sensitively through the interference of the
nuclear and Coulomb amplitudes.  This has been addressed in the past
in a number of papers, some recent references are
\cite{MargueronBB02,ChatterjeeS02,EsbensenB02a,EsbensenB02b}.

A simple (analytic) result is given in \cite{BertulaniB88} for a
neutron halo nucleus.  The amplitude of Coulomb excitation in first
order (Eq.~(3.4.16) of \cite{BertulaniB88}) 
is combined with the nuclear breakup amplitude of Akhiezer and Sitenko
\cite{AkhiezerS57} (Eq.~(3.4.14) of \cite{BertulaniB88}).  
In this way a good estimate of the interference effect can be found.
\begin{figure}[p]
\begin{center}
\resizebox{0.42\textwidth}{!}{%
\includegraphics{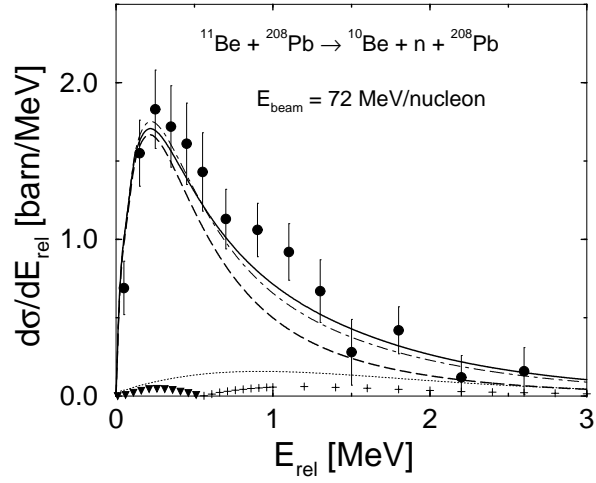}
}
\end{center}
\caption{Fingerprints of Coulomb-nuclear interference: effects of nuclear
and Coulomb-interference are shown for the breakup of $^{11}$Be on
$^{208}$Pb at 72~$A$MeV. Shown are the contribution from pure Coulomb
(dashed), nuclear (dotted) and their incoherent (dashed-dotted) and 
coherent sum (solid line). The magnitude of negative and positive 
Coulomb-nuclear interference terms are shown as inverted triangles and
plus signs, respectively. The experimental results are taken from
\protect\cite{Nakamura94}.  This figure is adapted from
\protect\cite{ChatterjeeS02}.}
\label{fig:chatterjee2}
\end{figure}
\begin{figure}[p]
\begin{center}
\resizebox{0.8\textwidth}{!}{%
\includegraphics{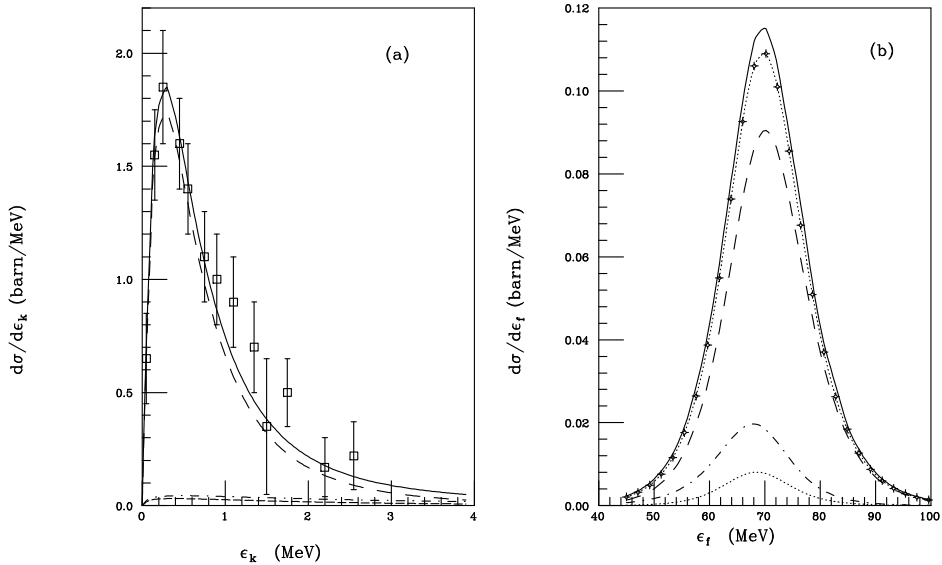}
}
\end{center}
\caption{Effects of nuclear and Coulomb-nuclear interference are shown
for the breakup of $^{11}$Be on $^{208}$Pb at 72~$A$MeV in the
projectile (left)and target (right) frame. Shown are the results for
nuclear breakup (dot-dashed), Coulomb (long-dashed) and
Coulomb-nuclear interference (dotted) and the sum (solid line).  The
experimental results are taken from \protect\cite{Nakamura94}.  This
figure is taken from \protect\cite{MargueronBB02}, Fig.~4 there.}
\label{fig:marguerson4}
\end{figure}

More detailed calculations 
have been done in the mean time. E.g., in \cite{MargueronBB02,MargueronBB03} 
a calculation within Glauber theory is presented. In \cite{EsbensenB02a} 
the time-dependent Schr\"odinger equation in the semiclassical
approximation is solved numerically and compared to a first order calculation 
and to the eikonal approximation.

In the case of halo nuclei the condition of {\em no nuclear
penetration} may not be well fulfilled and the decomposition 
of the Coulomb interaction into $r_<$ and $r_>$ parts (see 
Eq.~(\ref{eq:elmulti})) may not be reliable,
as shown in \cite{EsbensenB02a} for the case of $^8$B.

We illustrate the fingerprints of Coulomb-nuclear interference and
nuclear effects from two recent theoretical model calculations. The
results shown in Fig.~\ref{fig:chatterjee2} are based on a post-form
DWBA approach, see Sec.~\ref{sec:analytic}.  The results shown in
Fig.~\ref{fig:marguerson4} are based on a Glauber type calculation,
see Eq.~(\ref{eq:mpglauber}),
taking higher orders in the nuclear interaction together with
the Coulomb interaction in lowest order into account.

In both cases breakup of $^{11}$Be on a $^{208}$Pb target at 72~$A$MeV
are shown in comparison to the experimental data of
\cite{Nakamura94}. One sees that both nuclear and Coulomb-nuclear
effects are rather small in this case.

\section{Coulomb Dissociation and Nuclear Structure}
\label{sec:nstruc}

\subsection{\it Primakoff Effect}
We want to recall first at this point
that the strong nuclear Coulomb field 
has also been used in particle physics to study photon interactions
with (unstable) particles. Typically a high energy 
secondary beam, like a beam of $\Lambda$ particles
hits a heavy target nucleus. In the nuclear Coulomb field, 
high energy $\Sigma^0$ hyperons are produced in the reaction 
\begin{equation}
\Lambda+Z \rightarrow \Sigma^0 +Z.
\end{equation}
The cross section for this process is proportional to the $B(M1)$-
value for the $\Lambda \rightarrow \Sigma^0$ electromagnetic
excitation. From the measurement of this 
cross section the lifetime of the $\Sigma^0$ is 
obtained. For further details of the analysis see \cite{BertulaniB88}, 
where further references and other 
applications of the {\em Primakoff effect} (like the determination of the pion 
polarizability) are given. A more recent example is the 
radiative decay width measurement of neutral Kaon excitations 
using the Primakoff effect by the KTeV Collaboration 
\cite{AlaviHarati02}.
They used $K_L$ mesons in the 100-200 GeV energy range
to produce the axial vector $(1^+$) mesons $K_1(1270)$ and
$K_1(1400)$ in the nuclear Coulomb field of a Pb target.
In this way the radiative widths for the decay of these
particles into $K^0 +\gamma$ could be determined.
For further details we refer to this paper \cite{AlaviHarati02}.
This approach is very similar to the one used nowadays
in nuclear physics: an exotic (unstable) nucleus
is excited by the quasireal photons provided by the nuclear 
Coulomb field. 

\begin{figure}[tb]
\begin{center}
\resizebox{0.5\textwidth}{!}{%
\includegraphics{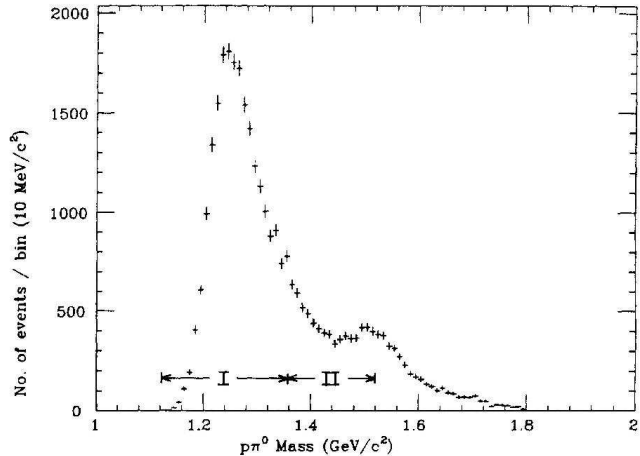}}
\end{center}
\caption{The invariant mass spectrum of the $\pi^0-p$ system in the reaction 
$p+\mbox{Pb} \rightarrow \pi^0 +p +\mbox{Pb}$ with a 185~GeV/$c$ proton 
for a momentum transfer of
$|t|<1\times10^{-3} (\mbox{GeV}/c)^2$ is shown.
These small momentum transfer values ensure that photon exchange is 
the dominant excitation mechanism.
 Peaks due to the $\protect\Delta^+(1232)$ and $N^*(1520)$ resonances
are clearly seen.  Reproduced from Fig.~2 of 
\protect\cite{CareyCC90}, to which we refer for further details.
Copyright (1990) by the American Physical Society.}
\label{fig:carrey}
\end{figure}

We mention a second example to illustrate the use of 
equivalent photons:  
At Fermilab the Primakoff effect is used for a determination  of 
the proton polarization of a 185~GeV/$c$ proton beam
by means of azimuthal asymmetry measurements \cite{CareyCC90}. 
This method could also be useful  for antiproton beams.
For further details we refer to this reference \cite{CareyCC90}.
From this work one can see that it is possible 
to study the photon-proton interaction
in the nucleon resonance region 
using equivalent photons.
In Fig.~\ref{fig:carrey} we show the invariant mass spectrum of 
the $\pi^0-p$ system in the reaction $p+Pb \rightarrow
\pi^0 +p +Pb$ for a momentum transfer $|t|<1\times10^{-3} (GeV/c)^2$. 
One can clearly recognize the nucleon-resonances (especially the 
$\Delta$-resonance) which are excited with the continuous
equivalent photon spectrum provided 
by the Coulomb field of the Pb target nucleus. Since the Lorentz factor
$\gamma$ of the proton is about 200 in this experiment,  the corresponding  
equivalent photon spectrum is quite hard: 
(see the discussion in Sec.~\ref{ssec:equiphot} about the cutoff in the photon
spectrum at $\omega_{max} \sim \gamma/R$) we can roughly 
take  $1/R \sim 30$~MeV
and obtain a maximum photon energy of about 6~GeV.

\subsection{\it Some Aspects of Electromagnetic Excitation in Relativistic
Heavy Ion Collisions}
\label{ssec:rhic}
Electromagnetic excitation is also used at (ultra-) relativistic heavy ion
accelerators to obtain nuclear structure information. Recent examples
are the Coulomb fission studies of radioactive nuclei 
at GSI \cite{Schmidt00,Heinz03}
and  Coulomb fission of $^{208}$Pb \cite{Abreu99} at SPS/CERN.
Due to the logarithmic rise of the cross-section with beam energy,
cross-sections for the excitation of the giant dipole resonance
({\em Weizs\"acker-Williams process}) at the  relativistic heavy ion colliders
RHIC and the forthcoming LHC(Pb-Pb) at CERN are 
huge \cite{BaurB89,BaurHT98,BaurHT02}, of the 
order of 100~b for heavy systems (Au-Au or Pb-Pb).

\begin{figure}[tb]
\begin{center}
\resizebox{0.5\textwidth}{!}{%
\includegraphics{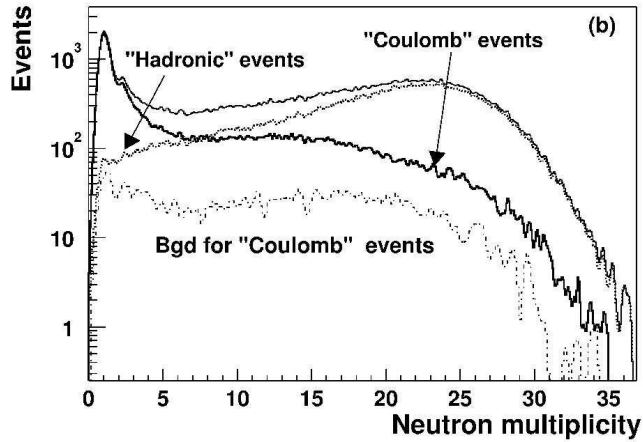}
}
\end{center}
\caption{Single arm ZDC (Zero Degree Calorimeter) neutron multiplicity 
spectrum for Coulomb and hadronic events. 
The two types of neutron emission can be separated from each other
as hadronic production
also leads to charged particle production in the more 
central parts of the detector.
See text and \protect\cite{ChiuEA01} for details.
Reproduced from Fig.~2 of \protect\cite{ChiuEA01}.
Copyright (2001) by the American Physical Society.}
\label{fig:chiu}
\end{figure}

The  neutrons from GDR decay were observed at RHIC \cite{ChiuEA01}. 
In this reference mutual Coulomb dissociation was measured in 
$\sqrt{s_{NN}}=130$ GeV Au-Au collisions.
This is shown in Fig.~\ref{fig:chiu} (taken from \cite{ChiuEA01}).

The two classes of events are defined 
according to the number $n_{BBC}$ of hits in the BBC ({\em beam beam counters})
photomultipliers. The PHENIX (one of the detectors at 
RHIC) BBC measures the relativistic charged particles
produced in cones around each beam with a rapidity range
$3.05<|\eta |<3.85$) with $2 \pi$ azimuthal coverage.
For {\em hadronic} events we have $n_{BBC}>1$
in each arm, for {\em Coulomb} events there is $n_{BBC}\le1$
in at least one arm. We see that the ``Coulomb'' 
events tend to have a low neutron multiplicity. This is easily understood 
qualitatively: Coulomb events are mainly due to 
the electromagnetic excitation of the GDR (in $^{197}$Au)
in ultraperipheral collisions ({\em UPC}\/). The GDR
decays subsequently  by the emission  of one (or only a few) 
neutrons. See \cite{ChiuEA01} for further details. 

In colliders, this effect leads to a loss of the beam, due to the 
particle (neutron) decay of the GDR. Even worse, this effect can lead 
to a localized beam pipe heating in Pb-Pb collisions  at LHC, as was noticed 
in \cite{Klein01}. (An even more severe process in this context, limiting the
maximum luminosity that can be achieved at the LHC, is the electromagnetic
process of bound-free electron 
pair production.) On the other hand, this effect can also be useful as a 
luminosity monitor by detecting the neutrons in the 
forward direction as first proposed in \cite{BaltzCW98}
and demonstrated in \cite{ChiuEA01}. 
One measures the neutrons which are produced
in the decay of the giant dipole resonance, which is excited
in each of the ions ({\em mutual excitation}).
Since this process has a steeper impact parameter dependence than the 
single excitation cross-section, there is more sensitivity to the 
cut-off radius and to nuclear effects. 
The neutrons from the GDR decay can also serve as a trigger on the 
ultraperipheral collisions. 
For details and further references, see \cite{BaurHT98,BaurHT02}.

UPCs are also of great interest in particle and hadron physics:
at RHIC and LHC the equivalent photon spectrum extends to about
500 GeV and 1 PeV ($10^3$ TeV), respectively,
in the rest frame of one of the ions.
This leads to many interesting applications like the coherent 
production of vector mesons \cite{HenckenW02,KleinN99}.
Coherent $\rho^0$ production with and without simultaneous GDR excitation 
has been observed at RHIC \cite{Adler02}.
This is a very hot topic, but it is outside the scope 
of the present article, we refer to the reviews in
\cite{KraussGS97,BaurHT98,BaurHT02,BaurBC02}.

\subsection{\it Intermediate Energy Coulomb Excitation 
of Discrete Levels and Gamma Decay}
\begin{figure}[tb]
\begin{center}
\resizebox{0.45\textwidth}{!}{%
\includegraphics{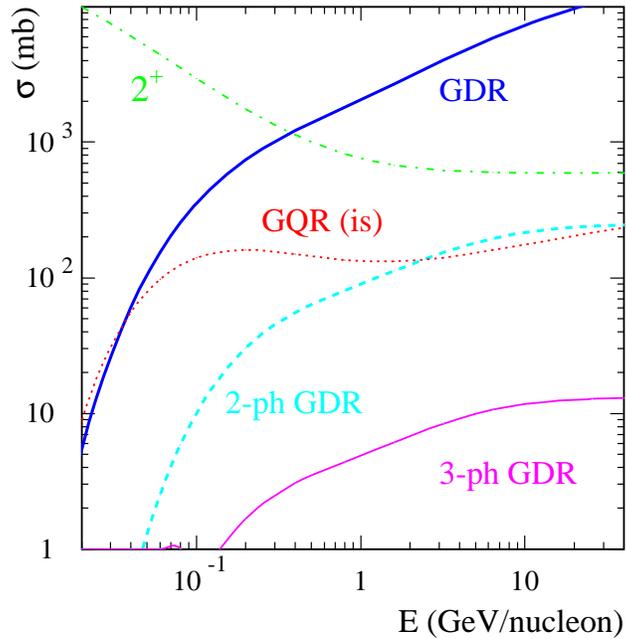}
}
\end{center}
\caption{Cross sections for the electromagnetic
excitation of collective states of an exotic nucleus 
on a Pb target. Typical nuclear structure parameters
for a medium mass nucleus were adopted. 
This figure is taken from \protect\cite{CDR}, p.~128. 
}
\label{fig:GDRCDR}
\end{figure}

Electromagnetic excitation is a very clean and 
efficient way to excite nuclear states. 
Interesting nuclear structure properties like the position
of energy levels, as well as, electromagnetic transition
matrix elements can be determined.
The general features of electromagnetic excitation have been 
given in Sec.~\ref{sec:theory}. Collective states, i.e., those with large 
electromagnetic matrixelements, are most strongly excited.
While in Coulomb excitation below the barrier one can only populate low 
lying states like rotational or low lying vibrational states (see discussion
in Sec.~\ref{sec:theory}),
it becomes possible at intermediate energies to excite also
high lying states, the prime example being the GDR. The
main features are  shown in Fig.~\ref{fig:GDRCDR}, taken from 
\cite{CDR}, see also Figs.~3.3 and~3.7 of \cite{BertulaniB88}. In 
these figures
the excitation cross section of an exotic nucleus 
with typical values for structure parameters (whose exact 
values do not matter now) is shown. 
The low lying $2^+$-collective level is excited most strongly
at the low energies, while the GDR takes over at the higher 
energies. Two- and three-phonon GDR's (labelled
2-ph GDR, and 3-ph GDR) are excited with smaller cross-sections
(see Sec.~\ref{sec:higher:dgdr}). They tend to a constant in the limit of 
high beam energies.
The one-photon GDR excitation cross section first rises quite steeply with
increasing energy, overcoming the adiabatic cut-off criterion;
eventually it will show the asymptotic rise with $\ln\gamma$, 
see Eq.~(\ref{eq:nomegaln}). The (high lying) 
isoscalar quadrupole excitation is also given.
The cross section for the low lying $2^+$ excitation shows 
a $1/v^2$-behaviour,
while its logarithmic rise will only occur at extremely high 
energies (see the remarks in Sec.~\ref{sec:theory}). These excited states
can decay by photon- (or, at sufficiently high excitation energies,
by particle-) emission.

Photons emitted in-flight from intermediate energy projectiles can be 
distinguished with modern gamma ray detectors from $\gamma$-rays originating 
from the target. In order to accomplish
this, the information on the emission angle is used
to correct for the Doppler broadening.
The development of position-sensitive $\gamma$-detectors was a 
key to the success of this method.
For a detailed description of this field we refer to the review by 
Glasmacher \cite{Glasmacher98} and more recently \cite{Glasmacher01}
and \cite{CDR}.

This method is very well suited for radioactive beams and very valuable
information about the nuclear structure of unstable nuclei has been obtained 
and is expected to be obtained in the future.

Many interesting examples of the investigation of 
nuclear structure effects of unstable nuclei with 
this method are given in  \cite{Glasmacher01}. We
refer to this reference for more details. Let us 
mention here only some points:

Electromagnetic
excitation of intermediate energy (exotic) beams has been
developed into a useful spectroscopic tool \cite{Motobayashi95,Scheit96}.
This method is ideal to study the behaviour and 
onset of collectivity (deformation) for $0^+ \rightarrow 2^+$
transitions in nuclei far from stability.
By measuring the excitation energies of the first $2^{+}$ 
states and the corresponding
$B(E2)$--values, nuclear structure effects like deformation,
can be studied in a unique way for nuclei far off stability. 

Electromagnetic excitation of the 1$^{st}$ excited state 
in ${}^{11}$Be has been studied experimentally at GANIL \cite{Anne95}, 
RIKEN \cite{Nakamura97} and MSU  \cite{Fauerbach97}. This is a good test 
case, since the $B(E1)$--value of the corresponding ground-state transition 
is known independently. Theoretical calculations \cite{BertulaniCH95,TypelB95}
show that higher order effects are small. They decrease
with increasing beam energy, as expected.

Subsequently, many interesting nuclear structure studies followed
at RIKEN, GANIL, MSU/NSCL and GSI with this 
method, which can be considered as well established by now.  

We mention here that the neutron-rich nucleus $^{34}$Mg has been studied 
recently via Coulomb excitation using a radioactive beam on a Pb target
\cite{Iwasaki01}.
The high $B(E2)$ value for the excitation of the $2^+_1$ state corresponds to 
a quadrupole deformation parameter $\beta_2$ of $0.58(6)$,
implying an anomalously large deformation of $^{34}$Mg.

The work done at MSU can be found at \cite{Glasmacherweb}.
Let us mention one of their latest results: 
The $0^+_{gs} \rightarrow 2^+$ excitations in the mirror
nuclei $^{32}$Ar and $^{32}$Si were compared to each other. 
This is a sensitive test of isospin symmetry which
could be extended to a $T=2$ isobaric multiplet
\cite{Cottle02}.

In this context it is worth mentioning that a computer
program for scattering at intermediate energies has 
recently been published in \cite{BertulaniCG02}.
Nuclear, as well as, Coulomb excitation processes are calculated 
using the eikonal approximation and semiclassical 
coupled channels methods. The eikonal
method is very appropriate for intermediate energies: it is both numerically 
accurate and fast (as opposed to DWBA methods, where also the relative 
center of mass motion, with its huge number of partial waves,
is treated in a quantum-mechanical way).
The general problem of the coupled channels method is 
to find the important states to be included. In many practical 
cases this proves to be  possible. In any case, as we 
have shown above, the strength parameter is proportional 
to $1/v$, i.e., higher 
order effects tend to be small for intermediate and 
high beam energies. On the other hand this also means that 
one cannot have efficient multiple Coulomb excitation for  
intermediate beam energies. 
The two-photon excitation of the strongly collective DGDR discussed above
is an interesting exception. For low energy Coulomb excitation,
the multiple excitation of rotational bands is a very well known 
feature. Angular distributions of elastically
and inelastically scattered particles, as well as,
angular distributions of $\gamma$-rays are also calculated
in \cite{BertulaniCG02}.

\subsection{\it Intermediate Energy Coulomb Dissociation,  
Invariant Mass Spectroscopy and Low Lying $E1$ Strength}
Nuclei in the valley of stability show a prominent 
collective mode, the Giant Dipole Resonance (GDR). 
In this mode, all protons oscillate against all neutrons. This 
state has an excitation energy of about $E_{GDR}\approx 
80\mbox{MeV}\ A^{-1/3}$ and 
exhausts the classical energy-weighted Thomas-Reiche-Kuhn (TRK) sum rule 
to a large extent. As a consequence of this, low lying $E1$-strength is
strongly hindered. In \cite{UchiyamaM85} it was shown that
this hindrance of $E1$-transition strength disappears when loosely 
bound nucleons are involved. This is, e.g., the case in neutron halo nuclei.
A classic case is the $E1$-strength 
in the prototype of a neutron halo nucleus: the deuteron.
Using simple zero range wave functions, the $E1$ strength is calculated
in \cite{BlattW79}. A similar calculation for neutron halo nuclei is given
in \cite{BertulaniB88}, which clearly shows how the 
low-lying $E1$-strength is obtained in a single particle model.
With the zero range bound state wave function, 
see Eq.~(\ref{eq:phigs}) above or Eq.~(3.4.11) of \cite{BertulaniB88}
\begin{equation}
\Psi_i(r)=\sqrt{\frac{\kappa}{2\pi}} \exp(-\kappa r)/r
\label{eq:phigs2}
\end{equation}
and the continuum wave function, see Eq.~(\ref{eq:phicontinuum}) or
Eq.~(3.4.12) of \cite{BertulaniB88},
(for the relevant partial waves ($l>0$) this corresponds 
to the spherical Bessel functions, which describe a free particle)  
we obtain for the electromagnetic matrix-element Eq.~(\ref{eq:melambda})
(see Eq.~(3.4.19) of \cite{BertulaniB88}):
\begin{equation}
M(Elm)=e \sqrt{2\pi \kappa} (-i)^l l! 2^{l+1}
(Z_1\beta_1^l+(-1)^l Z_2\beta_2^l)
\frac{q^l}{(\kappa^2+q^2)^{l+1}} Y_{lm}(\hat q),
\label{eq:MElmzerorange}
\end{equation}
with $\beta_1=m_2/(m_1+m_2)$ and $\beta_2=m_1/(m_1+m_2)$ and $Z_1$ and $Z_2$
denote the charge numbers of the two clusters.

\begin{figure}[tb]
\begin{center}
\resizebox{0.3\textwidth}{!}{%
\includegraphics{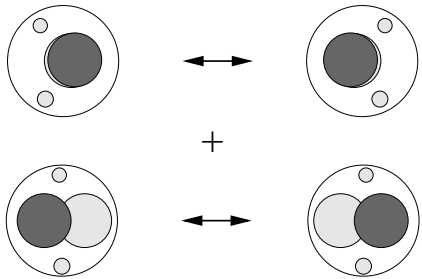}
}
\end{center}
\caption{In the GDR all neutrons move against all protons ({\em hard mode},
lower figure). In a nucleus
which contains also  loosely bound nucleons, 
e.g., in a 2-neutron halo nucleus like $^{11}$Li,
a new type of oscillation, called the {\em soft dipole mode} (upper figure), 
can occur. In this mode the two loosely bound neutrons oscillate against the 
inert core of strongly bound protons and neutrons. This gives rise  to
$E1$ strength at low energies. 
Figure adapted from \protect\cite{Ikeda92}.} 
\label{fig:softgdr}
\end{figure}

A general discussion of dipole strength in neutron rich nuclei
is given by Hansen and Jonson \cite{HansenJ87}
and Ikeda \cite{Ikeda92}. A significant low-lying $E1$ strength
is found. 
These modes are sometimes called {\em soft dipole state} or {\it 
``pigmy resonance''}, see Fig.~\ref{fig:softgdr}. 
An early discussion of the disappearance of the hindrance of $E1$-transitions 
involving loosely bound nucleons is given in \cite{UchiyamaM85}.

A measure of the $E1$-strength of certain configurations is given by the 
{\em cluster sum rule}: in the usual TRK sum-rule the relevant degrees of 
freedom of the nucleus are the neutrons and protons
\footnote{On a more microscopic scale the relevant degrees of freedom
are the quarks, see also
Exercise~3 of \protect\cite{Bertsch98}. The nucleon resonances,
see Fig.~\protect\ref{fig:carrey} are a manifestation of these
degrees of freedom}. In this case the TRK 
sum rule tells us that
\begin{equation}
\sum_n (E_n-E_0)| D_{n0}|^2=\frac{3\hbar^2}{2m_N} \frac{NZ}{A},
\label{eq:TRKsum}
\end{equation}
where $N$,$Z$, and $A$ denote the neutron, proton and mass number of the 
nucleus and
where $D_{n0}$ is the dipole matrix-element between the ground 
state $0$ and the excited state $n$. In this equation we have neglected
the effect of exchange forces in the nucleon-nucleon 
interaction, which is of the order of some 10\%.
We can determine the contribution of certain cluster configurations 
to the energy-weighted sum rule by a so-called cluster sum rule:
assuming an inert core with mass $M$ and charge $Z$ and
some valence neutrons with mass $m$ as the relevant 
degrees of freedom, one can write down a {\em cluster sum rule}, taking those
as the relevant degrees of freedom. 
In this sum rule the factor $NZ/A$ in Eq.~(\ref{eq:TRKsum}) is replaced
by $\frac{(Z_1 N_2 - Z_2 N_1)^2}{(A_1+A_2) A_1 A_2}$, where $N_i$,$Z_i$ and 
$A_i=(N_i+Z_i)$ are the neutron, proton, and nucleon number 
of each cluster \cite{AlhassidBG82}. In the case
of a neutron halo ($Z_2=0$, either single neutron or more) this gives a factor
$\frac{Z_1^2 N_2}{(A_1+A_2) A_1}$.

Effective field theories ({\em EFT}\/) are nowadays also used for 
the description of halo nuclei, see \cite{BertulaniHK02}.
The relative momentum $k$
of the neutron and the core is indeed much smaller
than the inverse range of their interaction $1/R$ and
$k R$ is a suitable expansion parameter. (In our model of the pure-Coulomb
breakup of a bound state bound by a zero range force, 
see Sec.~\ref{sec:analytic} 
above, we have $R=0$, i.e., $kR=0$ and we have the 
zero order contribution of the expansion). 
Effective range theory seems a natural starting point. This aspect was 
pursued in \cite{KalassaB94,KalassaB96}.
In \cite{KalassaB96} radiative capture cross sections into $s$-, $p$- and 
$d$-bound
states are calculated in simple models, and the cross sections depend
only on a few low energy parameters. The neutron halo effect
on direct neutron capture and photodisintegration of $^{13}$C 
was studied in \cite{MengoniOI95} and \cite{OtsukaIFN94}.
In their figures it can very well be seen that the radial integrals are 
dominated by the outside region. While they find a sensitivity on neutron
optical model parameters for $s \rightarrow p$ capture, this sensitivity
is strongly reduced for the $p \rightarrow s$ and $p \rightarrow d$ capture 
cases.
In \cite{BertulaniHK02} it is remarked that the EFT approach {\it ``is not 
unrelated to traditional single-particle models''} and that {\it 
``it remains to be 
seen whether these developments will prove to be a significant improvement 
over more traditional approaches.''}\/  With a wealth of data on halo nuclei 
to be expected from the future rare ion beams we can be confident that 
these questions will be answered. An interesting effect is known from
the deuteron, the mother (prototype) of all halo nuclei:
for very low energies,
there is an $M1$ $s \rightarrow s$ transition, which dominates over 
$E1$ \cite{BlattW79}. It remains to be seen whether a similar situation 
will be found in other halo nuclei.

\begin{figure}[tb]
\begin{center}
\resizebox{0.7\textwidth}{!}{%
\includegraphics{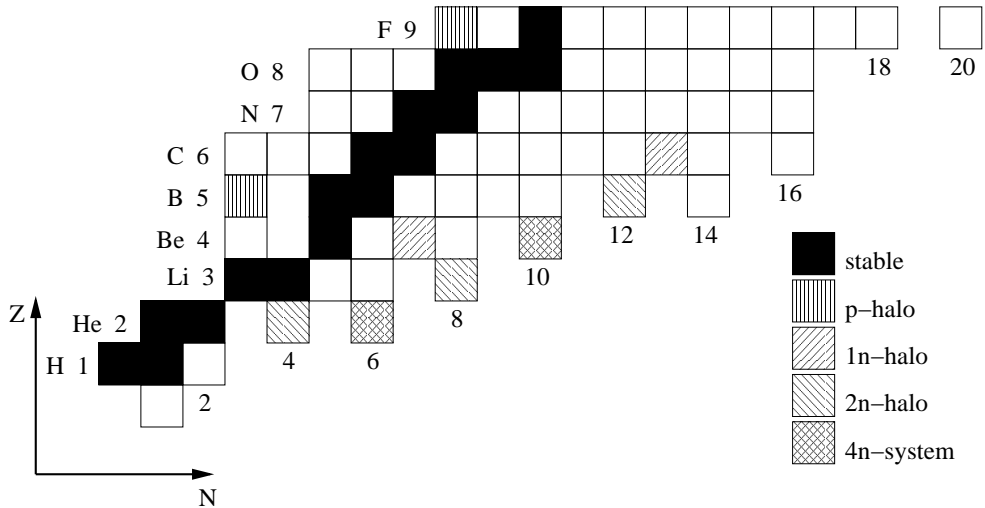}}
\end{center}
\caption{Chart of the (particle stable) light nuclei. Stable
nuclei, one- and two-neutron halo nuclei, four-neutron cluster nuclei
and proton halo nuclei are marked. The deuteron (marked in black)
is the prototype of a one-neutron halo nucleus.
Coulomb dissociation experiments of most of those nuclei are discussed in 
the present and the following sections.
}
\label{fig:halochart}
\end{figure}

In Fig.~\ref{fig:halochart} we give a part of the chart of nuclides.
We concentrate on the light nuclides and indicate 
one- and two-neutron halo nuclei. On the proton rich side,
$^{8}$B (and probably also $^{17}$F, at least its $1/2^+$ first excited state)
can be considered as a proton 
halo nucleus. $^8$He, as well as, $^{14}$Be can be regarded as four-neutron
cluster nuclei.
\begin{figure}[p]
\begin{center}
\resizebox{0.5\textwidth}{!}{%
\includegraphics{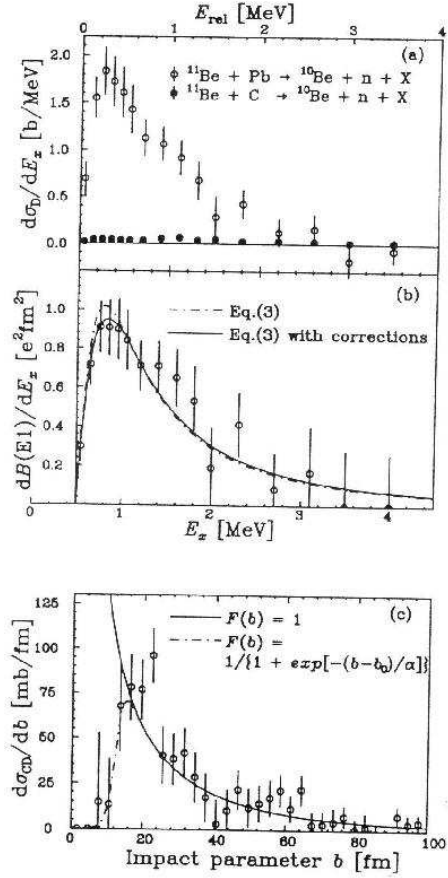}
}
\end{center}
\caption{Fig1(a) Dissociation cross sections of $^{11}$Be on a Pb target as 
a function of the excitation energy $E_x$ for the $^{10}Be+n$
system. Data from a C target are also indicated.
(b) Dipole strengths deduced from the $d\sigma_{CD}/dE_x$
 spectrum are shown by the open circles.
(c) Impact parameter dependence of the Coulomb dissociation cross section.
The low lying  $E1$ strength in  $^{11}$Be is clearly seen.
This figure is taken from \protect\cite{Nakamura94}, Fig.~1 there,
where further details can be found.
}
\label{fig:nakamura}
\end{figure}

Coulomb dissociation of exotic nuclei is 
just the right tool to study this nuclear structure problem.
By this method one determines the electromagnetic
matrix elements between the ground state and the nuclear continuum.
A classic example is the $^{11}$Be Coulomb dissociation. 
The excitation energy spectrum of the ${}^{10}$Be+$n$ system in the Coulomb 
dissociation of the one-neutron halo nucleus  ${}^{11}$Be on a Pb target at 
$72$~$A$MeV was measured \cite{Nakamura94}, from which we show 
Fig.~1a-c as Fig.~\ref{fig:nakamura}.

Low lying $E1$-strength was found.
It shows the 
shape of the $B(E1)$ distribution obtained from the simple zero 
range model of Eq.~(\ref{eq:MElmzerorange}). This distribution of low lying
$E1$-strength is one of the most convincing demonstrations of the 
neutron halo phenomenon.

The Coulomb dissociation of the extremely neutron-rich nucleus 
${}^{19}$C was recently studied in a similar way \cite{Nakamura99}. 
From the shape of the dipole distribution, the neutron separation 
energy of ${}^{19}$C could also be determined to be $530\pm130$~keV.
As can be seen from Eq.~(\ref{eq:MElmzerorange}), this shape depends directly on 
the binding energy parameter $\kappa$.

We note the similarity of the shape of the 
low lying $B(E1)$ strength in the case of $^{11}Be$ (Fig.~\ref{fig:nakamura}),
$^{19}$C (Fig.~\ref{fig:typelb01} above) and also for the deuteron
(see Fig.~4.1, p.~609 in \cite{BlattW79}).
The reason is that all these nuclei are halo nuclei: 
the ground state wave function is an $s$-wave function 
with a small binding energy parameter $\kappa$ and it is 
rather well described by Eq.~(\ref{eq:phigs2}) or~(\ref{eq:phigs}). 
The neutron spends most of 
its time in the classically forbidden region. The continuum is a $p$-wave 
function which differs little from the spherical Bessel function
which describes a free neutron. At the low relevant neutron 
energies there is only little interaction with the strong short range force
in the $p$- or higher partial wave channels. In this case, 
the electric transition dipole moment is given by Eq.~(\ref{eq:MElmzerorange}) 
above.

The $E1$-strength in other 
neutron-rich carbon isotopes has also been studied experimentally:  
Cou\-lo\-mb dissociation of $^{15}$C is again well described 
by a model with a $^{14}$C-core coupled to an $s_{1/2}$ neutron.
The analysis of \cite{Aumann01}
gave a spectroscopic factor of 0.75, consistent
with the one found from $(d,p)$-reactions. It would be of interest 
to compare their results obtained with a $606$~$A$MeV $^{15}$C
beam to those of \cite{Horvath02}, see also Sec.~\ref{sec:astro}.

The Coulomb dissociation of $^{17}$C was also studied. One can
adopt a model of a neutron coupled to a $^{16}$C core. The 
importance of the $2^+$-core excited state is directly evident 
from the Coulomb dissociation measurement: Coulomb 
dissociation is dominantly accompanied  by the emission of photons from
the 1.77~MeV $\gamma$-line from the deexcitation of the $2^+$ state in 
the $^{16}$C core. A sizeable $s$-component of the neutron coupled to the 
$2^+$ core is found,
and it is concluded that the ground state spin of $^{17}$C is either 
$3/2^+$ or $5/2^+$. For further details we refer to \cite{Aumann01}.

It is appropriate to mention at this point also a different exotic nucleus:
the hypernucleus
$^3_{\Lambda}$H. The binding energy $B_{\Lambda}$ of the 
hyperon to the deuteron core is $0.06 \pm 0.06$~MeV.
The Coulomb disintegration of this hypernucleus depends sensitively on 
its binding energy, see \cite{BohmW70,Juric73,Lyuboshits90}.
  
Quite similarly, the Coulomb
dissociation of the $2n$-halo nuclei can be studied experimentally
in order to investigate the low lying $E1$-strength.
Certainly, this is a richer field than the study of
single-neutron halo nuclei, where essentially  
single particle effects
show up. Now there are in addition interesting correlation effects.
E.g., $^{11}$Li and also other $2n$-nuclei like $^6$He are 
so-called {\em Borromean systems}:
the core-$2n$-system is bound, whereas none of the binary subsystems
($n$-$n$, or core-$n$) are bound. 

\begin{figure}[tb]
\begin{center}
\resizebox{0.5\textwidth}{!}{%
\includegraphics{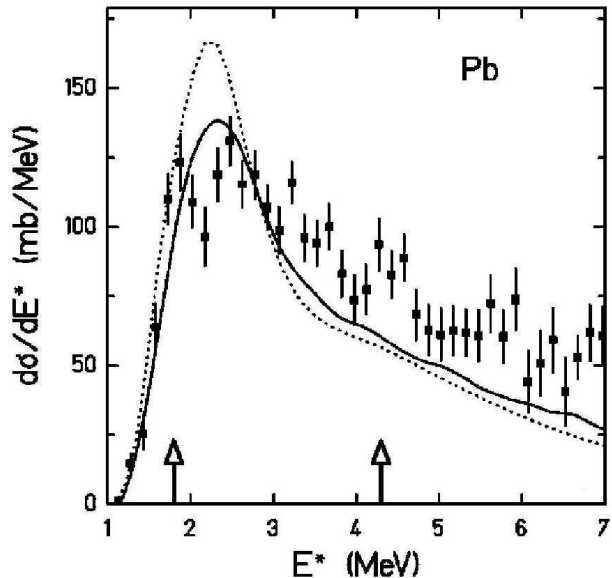}
}
\end{center}
\caption{
The excitation cross section $d\sigma/dE^*$ is shown for the 
breakup of $^6$He at 240~$A$MeV on a Pb target, where $E^*$ has been 
reconstructed from the invariant mass of the $\alpha + n + n$ fragments.
The dotted curve shows the result of a semiclassical perturbative 
calculation using the $dB(E1)/dE*$ distribution from the  three body model
of \protect\cite{DanilinTVZ98}, the solid one
shows the convolution of the dotted curve with the instrumental response. 
The two arrow show two $2^+$ resonances, a known one at an energy of
$E^*=$1.80~MeV and a predicted one at $E^*=$4.3~MeV 
\protect\cite{DanilinVEH97}. Reproduced from 
Fig.~3 of \protect\cite{Aumann99}, where further details can be found.
Copyright (1999) by the American Physical Society.}
\label{fig:aumannhe6}
\end{figure}
\begin{figure}[p]
\begin{center}
\resizebox{0.5\textwidth}{!}{%
\includegraphics{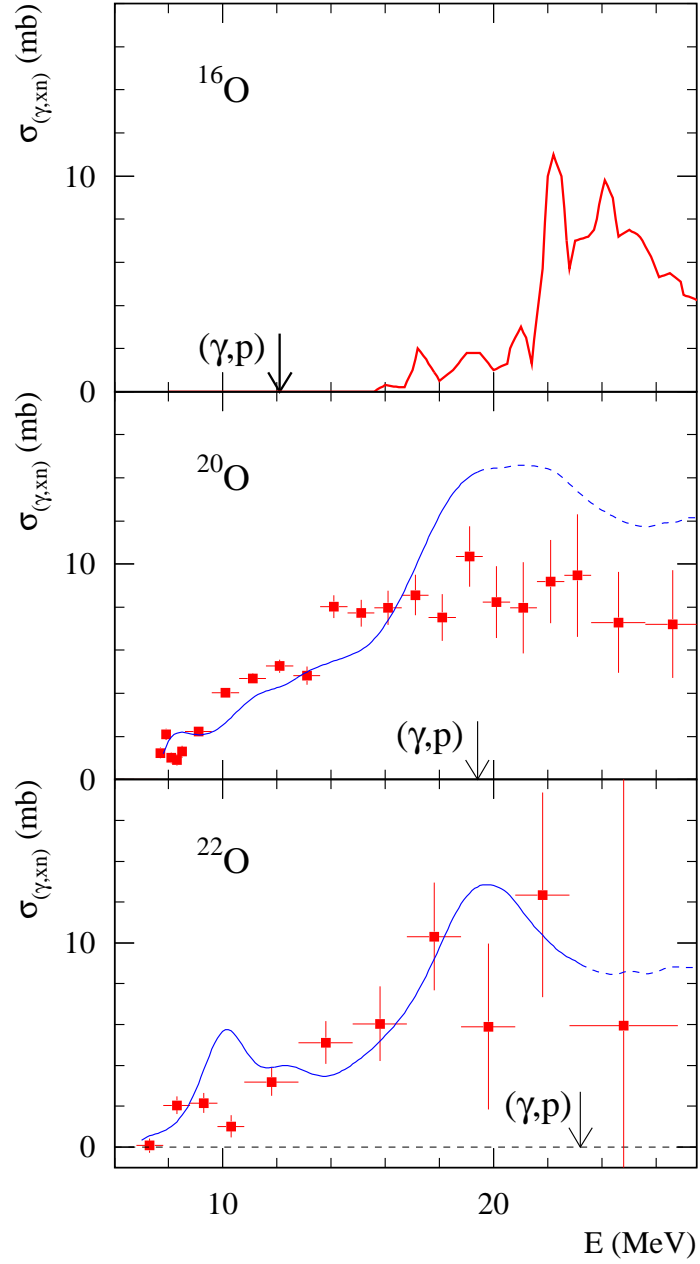}
}
\end{center}
\caption{Photo-neutron cross sections measured \protect\cite
{Leistenschneider01} for the unstable oxygen
isotopes $^{20,22}$O in comparison to that of the stable isotope
$^{16}$O. The data for $^{20,22}$O are compared to shell
model calculations \protect\cite{SagawaS99}.
The $(\gamma,p)$ thresholds are indicated by arrows.  
This figure is taken from \protect\cite{CDR} p.~86.}
\label{fig:leistenschneider}
\end{figure}
The most prominent example is ${}^{11}$Li (for $^6$He see below),
which was studied experimentally in various 
laboratories \cite{Kobayashi89,Shimoura95,Zinser97}.
The momentum distributions for the $(^{11}\mbox{Li},^9\mbox{Li}+n+n)$
breakup reaction were studied theoretically in \cite{EsbensenBI93}
and \cite{EsbensenB92}. In an experiment at 
MSU \cite{IekiG96}, the correlations of the outgoing neutrons were studied. 
Within the limits of experimental accuracy,
no correlations were found. (To be sure there must be 
correlations, as the Borromean effect itself shows, they are just 
hard to see in the breakup experiments). Further studies were reported in 
\cite{Zinser97}. In this reference invariant mass spectroscopy
of $^{10}$Li and $^{11}$Li was performed. It was concluded that
{\it ``a strong di-neutron correlation can be disregarded''}. Low lying $E1$-
strength was found, it exhausts up to about 8\% of the TRK-sum rule
below 4~MeV excitation energy
which corresponds to 96\% of the cluster sum rule assuming a $2n$-halo.
Two low lying structures
were observed for $^{10}$Li ($n$-$^9$Li final state interaction),
see also \cite{BertschHE98}.

The Coulomb breakup of the neutron-rich He-isotopes $^6$He and $^8$He
has also been investigated experimentally \cite{Meister02,Aumann99},
see also Fig.~\ref{fig:aumannhe6}.

The Coulomb dissociation cross-section
at 227~$A$MeV on Pb is about a factor of three smaller in $^8$He than
in $^6$He. This indicates that $^6$He is a halo-nucleus, with
low-lying $E1$-strength, whereas $^8$He is not. 
The low lying strength in $^6$He (up to an excitation energy of 5~MeV) was 
found to exhaust $10\pm2$\% of the TRK sum rule, which corresponds to about
40\% of the cluster sum rule.

The dissociation reaction  $^6\mbox{He} \rightarrow \alpha+2n$ was studied  
at $23.9$~$A$MeV on 
C, Al, Cu, Sn and Pb targets \cite{WangGK02}. For U the Coulomb part 
accounts for $2/3$ of the total two-neutron removal cross section 
$\sigma_{-2n}$. Only a small correlation between the 
two neutrons
was found. The $B(E1)$-strength found is in agreement with \cite{Aumann99},
an experiment with about ten times the beam energy of \cite{WangGK02}.

Similar studies were also carried out in GANIL \cite{Marquez01}.
The neutron-neutron correlation there was analysed in terms of the radius of
the {\em producing source} and using Dalitz plots,
see also \cite{Aumann01}.
Final state effects are taken into account. Differences between the Coulomb
and nuclear breakup are found, which are interpreted as being due to the
direct breakup in nuclear dissociation and a resonant breakup in Coulomb 
dissociation.

Photoneutron cross sections for unstable neutron-rich oxygen isotopes
were recently studied in \cite{Leistenschneider01}. Their results 
for the isotopes $^{20,22}$O are shown in Fig.~\ref{fig:leistenschneider}, 
where also the 
corresponding results for the stable isotope $^{16}$O are given. 
It was found that
there is systematically a considerable fraction of low-lying dipole strength.
Low lying $E1$ strength was also found in the proton-rich isotope $^{13}$O
by a $^{13}\mbox{O} \rightarrow p + ^{12}\mbox{N}$ Coulomb dissociation
measurement \cite{MinemuraMSM02b}.
It is very promising to extend such types of studies to even heavier
unstable neutron- (and also proton-) rich nuclei.  
This is also a challenge for modern nuclear structure theory. 
While we just discussed the single particle aspect, which is the main point 
for 1-$n$, or 2-$n$ halo nuclei, the onset of a {\em pigmy collectivity} is a 
question for more involved theories. 
The exploration of the structure of unstable nuclei is a large 
field for modern nuclear structure theory, which benefits also from 
the modern high computing power.
This field is rapidly expanding, as can be seen from the many conferences which
are taking place on this topic; as recent ones let us mention the
{\it 
``Hirschegg workshop 2003 on Nuclear Structure and Dynamics at the Limits''}
\cite{hirscheggweb} or
{\it 
``INT workshop on Reaction Theory for Nuclei Far From Stability''}\cite{intweb}.
We can refer here only to further references which may, e.g., 
be found in \cite{CDR}.

\section{Nuclear Astrophysics}
\label{sec:astro}

\begin{verse}
\it Du verstehst nicht die Sterne\\
ohne die Kerne.\\
(unknown poet, last century)
\end{verse}

In nuclear astrophysics, radiative capture reactions of the type
\begin{equation}
b + c \to a + \gamma
\label{eq:astro1}
\end{equation}
play a very important role. They can also be studied by the
time-reversed reaction
\begin{equation}
\gamma + a \to b + c \:,
\label{eq:astro2}
\end{equation}
at least in those cases where the nucleus $a$ is in its ground state.

Such a photodissociation experiment has been recently performed with a
real photon beam by Utsunomiya et al.  \cite{UtsunomiyaYA01}.  They
investigated the photodisintegration of $^9$Be with laser-induced
Compton backscattered $\gamma$ rays.  We refer to this work for
further details.

The two cross sections Eqs.~(\ref{eq:astro1}) and~(\ref{eq:astro2})
are related to each other by detailed balance
\begin{equation}
\sigma(b+c \to a +\gamma)=\frac{2(2j_a+1)}{(2j_b+1)(2j_c+1)}
\frac{k_{\gamma}^2}{k_{CM}^2} \sigma(\gamma +a \to b+c\:),
\end{equation}
where the wave number in the $(b+c)$-channel is given by $k_{CM}^2=
2\mu_{bc}E_{CM}/\hbar^2$, with $\mu_{bc}$ the reduced mass; the photon
wave number is given by $k_{\gamma}= (E_{CM}+Q)/(\hbar c)$, where $Q$
is the Q-value of the capture reaction Eq.~(\ref{eq:astro1}).  One
typically has $k_{\gamma}/k_{CM} \ll 1$ for energies not immediately
in the threshold region $k_{CM} \approx 0$ .  Due to this phase space
factor there is a strong enhancement of the dissociation cross section
Eq.~(\ref{eq:astro2}) as compared to the capture cross section
Eq.~(\ref{eq:astro1}).

As a photon beam, we use now the equivalent photon spectrum, which is
provided by the Coulomb field of the target nucleus in the fast
peripheral collision \cite{BaurBR86}. There are previous reviews,
where both experimental, as well as, theoretical aspects have been
given, see \cite{BaurR94,BaurR96}. We want to concentrate here on
theoretical aspects, but also discuss experimental issues, at least
from our point of view as theoreticians. The last review has been
written in 1996 and much progress, on the theoretical, as well as,
experimental side, has been made in the meantime.

In \cite{Austin02} Austin gives a minireview of various indirect
methods in nuclear astrophysics. He remarks that {\it ``it is a common
perception that experimental nuclear astrophysics involves long
measurements of small cross sections at lower and lower energies, so
as to permit a reliable extrapolation to actual astrophysical
energies. This perception is only partially correct.  Recent
developments, especially of radioactive beams, often permit one to
obtain equivalent information with higher energy beams. The high
energy experiments commonly yield higher event rates and sometimes
yield information not available in the classical approach.''}  Coulomb
breakup is a good example.

In non-resonant charged particle reactions the energy dependence of
the cross section is dominated by the penetration of the Coulomb
barrier. This energy dependence is usually factored out by defining
the astrophysical $S$-factor
\begin{equation}
S(E_{CM})= E_{CM} \sigma(E_{CM}) \exp(2\pi \eta),
\end{equation}
where $\eta$ is the Coulomb parameter Eq.~(\ref{eq:etadefined}).

We want to give here some examples of Coulomb breakup experiments of
astrophysical relevance. We also (briefly) explain their astrophysical
relevance and the theoretical background of the experiments. In
judging the conditions of an actual experiment, it is always useful to
keep in mind the values of the basic parameters $\eta$, $\xi$, and
$\chi$ of electromagnetic excitation, see Eqs.~(\ref{eq:xidefined})
and~(\ref{eq:chidefined}).
In this way it is, e.g., quite clear that Coulomb dissociation of
$^8$B is a more favourable case than that of $^{16}$O, with its
comparatively high $\xi$-values (see also Fig.~3 of
\cite{MuendelB96}).

\subsection{\it Li Isotopes}

\subsubsection*{$^6\mbox{Li}\rightarrow \alpha+d$}

The ${}^{6}$Li Coulomb dissociation into $\alpha$+d has been a test
case of the Coulomb breakup method, see \cite{KienerGR91} and
\cite{BaurBR86,BaurR94,BaurR96}. These experiments were carried out
with a 156 MeV $^6$Li beam at the Karlsruhe Isochronous Cyclotron
using the magnetic spectrograph ``Little John''. This reaction is of
astrophysical importance since the $d(\alpha,\gamma)^6$Li radiative
capture is the only process by which $^6$Li is produced in the
standard primordial nucleosynthesis models.  There has been renewed
interest in $^6$Li as a cosmological probe in recent years, mainly
because the sensitivity of searches for $^6$Li has been
increased. $^6$Li has been found in metal-poor halo stars at a level
exceeding even optimistic estimates of how much of it could have been
made in the standard big bang nucleosynthesis. For more discussion on
this, see \cite{NollettLS97}.

A $^6$Li-breakup experiment on $^{208}$Pb at 60~MeV was performed at
the Heidelberg tandem \cite{HesselbarthK91}.  These authors found that
the measured breakup cross section could not be related directly by
first-order Coulomb excitation theory to the astrophysically relevant
$^4\mbox{He}(\alpha ,\mbox{d})^6\mbox{Li}$ capture reaction.  It would
be interesting to redo the analysis with a modern reaction code and
see the influence of higher order electromagnetic and nuclear effects.

An experiment at the much higher beam energy of 150~$A$MeV than at
Karlsruhe \cite{KienerGR91} or Heidelberg \cite{HesselbarthK91} is
under way at GSI.  Results from this experiments are eagerly awaited.

For $(N=Z)$-nuclei electric dipole transitions obey the selection rule
$\Delta T=1$. Since the two fragments $d$ and $\alpha$ have isospin
$T=0$ the electric dipole transition is forbidden by isospin selection
rules.  The contribution from the $E2$ transition should dominate.
$E1$ transitions can still occur due to isospin violations.

$\mbox{d}(\alpha,\gamma)^6$Li radiative capture experiments were also
performed: Robertson et al. \cite{Robertson81} made measurements down
to a relative energy of 1~MeV by detecting the recoil $^6$Li
ions. Cecil et al. \cite{Cecil96} searched for the capture reaction at
a relative energy of $E_{cm}=53$~keV; they obtained an upper limit on
the reaction $S$-factor of $2\times10^{-7}$~MeV$b$. The direct capture
into the $3^+$ resonance was measured also by Mohr et
al. \cite{Mohr94}.

\subsubsection*{$^7\mbox{Li}\rightarrow \alpha + t$}

The reaction $t(\alpha,\gamma)^7$Li is also astrophysically relevant,
and many experimental groups have measured this process in direct
capture experiments.  The nuclide $^7$Li is produced in the early
universe via this radiative capture reaction. The Coulomb dissociation
method has also been applied under various conditions.  We mention
here the direct breakup of 70~MeV $^7$Li scattered from a $^{120}$Sn
target \cite{Shotter84}.  It was found that the breakup is dominated
by the Coulomb interaction for scattering angles inside the grazing
angle.  $^7$Li breakup measurements on $^{197}$Au at 54~MeV were
performed in \cite{Gazes92}.  No straightforward analysis in terms of
first order Coulomb breakup was found to be possible.

A rather detailed measurement of the $^7\mbox{Li} \rightarrow \alpha +
t$ breakup was recently performed by Tokimoto et
al. \cite{TokimotoEA01} with improved experimental techniques.  This
measurement was also accompanied by a theoretical analysis, where the
time dependent Schr\"odinger equation was solved numerically, see
\cite{TypelW99} and also the discussion in Sec.~\ref{sec:higher}.  In
this way, valuable insight into the Coulomb dissociation process, the
$E1$ and $E2$ mixtures, its higher order effects ({\em
postacceleration}) and the time-dependence of the tunneling process
became possible. We refer the reader to this reference for further
details.

\subsubsection*{$^9\mbox{Li} \rightarrow ^8\mbox{Li} + n$}

The cross section for the radiative capture reaction
$^8Li(n,\gamma)^9Li$ was studied with the Coulomb dissociation of a
$^9Li$ beam of $28.53$~$A$MeV at MSU \cite{Zecher98}.

This reaction is of importance for the nucleosynthesis in
inhomogeneous big bang models and in Type~II supernovae. While the
standard big bang nucleosynthesis ends with $^7$Li, nucleosynthesis in
neutron-rich regions could produce an observable amount of $A>8$
nuclei. The $^8$Li$(n,\gamma)^9$Li reaction is in competition with the
$^8$Li$(\alpha,n)^{11}$B reaction in determining the reaction path in
the evolution network.  This reaction may also play an important role
in the explanation of the origin of light neutron-rich nuclei like
$^{36}$S, $^{40}$Ar, $^{46}$Ca and $^{48}$Ca.  Recently, neutron-star
mergers have been proposed as possible alternative sites for an
$r$-process \cite{RosswogFT01}, where similar reaction chains occur.

This reaction is also of importance in order to determine the
different primordial abundances of Li, Be, B, and C, in order to
confine the possibility of inhomogeneous big bang nucleosynthesis. In
such models the neutron-to-proton ratio is different in different
regions in the early universe. In the neutron-rich region this
reaction could therefore lead to a different bridging of the mass
number $A=8$.

The half-life of $^8$Li is less than a second (838ms), this probably
makes a direct measurement of the capture cross section
impossible. One has to rely on indirect methods, like the Coulomb
dissociation of $^9$Li (with a half-life of 178~ms, but the Coulomb
dissociation still works for such a case, as the $^9$Li beam is
produced in a fragmentation reaction). In this experiment
\cite{Zecher98}, only an upper limit could be put on the relevant
cross section.  Still, this is interesting since it directly rules out
two theoretical calculations \cite{MaoC91,MalaneyF89} while the found
upper limit is still consistent with the theoretical results of
\cite{Rauscher94,Descouvement93}.  A new attempt was recently made at
MSU \cite{KobayashiIH02} to measure the $^8$Li ($\gamma$,n)$^9$Li
cross-section by the Coulomb dissociation of $^9$Li with a more
sensitive equipment than in \cite{Zecher98}. It considerably lowers
the upper bounds found there. We refer to this reference
for details.

\subsubsection*{$^{11}\mbox{Li} \rightarrow ^9\mbox{Li} + 2 n$}

This reaction is mainly of interest for nuclear structure studies due
to the Borromean nature of $^{11}$Li. It is discussed in
Sec.~\ref{sec:nstruc} above. Of general interest with respect to
astrophysics is the question of the correlation of the two halo
neutrons. This influences the rate of the two-neutron capture
reaction.
An understanding of this correlation is also of importance for other
two-neutron capture reactions to be discussed in the Sec.~\ref{ssec:2pcapture}.
\subsection{\it $^{14}\mbox{C}(n,\gamma)^{15}\mbox{C}$}
The nuclear structure aspect of the Coulomb dissociation of $^{15}$C
was touched upon in Sec.~\ref{sec:nstruc}, where a measurement at GSI
with a 606~$A$MeV $^{15}$C beam is discussed.  In this experiment, the
$B(E1)$-distribution in the range of excitation energies from the
threshold ($E_{threshold}=1.2184$~MeV) up to about $E=8$~MeV was
studied.

\begin{figure}[tb]
\begin{center}
\resizebox{0.7\textwidth}{!}{%
\includegraphics{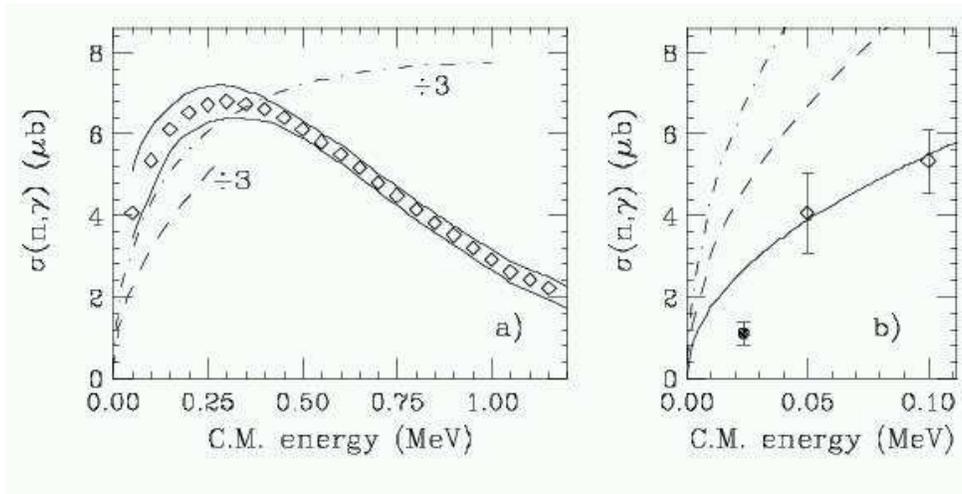}
}
\end{center}
\caption{
(a) The $(n,\gamma)$ capture cross sections on $^{14}C$ are shown as a
function of the c.m. energy.  The diamonds enclosed by two solid
curves define the region of the experimental results. Plotted as 1/3
of the actual values the dashed curve is a prediction of\
\protect\cite{WiescherGT90}, the dot-dashed curve is a prediction of
\protect\cite{Descouvement00}.
(b) Blowup of the low energy region showing the same theoretical
curves (not divided by 3), the experimental result of
\protect\cite{Beer92} (solid point) and the lowest energy points
(diamonds) of \protect\cite{Horvath02} with a kinematical fit (solid
curve).  This is Fig.~8 of \protect\cite{Horvath02}.
}
\label{fig:horvath}
\end{figure}
Now we are interested in thermonuclear energies, i.e., relative
$n+^{14}$C energies from about 10~keV up to about 300~keV
(corresponding to temperatures between 0.1 --- 3$\times 10^9$~K).  The
Coulomb dissociation of $^{15}$C was discussed in \cite{BaurR96} as a
way to investigate the radiative capture reaction
$^{14}\mbox{C}(n,\gamma)^{15}\mbox{C}$. This reaction is important in
neutron-induced CNO cycles of stellar evolution phases beyond the main
sequence. It is also relevant in inhomogeneous big bang scenarios. A
$^{15}$C Coulomb dissociation experiment was performed at MSU
\cite{Horvath02} where details and further references can be found.
The present status is not entirely conclusive, as one can see from
Fig.~\ref{fig:horvath} (Fig.~8 of \cite{Horvath02}).  The excitation
function for the $^{14}\mbox{C}(n,\gamma)$ cross section was measured
in \cite{Horvath02} with an $E/A=35$~MeV beam.  This energy is much
lower than the energy of 606~$A$MeV used at GSI. It is still high
enough to break up the $^{15}$C nucleus and excite the low energy
continuum relevant for astrophysics. The deduced excitation function
is shown as diamonds in Fig.~\ref{fig:horvath}(a). They are enclosed
by the two solid curves, which define the region of the experimental
results of the Coulomb dissociation experiment of \cite{Horvath02}.
Plotted as $1/3$ of the actual values, the dashed curve is a
theoretical prediction of Wiescher et al. \cite{WiescherGT90} and the
dot-dashed curve is a prediction of \cite{Descouvement00}.  In
Fig.~\ref{fig:horvath}(b) a blowup of the low energy region is shown.
The result from the direct capture experiment at~23~keV of the
Karlsruhe group \cite{Beer92} (solid point) and the low energy points
of the MSU experiment with a kinematical fit (solid curve) are
shown. For further details see \cite{Horvath02}.

The excitation function for the $^{14}\mbox{C}(n,\gamma)$-reaction, as
deduced from the Coulomb breakup, is consistent with an $E_n^{1/2}$
rise, as expected for $p$-wave capture.  A fit gives
$\sigma_{n,\gamma}(23\mbox{keV})= 3.2\pm 0.9 \mu$b, about three times
higher than the result of \cite{Beer92}.  Also, the shape of the
excitation curve has a peak at around $E_n=200$~keV. On the other
hand, theoretical predictions by Wiescher et al. \cite{WiescherGT90}
and Descouvemont \cite{Descouvement00} increase monotonically in this
energy range.  Another issue is the radiative capture contribution to
the $5/2^+$-excited state at 740~keV. This contribution is not
included in the Coulomb breakup measurement.  It is estimated to be
small on theoretical grounds: the $E_{\gamma}^3$-dependence of the
dipole transitions and the measured spectroscopic factors favour the
ground state transition.  We conclude that more work is necessary to
resolve these questions.

\subsection{\it $^{16}\mbox{O} \rightarrow \alpha+^{12}$C Coulomb
Dissociation}

The helium burning reaction
$^{12}\mbox{C}(\alpha,\gamma)^{16}\mbox{O}$ at thermonuclear energies
is a key process for the evolution of massive stars and for the
nucleosynthesis of $^{16}$O and heavier elements up to Fe. Despite its
importance for nuclear astrophysics, the cross section for this
reaction is still quite uncertain in the stellar energy domain
($E_{cm} \sim 300$~keV). The reason is the extremely small value of
this cross section and the superposition of $E1$- and $E2$-multipole
contributions in the capture process.  Several direct measurements of
the radiative capture reaction, elastic $\alpha+^{12}$C scattering and
$^{16}$N-decay studies have succeeded to determine reasonably well the
$E1$-part of the astrophysical $S$-factor. However, the $E2$-part,
which is thought to be of about the same magnitude as the $E1$-part,
is still quite uncertain.

In this situation, Coulomb breakup seems to be an interesting method,
since $E2$-excitations are enhanced as compared to $E1$, see
Sec.~\ref{sec:theory}, especially Eq.~(\ref{eq:afirelat}).  However,
several points render this method quite challenging. The $\alpha+
^{12}$C-breakup threshold $E=7.162$~MeV is quite high, which means
that the $\xi$-values, even at high $^{16}$O beam energies are not
small (for the experiment mentioned below we get a value of about
$\xi=0.8$). The relatively low flux of equivalent photons at these
high energies leads to a serious competition with nuclear excitation
effects. Also, there is $E1$- and $E2$-interference.  Undaunted by all
this, a breakup experiment was performed at GANIL at a projectile
energy of 95~$A$MeV. Assuming only $E2$ Coulomb and $\Delta L=2$
nuclear excitation the measured breakup cross sections were compared
to optical model calculations and an $E2$ $S$-factor was extracted in
\cite{Kiener01}.  This analysis is based on theoretical calculations
with the ECIS-code in \cite{Tatischeff95} and calculations of the
fragment angular correlations in \cite{BaurW89}. Further experimental
studies were done recently at KVI , Groningen at 80~$A$MeV
\cite{Fleurot02}.  The large angular and momentum acceptances make the
spectrometer BBS a very suitable device for such coincidence studies.
A further experiment is planned at MSU which can use all the
experience gained from the previous experiments.  The difficulties of
the experiment were identified in the previous approaches and can
hopefully be overcome.  Much is at stake. As Wolfgang Pauli said (when
he invented the neutrino): {\it ``Nur wer wagt, gewinnt!''  (``Nothing
ventured, nothing gained'')}.

\subsection{\it Some $(p,\gamma)$-Reactions studied with Coulomb
Dissociation}
The Coulomb breakup of $^{14}$O has been studied at RIKEN
\cite{Motobayashi91} and GANIL \cite{Kiener93} in order to determine
the $S$-factor of the $^{13}$N$(p,\gamma) ^{14}$O capture reaction.
It is astrophysically relevant for the hot CNO cycle. The $S$-factor
is dominated by the $1^-$ resonance in $^{14}$O which can be reached
from the ground state in an $E1$ transition. Such a case is certainly
simpler to study than a transition into the (flat) continuum. We refer
to previous reviews \cite{BaurR94,BaurR96}. We note that there is good
agreement with the direct radiative capture measurement at Louvain-la
Neuve \cite{Decrock91}.  This consistency can also be considered as a
test of the Coulomb dissociation approach.  It is also worth
mentioning that as a by-product of the $^{14}$O Coulomb dissociation
approach the transition from the $\frac{1}{2}^-$ ground state to the
first excited $\frac{1}{2}^+$- state in $^{13}$N was studied in
\cite{Motobayashi91}.  The corresponding value of the radiative width
$\Gamma_{\gamma}$ was found to be in good agreement with values from a
different previous experiment.  Again, this can be considered as a
confirmation of the Coulomb dissociation method.  The accuracy of the
RIKEN measurements \cite{Motobayashi91} was improved in new
experiments at RIKEN \cite{SerataMSA02}.  There is agreement with the
previous studies; the reliability of the Coulomb dissociation method
has been tested to an accuracy of the order of 10 percent in the
$^{12}\mbox{C}(p,\gamma)^{13}\mbox{N}$ case.

The $^{12}$N $\rightarrow {}^{11}\mbox{C}+p$ breakup reaction was studied at
GANIL \cite{Levebvre95} in order to study the
$^{11}$C$(p,\gamma)^{12}$N radiative capture process. This reaction is
relevant for the hot $pp$-chain. From the experimental breakup yield,
the radiative width of the 1.19~MeV level in $^{12}$N and the
spectroscopic factor for the direct proton capture on $^{11}$C have
been extracted. The radiative width of the 1.19~MeV level is found to
be smaller by more than one order of magnitude compared to a recent
theoretical calculation but in rough agreement with an estimate by
Wiescher et al. \cite{WiescherT89}.  We refer to this reference for
further details.  The $^{11}\mbox{C}(p,\gamma)^{12}\mbox{N}$ reaction
was also studied at RIKEN by the Coulomb dissociation method
\cite{Motobayashi02,MinemuraMSM02a}.  Their result is consistent with
the GANIL measurement \cite{Levebvre95}, the accuracy of the value for
the radiative width of the 1.19~MeV level in $^{12}$N is improved; for
further details see these references.  An experimental study of the
$^{22}\mbox{Mg}(p,\gamma)^{23}\mbox{Al}$ reaction by the Coulomb
dissociation method is also reported at RIKEN \cite{GomiMYK02}.  This
reaction is relevant for the nucleosynthesis of $^{22}$Na in Ne-rich
novae.

A very instructive test case would be $^{17}$F-Coulomb dissociation.
There are very accurate radiative capture experiments \cite{Rolfs73}
and \cite{Morlock97} to which one could compare possible Coulomb
breakup measurements. The radiative capture reaction goes to the
ground state and the $\frac{1}{2}^+$ excited state, with the Coulomb
breakup method one can only study the ground state transition. There
are data for the 170~MeV $^{17}$F dissociation into $p+^{16}\mbox{O}$
\cite{Liang00}.  They were analyzed theoretically by Esbensen and
Bertsch \cite{EsbensenB02}; substantial higher order effects were
found at this energy.  Higher $^{17}$F beam energies would reduce
these effects and greatly facilitate the analysis.  A recent work
\cite{BertulaniD02} found large nuclear effects in the dissociation of
$^{17}$F on $^{208}$Pb at an energy of 65~$A$MeV. This would make an
extraction of the electromagnetic transition matrixelement quite
difficult.
  
\subsection{\it $^8B$ Coulomb Dissociation and the Solar 
Neutrino Problem}  
 
\subsubsection*{Astrophysical motivation}

Two kinds of extraterrestrial neutrinos, solar neutrinos and
supernovae neutrinos, have been observed and their measured fluxes, as
well as, energy spectra, shed a new light on astrophysics and particle
physics \cite{nobelprice02}.  In this context, the
${}^{7}$Be(p,$\gamma$)$^{8}$B radiative capture reaction is relevant
for the solar neutrino problem.  It determines the rate of production
of ${}^{8}$B which leads to the emission of high energy neutrinos. The
SNO experiment \cite{Ahmad01} has measured the $\nu_e$ flux in the
{\em charged current} (CC) reaction
\begin{equation}
d+\nu_e \rightarrow p+p+e^- ,
\label{eq:astroCC}
\end{equation}
and also the {\em neutral current} (NC) reaction
\begin{equation}
d+\nu_x \rightarrow p+n+\nu_x
\label{eq:astroNC}
\end{equation}
was recently measured \cite{Ahmad02a,Ahmad02b}.  The CC reaction is
only sensitive to electron neutrinos, whereas the NC reaction is
equally sensitive to all neutrino flavours $\nu_x$ where $x=e$, $\mu$,
and $\tau$.

The Kamiokande and SNO detectors also measure neutrino {\em elastic
scattering} (ES) on the electrons:
\begin{equation}
\nu_x +e^- \rightarrow \nu_x + e^-.
\label{eq:astroES}
\end{equation}
This reaction is sensitive to all kinds of neutrino flavours, mainly
to electron neutrinos but also to some extent (to about 14\%) to
$\nu_{\mu}$ and $\nu_{\tau}$ via the neutral current interaction.
Comparison of the two neutrino fluxes from SNO and Kamiokande and of
the NC and CC results of SNO are a direct proof that there are
neutrino oscillations.  Some of the Kamiokande and SNO neutrinos come
from flavours other than the electron flavour
\footnote{ Oscillations of antineutrinos (coming from quite a few
nuclear power reactors about 180~km from the detector) were recently
also found by the KAMLAND experiment \protect\cite{Eguchi02}.}.

In this argument the astrophysical $S$-factor $S_{17}$ does not enter,
since one considers only the ratios of fluxes.  However, this
$S$-factor is still of vital importance: it determines the absolute
value of theoretical calculations of the solar $^8$B $\nu$ flux and
therefore many efforts were undertaken to study the
$^7\mbox{Be}(p,\gamma)^8\mbox{B}$ reaction \cite{BahcallPB01,BahcallBP98}.

In scenarios involving sterile neutrinos it is important to know the
standard solar model $^8$B neutrino flux in order to derive which part
of the flux has oscillated into the sterile ones.  We quote from
\cite{BargerMW02}: {\it ``Thus, the $^8$B plays a crucial role in the
interpretation of these experiments.  Unfortunately, the predicted
value of the $^8$B flux normalization is quite uncertain, mainly due
to poorly known nuclear cross sections at low energies.''}

For recent developments we refer to \cite{bahcallweb}.  The situation
is summarized in \cite{JunghansMS02}: {\it Improved (\/${}^8$B) production
rate predictions are very important for limiting the allowed neutrino
mixing parameters, including possible contributions of sterile
neutrinos. The astrophysical $S$ factor $S_{17}(0)$ for this reaction
must be known to $\pm 5\%$ in order that this uncertainty not be the
dominant error in predictions of the solar $\nu_e$ flux.}

\subsubsection*{Direct measurements of the 
$^7\mbox{Be}(p,\gamma)^8\mbox{B}$-capture cross section}

There are direct reaction measurements, for the recent ones see
Ref.~\cite{Hammache98,Strieder01,JunghansMS02,BabyBG02}. 
The target is the radioactive nucleus $^7$Be
and a major problem is the determination of the thickness of this
target. The reaction has been measured at $E_{c.m.}=185.8, 134.7$ and
$111.7$~keV and the zero energy $S$-factor inferred from these data is
18.5$\pm 2.4$~eVb. This reaction has been studied recently
in \cite{Terrasi01} with a $^7$Be radioactive beam. In this way the
target thickness problem is overcome, see also the discussion in
\cite{Austin02}.

There is an experiment by Junghans et al. from Seattle and TRIUMF
\cite{JunghansMS02} with a somewhat high value of the $S$-factor. The
cross section values of this paper are currently being revised, see
Ref.~11 of \cite{BabyBG02}.  A precision measurement at
$E_{lab}=991$~keV with an implanted $^7$Be target was recently reported
in \cite{BabyBG02}.

\subsubsection*{Coulomb dissociation of ${}^{8}$B}
While there are certainly experimental problems with the direct
capture experiments, there are (in addition to the specific
experimental problems of the Coulomb dissociation method) problems of
the theoretical analysis of the experiments in terms of the
astrophysical $S$-factor.  These theoretical problems can be solved in
principle, as peripheral reactions are the ones best understood on a
quantitative basis.  They are mainly related to higher order
electromagnetic effects (see also Sec.~\ref{sec:higher} above),
nuclear effects (see Sec.~\ref{sec:nuclear}) and the mixture of
different multipoles (especially $E1$ and $E2$). Yet, the Coulomb
dissociation of $^8$B offers some advantages: at the low binding
energy of $^8$B of 137~keV, the relevant photon fluxes are very
high. Experiments under various conditions have been performed, which
are affected in different ways by the different problems mentioned
above. With all these problems identified a consistent picture emerges
and an $S$-factor is obtained in an entirely different way as compared
to the direct measurements.  This in itself is valuable.

The story of the mid-nineties has been vividly told by Taube
\cite{Taube94} and efforts to determine the astrophysical $S$-factor
$S_{17}$ by many different methods including Coulomb dissociation
continue up to now.

We now discuss the various experimental and theoretical contributions:
Higher order perturbation theory using the semiclassical approximation
was studied in \cite{TypelB94b,TypelWB97}.  The semiclassical
time-dependent Schr\"odinger equation was solved in
\cite{EsbensenB96,EsbensenB95}. CDCC calculations were performed in
\cite{MortimerTT02}, which become increasingly cumbersome for higher beam
energies.  The reactions at the different radioactive beam facilities
have been done under different kinematical conditions, the highest
$^8$B beam energy is at GSI, 254~$A$MeV \cite{Iwasa99}, at MSU it is
about 40 to 83 $A$MeV \cite{Kelley96,Davids01}.  At RIKEN it is about
50~$A$MeV \cite{Motobayashi94}.

The measurement at Notre Dame is at a very low $^8$B beam energy of
26~MeV \cite{Schwarzenberg96}. For such beam energies higher order
effects are very important and it is not a simple task to extract a
model independent $S$-factor. Nevertheless, these experiments are of
great interest since they allow to test the dynamics of the Coulomb
breakup process. With modern reaction codes (e.g. solving of the
time-dependent Schr\"odinger equation
\cite{TypelW99,EsbensenB02,EsbensenB02b}, see also
Sec.~\ref{sec:higher}).  One can well describe such reactions,
however, the connection to the astrophysical $S$-factor is becoming
somewhat indirect.
\begin{figure}[p]
\begin{center}
\leavevmode\psfig{file=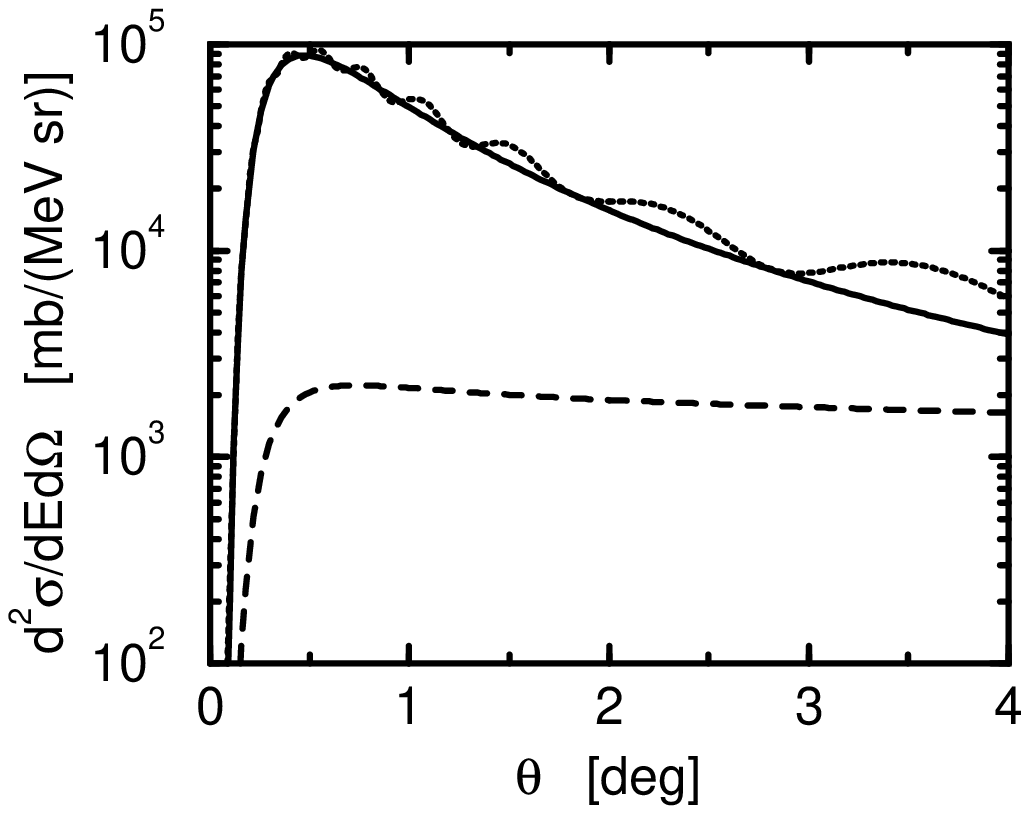,height=5.5cm,width=5.5cm}
\leavevmode\psfig{file=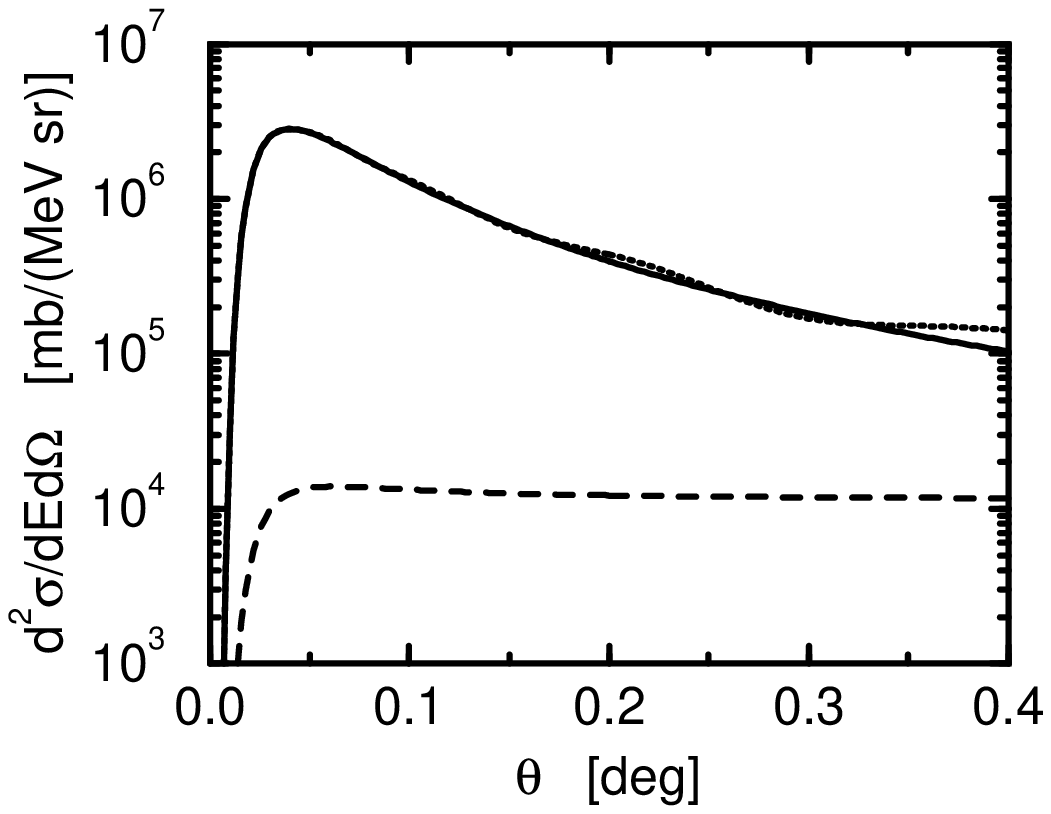,height=5.5cm,width=5.5cm}
\end{center}
\caption{
Coulomb dissociation cross section of ${}^{8}$B scattered on
${}^{208}$Pb as a function of the scattering angle for projectile
energies of 46.5~$A$MeV (left) and 250~$A$MeV (right) and a
${}^{7}$Be-$p$ relative energy of 0.3~MeV.  First order results $E1$
(solid line), $E2$ (dashed line) and $E1+E2$ excitation including
nuclear diffraction (dotted line).  (From Figs.~4 and~5 of
Ref.~\protect\cite{TypelWB97}.)  }
\label{fig:typel97}
\end{figure}
\begin{figure}[p]
\begin{center}
\resizebox{0.35\textwidth}{!}{%
\includegraphics{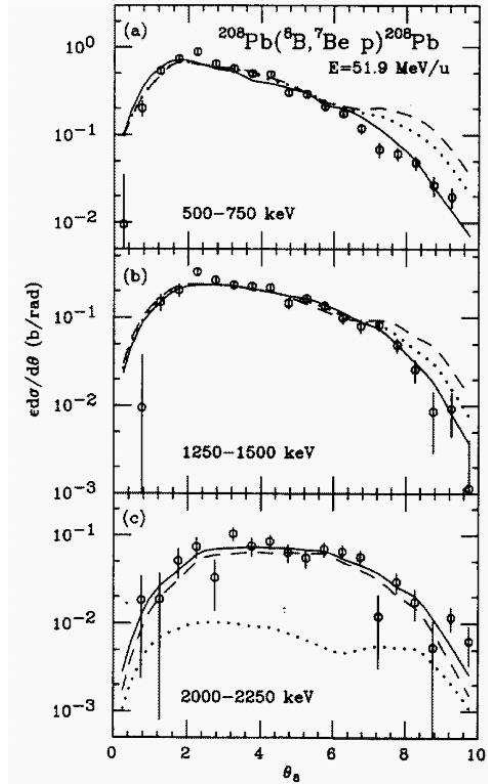}
}
\end{center}
\caption{Observed cross sections $\epsilon d\sigma /d\theta$
are shown as a function of the outgoing $^8\mbox{B}^*$ scattering
angle for the three indicated relative energy bins. The $y$ axis shows
the product of the detection efficiency $\epsilon$ and the
differential cross section.  The solid curve represents the best fit
obtained with calculated $E1$ (dashed curve in (c)) and $E2$ (Coulomb
and nuclear: dotted curve in (c)) amplitudes. For the first two energy
bins shown in (a) and (b), the best fits result in pure $E1$
transitions.  Dashed and dotted curves in (a) and (b) correspond to
the results calculated with $l=1$ and $l=2$ components predicted by
some theoretical models.
This is Fig.~2 from \protect\cite{Kikuchi97}, where further details
can be found.  }
\label{fig:plb391}
\end{figure}
\begin{figure}[p]
\begin{center}
\resizebox{0.44\textwidth}{!}{%
\includegraphics{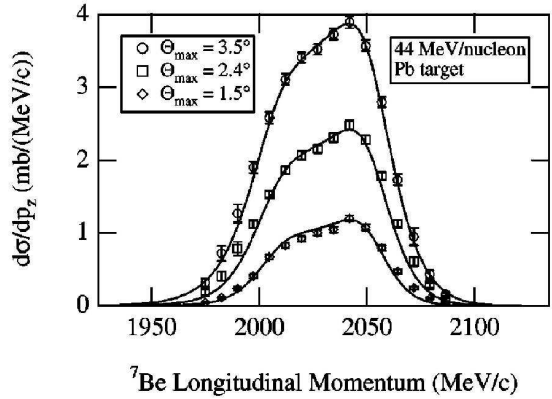}
}
\end{center}
\caption{The measured longitudinal momentum distributions of $^7$Be 
fragments from the Coulomb dissociation of 44 $A$MeV $^8$B on Pb with
several maximum $^7$Be scattering angle cuts are shown. The
measurements are compared to a first order perturbation theory
calculation convoluted with the experimental resolution. The asymmetry
due to the $E1$-$E2$ interference is clearly seen. The overall
normalization and the $E2$ matrix element of this calculation are
scaled. Reproduced from Fig.~8 of \protect\cite{Davids01}, to which we
refer for further details.
Copyright (2001) by the American Physical Society.}
\label{fig:Davids8}
\end{figure}
\begin{figure}[p]
\begin{center}
\resizebox{0.49\textwidth}{!}{%
\includegraphics{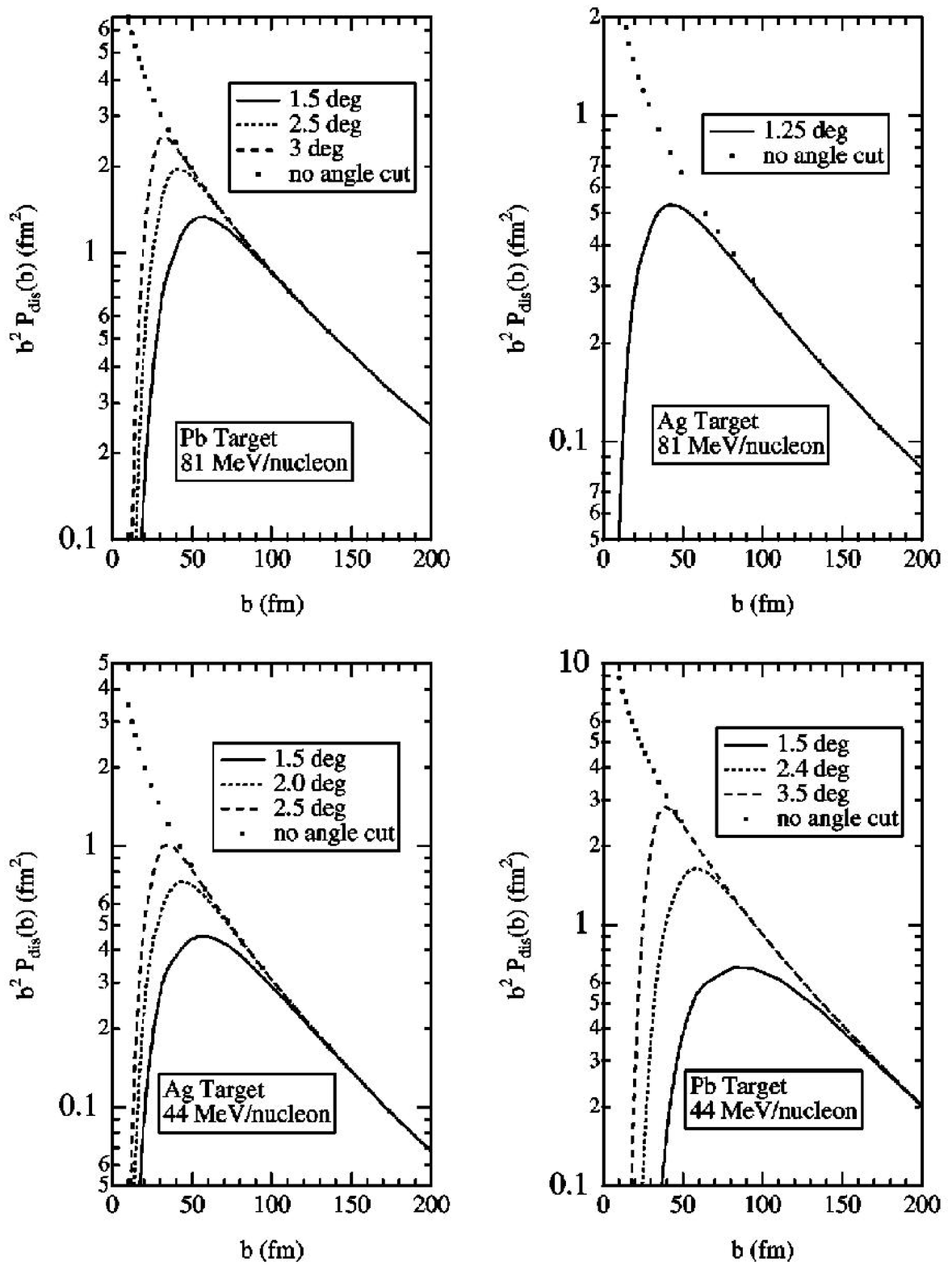}
}
\end{center}
\caption{The product of the Coulomb dissociation probability
$P(b)$ and $b^2$, where $b$ is the impact parameter are shown as a
function of $b$. Various $^7$Be scattering angle cuts are applied, as
indicated in the figures.  Reproduced from Fig.~10 of
\protect\cite{Davids01}.
Copyright (2001) by the American Physical Society.}
\label{fig:Davids01}
\end{figure}

One of the main issues is the contributions of the $E1$ and $E2$
multipoles and their separation.  As can be shown from
Eq.~(\ref{eq:afirelat}) the $E2$ photon number at intermediate and
lower energies is enhanced, therefore they can contribute considerably
to the Coulomb breakup cross section, whereas they are a small
contribution to the photodissociation (or radiative capture) cross
section.  At RIKEN, one uses the different angular behaviour of the
$E1$ and $E2$ multipolarities in order to disentangle the
contributions, see \cite{Kikuchi97,Iwasa96,Kikuchi98}.  The $E2$
contribution shows up predominantly at smaller impact parameters,
i.e., at larger scattering angles.  In Fig.~\ref{fig:typel97} we show
some theoretical calculations related to this point
\cite{TypelWB97}. The $E1$ contribution shows a
$(\sin\theta)^{-2}$-behaviour, whereas the $E2$ contribution is almost
constant, down to a minimum angle $\theta_{adiabatic}$.
Nuclear diffraction effects are small.  This is used in the RIKEN
experiment to determine the $E2$ contribution. We show their result in
Fig.~\ref{fig:plb391}. These authors conclude that {\it ``although
systematic effects could be large, the extracted $E2$ components
appear to be quite small''.}

In the MSU method, one uses the asymmetries, which are induced by the
$E1$-$E2$ interference effects in order to sensitively determine the
$E2$ $S$-factor \cite{DavidsAA98,Davids01,DavidsEA01}.  We show their
longitudinal momentum distribution here in Fig.~\ref{fig:Davids8}.

An intuitively appealing picture of the Coulomb dissociation process
can be gained \cite{Davids01} by plotting $b^2 P_{diss}(b)$, where
$P_{diss}(b)$ is the Coulomb dissociation probability with various
$^7$Be scattering angle cuts.  This is shown in
Fig~\ref{fig:Davids01}; one can clearly see that Coulomb dissociation
essentially takes place outside the nuclear interaction region.

\begin{figure}[tb]
\begin{center}
\resizebox{0.35\textwidth}{!}{%
\includegraphics{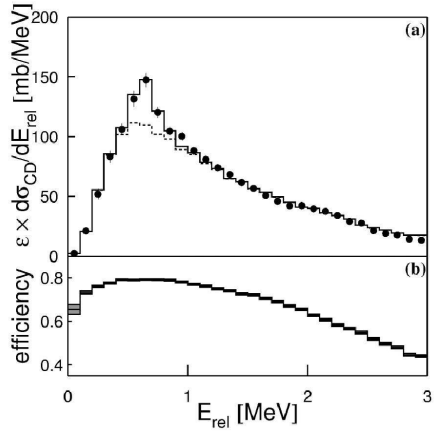}
}
\end{center}
\caption{Yields of breakup events 
(cross section times efficiency) plotted as a function of the relative
energy. The solid and dashed histograms denote simulated $E1+E2$ and
$E1$ yields respectively. Although the $M1$ contribution is reduced by
about a factor of $(v/c)^2$ in the Coulomb dissociation the $M1$ peak
(smeared out by the experimental resolution) is clearly seen in the
GSI data. Reproduced from  Fig.~1 of \protect\cite{Iwasa99}.
Copyright (1999) by the American Physical Society.}
\label{fig:Iwasa01}
\end{figure}

In a particle-core model one can see qualitatively the relative
importance of the various multipolarities.  The effective charges are
given by
\begin{equation}
Z_{eff}^{(\lambda)}=Z_b\left(\frac{m_c}{m_b+m_c}\right)^{\lambda}+
Z_c\left(\frac{-m_b}{m_b+m_c}\right)^{\lambda},
\end{equation}
with the charge numbers $Z_b$ and
$Z_c$ and the masses $m_b$ and $m_c$ of the fragments. We see that
$E1$ charges are relatively small, $E2$ charges are large: for
$^8\mbox{B}=^7\mbox{Be}+p$ we have $Z_{eff}^{(1)}=3/8$ and$
Z_{eff}^{(2)}=53/64$. This can be compared to a neutron-core system:
for $^{11}\mbox{Be}=^{10}\mbox{Be} + n$ we find $Z_{eff}^{(1)}=4/11$
and $Z_{eff}^{(2)}=4/121$. Furthermore, $M1$ transitions are
suppressed by about a factor of $(v/c)^2$.  This $M1$ peak is nicely
seen in the GSI experiment. It is somewhat smeared out due to the
resolution in $E_{rel}$ in the $p+^7$Be invariant mass, see
Fig.~\ref{fig:Iwasa01}.
\begin{figure}[tb]
\begin{center}
\resizebox{0.66\textwidth}{!}{%
\includegraphics{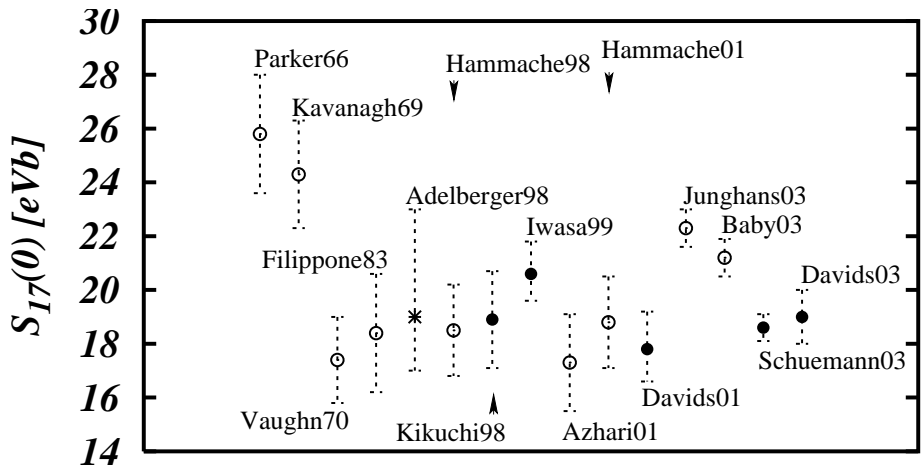}
}
\end{center}
\caption{
The zero-energy astrophysical $S$-factor $S_{17}$ for the
$^7\mbox{Be}(p,\gamma)^8\mbox{B}$ reaction is shown from selected direct and
indirect methods.
The open circle denote results obtained from direct measurements, the full
circle those of Coulomb breakup experiments. The star is the value given
in \cite{Adelberger98} as proposed reference value.
Adapted from Fig.~19 of \protect\cite{Davids01}.
}
\label{fig:davids19}
\end{figure}

A new experiment on $^8$B Coulombdissociation was recently completed at GSI
\cite{Schuemann03}.
They also pointed out that most extrapolations currently done are based on 
the cluster model of Descouvement \cite{DescouvementB94}, which does not 
seem to fully reproduce the energy dependence. This could partly explain the 
discrepancies of 
the different experiments done at different relative energies.
This idea is pursued further in \cite{DavidsT03} taking into account a full 
dynamic reaction calculation, as well as, an improved potential model for 
$^{8}B$. 
A recent minireview can also be found in  \cite{Gai03}, where also a future
CERN/ISOLDE measurement using a radioactive $^7$Be beam is briefly discussed.

The experimental results was summarized in Fig~19 of \cite{Davids01}. We have
updated this figure with the latest results and show it in
Fig.~\ref{fig:davids19}.
Altogether it is quite remarkable that completely different
experimental methods with possibly different systematic errors lead to
results that are quite consistent.

\section{Possible New Applications of Coulomb Dissociation
 for Nuclear Astrophysics}
\label{sec:future}

Radioactive beam facilities all over the world have considerably
widened the scope of nuclear physics in the past years. 
In the previous sections we described the applications
of Coulomb excitation and dissociation to nuclear structure
and astrophysics. For example halo nuclei and new 
regions of deformation could be studied in detail  
for light exotic nuclei. With the  
new forthcoming generation of RIA's also medium and 
heavy nuclei will be produced with sufficient intensities,
with further opportunities. We  mention 
the project in Europe \cite{CDR}, 
where a substantial part of the program is in Rare Isotope 
Beams research.
In the US the scientific opportunities with fast fragmentation 
beams from the Rare Isotope Accelerator are intensively investigated
see \cite{riawhitepaper}, and also \cite{townmeeting99}.
In Japan, at RIKEN, Wako the construction of a 
Radio Isotope Beam Factory (RIBF) is underway, see e.g.
\cite{ribfweb}.

A status report is also provided in \cite{Nupecc00}.
The astrophysical scenario is presented 
by K\"appeler et al. in \cite{KaeppelerWT98}.
In the following we indicate some of 
the future opportunities in this field related to nuclear 
astrophysics, especially  to the $r$- and $rp$-processes.

\subsection{\it Electromagnetic Properties of $r$-Process Nuclei} 

Nucleosynthesis beyond the iron peak proceeds mainly by the 
$r$- and $s$-processes (rapid and slow neutron capture). 
This is widely discussed 
in the literature, see e.g. \cite{RolfsR88,CowanTT91} or also \cite{tommyweb}.
Concise {\em executive summaries} are provided in \cite{CDR} (p.~96f)
or \cite{riawhitepaper} (p.~54).

In order to assess
the $r$-process path, it is important to know the nuclear properties like
$\beta$-decay half-lives and neutron binding energies. In the
waiting point approximation \cite{RolfsR88,CowanTT91} an (n,$\gamma$)-
and ($\gamma$,n)-equilibrium is assumed in an isotopic chain.
As the nuclei inside the isotopic chain are assumed to be in a thermal 
equilibrium, only binding energies and $\beta$-decay half{}lives are needed.

In general
the waiting point approximation should be replaced by a dynamic
$r$-process flow calculation, taking into account (n,$\gamma$), ($\gamma$,n)
cross sections
and $\beta$-decay rates, as well as, time-varying temperature and neutron
density for the astrophysical scenario. 
In this case the knowledge of (n,$\gamma$) cross sections is of importance. 

To establish the quantitative details
of the $r$-process, accurate energy-averaged neutron-capture cross
sections are needed. Such data provide information on the mechanism
of the neutron-capture process and time scales, as well as, temperatures
involved in the process. The data should also shed light on neutron
sources, required neutron fluxes and possible sites of the
processes (see Ref.~\cite{RolfsR88}).
\begin{figure}[p]
\begin{center}
\resizebox{0.6\textwidth}{!}{%
\includegraphics{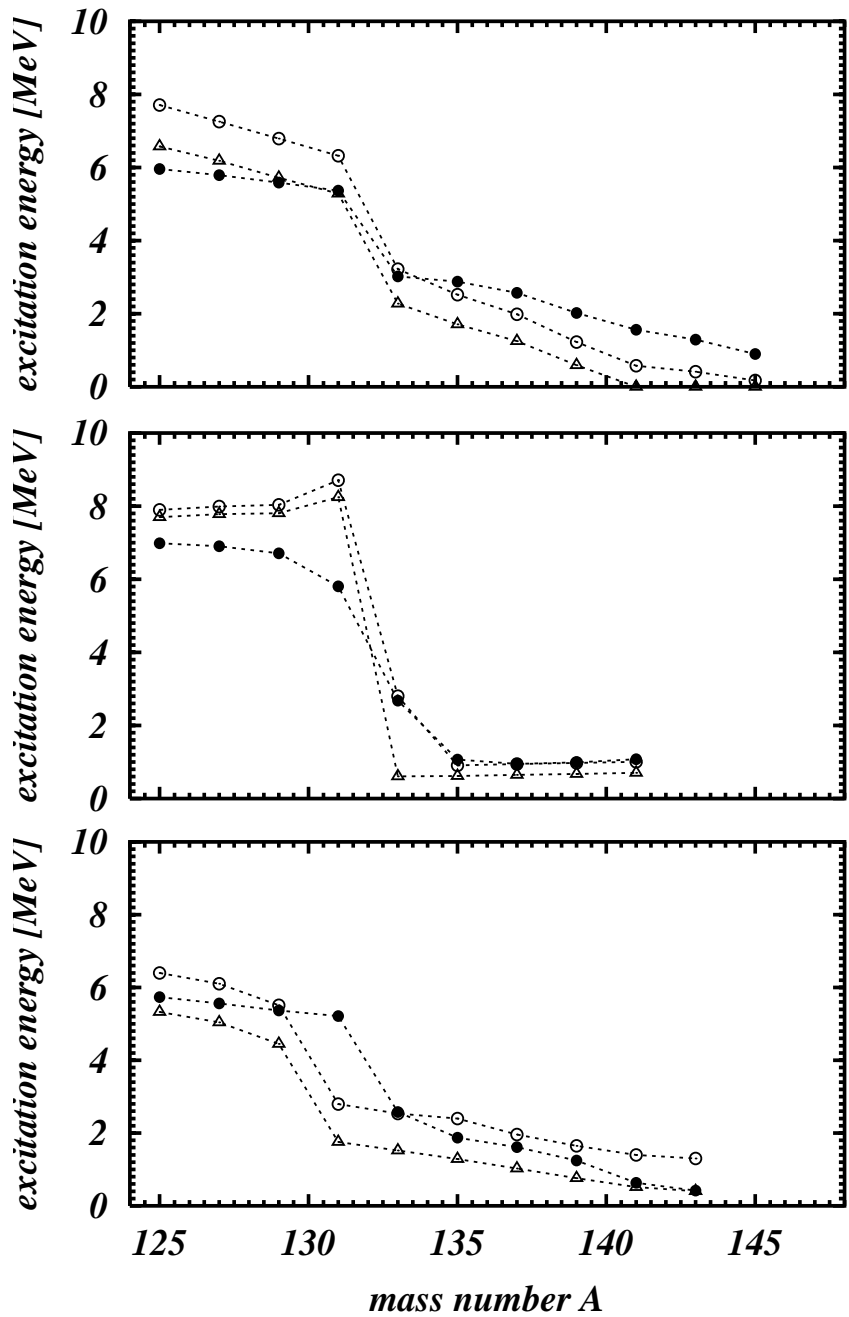}}
\end{center}
\caption{
Dependence of level energies on mass number for even-odd
Sn isotopes calculated in the Hartree-Fock-Bogolyubov
model (HFB) (upper panel), in the Relativistic Mean Field approach (RMF) 
(middle panel) and in the Finite Range Droplet Model (FRDM) (lower panel). 
The $1/2^-$state (open
circles), the $3/2^-$ state (triangles) and the calculated
neutron separation energy (full circles) are shown. The lines are drawn
to guide the eye. These are adapted from the Figs.~10--12 of \cite{Rauscher98}.
} 
\label{fig:Rauscher10}
\end{figure}
In this situation, it is very instructive to look at the model studies of 
\cite{Rauscher98} and the scenario described in \cite{Goriely98}.
The dependence of direct neutron capture
on nuclear structure models was investigated in \cite{Rauscher98}. 
The investigated models yield capture cross sections sometimes differing 
by orders of magnitude.
The $A$-dependence of level energies and neutron separation energies is
illustrated by comparing Figs~10--12 of \cite{Rauscher98}, shown
in Fig.~\ref{fig:Rauscher10}.

The dependence of the level energies and neutron separation energies 
on  various theoretical models (HFB, RMFT and 
FRDM) can strikingly be noticed there.
This may also lead to differences in the predicted astrophysical $r$-process
paths.  Because of the low level densities, the compound nucleus model may
not be applicable. 

\begin{figure}[tb]
\begin{center}
\resizebox{0.4\textwidth}{!}{%
\includegraphics{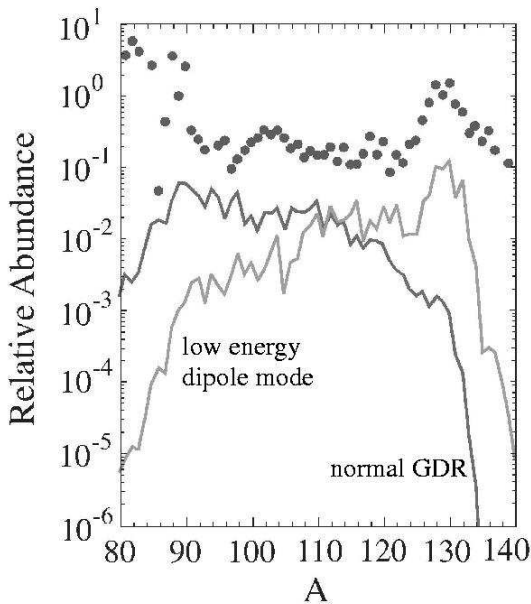}}
\end{center}
\caption{
The measured and calculated abundances of nuclei are compared with each
other. The dots show the measured abundance which are compared to the one
of a calculation including a low energy dipole mode and one with a normal
GDR mode. The presence of a low energy GDR leads to higher abundances for
larger $A$ (lighter curve), whereas the one with a standard GDR mode has
a higher abundance at lower $A$ (darker curve)
This is Fig.~24 of \protect\cite{riawhitepaper}, which is an adaption
from \protect\cite{Goriely98}.
}
\label{fig:ria24}
\end{figure}

The dependence of the 
$r$-process abundance on nuclear properties and astrophysical 
conditions is investigated in \cite{Goriely98}. The influence of the normal 
GDR and a GDR, which includes low energy dipole contributions,
is strikingly summarized  in Fig.~\ref{fig:ria24}.
Specifically 
the possible existence of a low-enery $E1$ pigmy resonance is studied.
There can  also be a problem with the application of 
the statistical (Hauser-Feshbach) model to neutron rich nuclei:
since the number of available resonance states may not be 
large enough for the application of this model, there could be a possible 
overestimate of the 
statistical (Hauser-Feshbach) predictions for resonance-deficient 
nuclei. The influence
of  nuclear structure properties on the  
$r$-process abundance distributions is shown in Fig.~6 of
\cite{Goriely98} for the astrophysical scenarios 
$T=1.0\times 10^9$~K, $N_n=10^{20}$~cm$^{-3}$, $\tau=2.4$~s and 
$T=1.5\times 10^9$~K, $N_n=10^{28}$~cm$^{-3}$ and $\tau=0.3$~s,
where $N_n$ is the neutron density and $\tau$ the irradiation time.
We recall that the astrophysical site of the $r$-process is still under
debate.  
From this study it can be concluded that \cite{Goriely98}
{\it ``there is an urgent need to carry on investigating theoretically
as well as experimentally, the radiative neutron captures by 
exotic nuclei in order to improve our understanding of the 
r-process nucleosynthesis.''}

In such a situation, the Coulomb dissociation can be a very useful tool
to obtain information on ($n$,$\gamma$)-reaction cross sections on
unstable nuclei, where direct measurements cannot be done. Of course,
one cannot and need not study the capture cross section on all the nuclei
involved; there will be some key reactions of nuclei close to
magic numbers. It was proposed in \cite{Gai97} to use the Coulomb
dissociation method to obtain information about (n,$\gamma$) reaction
cross sections, using nuclei like ${}^{124}$Mo, ${}^{126}$Ru, ${}^{128}$Pd
and ${}^{130}$Cd as projectiles. 

Since the flux of equivalent
photons has essentially a $1/\omega$ dependence, low neutron
thresholds are favourable for the Coulomb dissociation method.
The optimum choice of beam energy is an important issue. It will
depend essentially on the actual neutron binding energy. 
There is a factor of $(c/v)^2$ on the photon flux 
(see Eq.~(\ref{eq:nomegaexact}))
favoring the low beam velocities. On the other hand,
the beam energy should not be so low that higher order effects become 
important (the strength paramer $\chi$ has an $1/v$-dependence, 
see Eq.~(\ref{eq:chidefined}))
or that the adiabaticity criterion is not fulfilled
(see Eq.~(\ref{eq:xidefined})). Note
that only information about the ($n$,$\gamma$) capture reaction to the
ground state is possible with the Coulomb dissociation method. The
situation is reminiscent of the loosely bound neutron-rich light nuclei,
like ${}^{11}$Be, ${}^{11}$Li and ${}^{19}$C,
which were discussed in Sec.~\ref{sec:nstruc}.
We recall that it was possible to determine the neutron
binding energy of $^{19}$C from the shape of the low-lying
$E1$ strength distribution, see \cite{Nakamura99} 
or Sec.~\ref{sec:nstruc}.
The relevant level densities are low, which is also favourable for the 
Coulomb dissociation method. To first order
Coulomb dissociation probabilities are independent of the mass 
of the core (the recoil is proportional to $1/A$, but the Coulomb
force is proportional to $Z$, see Eqs.~(\ref{eq:ydefined}) and~(\ref{eq:PLO})),
so one can expect that 
the Coulomb dissociation method works essentially in a quite similar 
way as for the light ions like $^{11}Be$, or $^{15,17,19}C$.

In \cite{TypelB94a}
the 1$^{st}$ and 2$^{nd}$ order Coulomb excitation amplitudes
are given analytically in a zero range model for the neutron-core
interaction (see also Sec.~\ref{sec:higher}). 
We propose to use this handy formalism 
to assess, how far one can go down in beam energy
and still obtain meaningful results with the Coulomb
dissociation method, i.e., where the $1^{st}$ order amplitude
can still be extracted experimentally without being too much disturbed
by corrections due to higher orders. For radioactive
beam facilities, like ISOL or SPIRAL, the maximum beam energy is an
important question for possible Coulomb dissociation experiments. 
(For Coulomb dissociation with two charged fragments in the final
state, like in the ${}^{8}$B $\to$ ${}^{7}$Be$ + p$ experiment with
a 26~MeV ${}^{8}$B beam  \cite{Schwarzenberg96} 
it seems to be impossible to obtain such a simple 
analytical formula and one should resort to 
the more involved approaches mentioned in Sec.~\ref{sec:higher}.)

There is also another aspect: the radiative neutron capture 
by neutron rich nuclei at astrophysical energies is given by the 
low energy tail of the $E1$ strength. Standard values for the 
Lorentz parameters of the GDR are normally used. The possible existence of 
a low energy $E1$ pigmy resonance can strongly influence this picture,
see \cite{Goriely98}. Coulomb dissociation is { \it the} tool to
study this question experimentally; we refer to our discussion
of the light ions in Sec.~\ref{sec:nstruc}. Especially, at the GSI 
the dipole response of the oxygen isotopes has been studied 
\cite{Leistenschneider01}. It will be very interesting to see the 
corresponding results for the (medium) heavy nuclei. 

With the new radioactive beam facilities (either fragment separator or
ISOL-type facilities) some of the nuclei far off the valley of stability,
which are relevant for the $r$-process, can be produced. An impression
of the future possibilities can be obtained, e.g., from
Fig.~2 p.10 of \cite{riawhitepaper}. In the chart of nuclides
the production rates predicted for RIA can be seen, along with
a line which indicates the $r$-process path. The rates 
for the future accelerator at the GSI are given in Fig~1.17, p.~106 
and on p.~156 in \cite{CDR}.

The estimated minimum intensities of high-energy secondary high energy beams
necessary for a certain type of reaction studies
are given in Table~1.2, p.~125 of \cite{CDR}. For Coulomb breakup
a rate of the order of $10^3$~ions$/s$ is given as a rough general
estimate. 

\subsection{\it Two-Particle Capture Reactions and Cross Sections 
Relevant for the $rp$-Process}
\label{ssec:2pcapture}
A new field of application of the Coulomb dissociation method can be
two particle capture reactions of the type 
\begin{equation}
A+p_1+p_2 \rightarrow B + \gamma .
\end{equation}
The most famous reaction of this type is certainly the 
triple $\alpha$ capture reaction leading, via an $^8$Be
unstable state, to a resonance in $^{12}$C, which was predicted by Hoyle 
\cite{Hoyle54}.
Evidently, such a type of reaction cannot be studied
in a direct way in the laboratory. Sometimes this is not necessary, when the
relevant information about resonances involved can be obtained by
other means (transfer reactions, etc.), like in the triple $\alpha$-process.
Another way to investigate such processes can be the Coulomb
dissociation method, where the time reversed process
$ \gamma + B(\mbox{gs}) \rightarrow A+p_1+p_2 $ is 
studied using equivalent photons. Let us give some examples:

Two-neutron capture reactions in supernovae neutrino bubbles are studied
in Ref.~\cite{GorresHTW95}. 
In the case of a high neutron abundance, a sequence of two-neutron
capture reactions, ${}^{4}$He(2n,$\gamma$)${}^{6}$He(2n,$\gamma$)${}^{8}$He
can bridge the $A=5$ and $8$ gaps. 
The ${}^{6}$He and ${}^{8}$He nuclei
may be formed preferentially by two-step resonant processes through their
broad $2^{+}$ first excited states  \cite{GorresHTW95}. Dedicated Coulomb 
dissociation experiments can be useful. The $^6$He Coulomb 
dissociation is measured in  \cite{Aumann99}. Astrophysical
aspects related to the two-neutron capture on $^4$He are also discussed there.
One can envisage two mechanisms: a nonresonant two-neutron
capture and a capture with an unstable $^5$He resonant intermediate
state \cite{EfrosBH96}.
A dedicated Coulomb dissociation experiment could shed light 
on the question which mechanism is dominant.

Another key reaction can be the
${}^{4}$He($\alpha$n,$\gamma$)$^9$Be reaction \cite{GorresHTW95}. 
The ${}^{9}$Be($\gamma$,n) reaction
has been studied directly (see Ref.~\cite{AjzenbergSelove88})
and the low energy $s_{\frac{1}{2}}$ resonance is clearly established. 
Despite this, a ${}^{9}$Be Coulomb
dissociation experiment could be rewarding (see also Ref.~\cite{KalassaB96}).
Other useful information is obtained from $(e,e')$ and $(p,p')$ reactions on
$^9$Be \cite{KuechlerRW87}. Recently the photodissociation of $^9$Be
was studied by \cite{UtsunomiyaGa01,UtsunomiyaYA01}, as was already mentioned.
\begin{figure}[tb]
\begin{center}
\resizebox{0.65\textwidth}{!}{%
\includegraphics{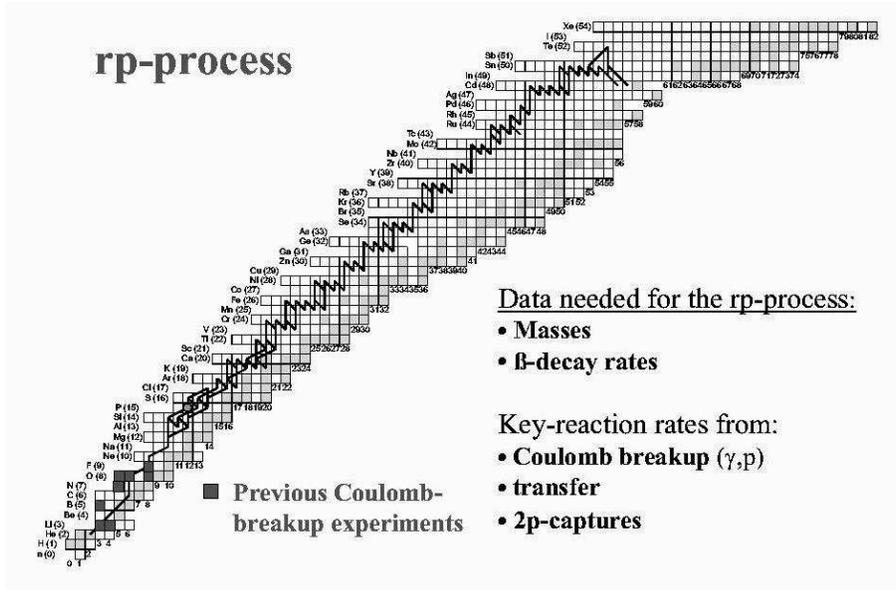}}
\end{center}
\caption{The $rp$-process path, including $2p$-captures, 
for temperatures of 1.9$\times 10^9$K and densities of 
$10^6 g cm^{-3}$ \cite{Schatz98}. Also shown are stable nuclei
and the position of the proton-drip line.
This is Fig.~1.14 of \protect\cite{CDR}.}
\label{fig:rpprocess}
\end{figure}

Another important application is for reactions relevant for the $rp$-process: 
neutron-deficient nuclei close to the 
proton drip line play an important role in a variety of astrophysical
scenarios such as Nova explosions, $X$-ray bursts, or $X$-ray pulsars.
In those scenarios, hydrogen is burnt via a sequence of rapid
proton captures and $\beta^+$-decays close to the proton drip line
(called the $rp$-process).
The impact and perspectives of radioactive beam experiments for 
the $rp$-process was recently discussed by 
Wiescher and Schatz \cite{WiescherS01}.
It is noted there that $(p,\gamma)$ capture experiments 
become increasingly difficult with increasing charge $Z$, and
Coulomb breakup, Coulomb excitation and particle transfer reactions
in inverse kinematics become promising indirect methods to access the 
relevant information. We recall that $(p,\gamma)$ reactions on light 
nuclei were studied by  the Coulomb dissociation of 
$^8$B, $^{12}$N, and $^{14}$O radioactive beams, see Sec.~\ref{sec:astro}. 
For further discussion we refer to \cite{CDR} p.~98ff and especially 
Fig.~1.14, shown here as Fig.~\ref{fig:rpprocess}.

In the $rp$-process, two-proton capture reactions can bridge the waiting 
points \cite{Baraffe97,GorresWT95,Schatz98}. 
From the ${}^{15}$O(2$p,\gamma$)${}^{17}$Ne, 
${}^{18}$Ne($2p,\gamma$)${}^{20}$Mg and ${}^{38}$Ca(2$p,\gamma$)${}^{40}$Ti
reactions considered in Ref.~\cite{GorresWT95}, 
the latter can act as an efficient reaction
link at conditions typical for $X$-ray bursts on neutron stars.
A ${}^{40}\mbox{Ti} \rightarrow  p + p + {}^{38}$Ca Coulomb dissociation 
experiment should be feasible. The decay with two protons is expected to be
sequential rather than correlated (``${}^{2}$He''-emission, 
$2p$-radioactivity). The relevant resonances are listed in Table~XII
of Ref.~\cite{GorresWT95}.
In Ref.~\cite{Schatz98} it is found that in $X$-ray bursts $2p$-capture 
reactions accelerate the reaction flow into the $Z \geq 36$ region 
considerably. In Table~1 of Ref.~\cite{Schatz98} nuclei, on which 
$2p$-capture reactions may occur,
are listed; the final nuclei are ${}^{68}$Se, ${}^{72}$Kr, ${}^{76}$Sr,
${}^{80}$Zr, ${}^{84}$Mo, ${}^{88}$Ru, ${}^{92}$Pd and ${}^{96}$Cd
(see also Fig.~8 of Ref.~\cite{Baraffe97}). It is proposed to study the Coulomb
dissociation of these nuclei in order to obtain more direct insight
into the $2p$-capture process.

\section{Conclusions}
\label{sec:conclusions}

Peripheral collisions of medium and high energy nuclei (stable or
radioactive) passing each other at distances beyond nuclear contact
and thus dominated by electromagnetic interactions are important tools
of nuclear physics research. The intense source of quasi-real (or
equivalent) photons present in such collisions has opened a wide
horizon of related problems and new experimental possibilities
especially for the present and forthcoming radioactive beam facilities
to investigate efficiently photo-interactions with nuclei (single- and
multiphoton excitations and electromagnetic dissociation).  We have
described the basic points of the theory, which rests on very solid
grounds: quantum mechanics and the electromagnetic interaction.
Problems in the technical description of the process, problems due to
nuclear interactions between the ions are identified and can be
considered as well understood. Modern computational methods are of
great help.

It has sometimes been said about the Coulomb dissociation method, see
e.g. \cite{JunghansMS02}, that {\it ``it is difficult to determine all
of their important systematic errors''}.  Certainly, there {\it are}
systematic effects which one has to take carefully into account. This
is the task of nuclear reaction theory which is described in this
review. We hope that it has become clear that these effects are well
identified and can be calculated. After all, it is essentially the
electromagnetic interaction which enters, and there is QED, the best
theory we have.  We hope that we have shown here that the view
expressed in \cite{JunghansMS02} is overly pessimistic.

Certainly, there are also experimental problems: one of them is the
accurate determination of the relative energy of the fragments. Due to
the influence of the Coulomb barrier, the astrophysical $S$-factors
depend very sensitively on this quantity.  However, we have to leave
such questions to our experimental colleagues.

After these theoretical considerations we discussed experimental
results in the field of nuclear structure and nuclear
astrophysics. Due to the good theoretical understanding of the
electromagnetic excitation, unambiguous conclusions have been drawn
from them.  Let us mention here the discovery of low lying $E1$
strength in neutron-rich nuclei and the determination of astrophysical
$S$-factors of radiative capture processes, like
$^7$Be(p,$\gamma)^8$B.

Fast moving heavy nuclei are a bright source of equivalent photons,
and we expect a bright future of this subject, especially at the
future rare isotope facilities.  Coulomb excitation and dissociation
is an opportunity to study the interaction of exotic nuclei with
photons.  It is unique, until electron-ion colliders will become
available as a complementary tool.

\section{Acknowledgments}
We have enjoyed collaboration and discussions on the present topics
with very many people, theorists, experimentalists and astrophysicists
alike.  We are especially grateful to (in order of their appearance in
this review)
K.~Alder,
A.~Winther,
C.~A.~Bertulani, 
H.~Emling, 
S.~Typel, 
H.~Wolter,
H.~Rebel, 
H.~Esbensen,
H.~Utsunomiya,
A.~Galonsky, 
F.~R\"osel, 
R.~Shyam,
T.~Motobayashi,
M.~Gai, 
J.~Kiener, 
T.~Rauscher,
F.~Fleurot, 
F.~K.~Thielemann,
K.~S\"ummerer, and
B.~Davids.

\appendix
\section{
The Electromagnetic Interaction: the Condition of No Nuclear Contact
and Expansion into Electromagnetic Multipoles }
\label{sec:appendix}

Although the following is quite familiar, it seems of interest to deal
with some basic physical points which arise in the special case of
electromagnetic excitation in nucleus-nucleus collisions. We
especially wish to clarify the difference of it to the treatment of
electron scattering, which is perhaps even more familiar to most
readers.

As was already emphasized, in hadron-hadron collisions the important
condition for the dominance of the electromagnetic interaction is
$r<r_p$, i.e., there are two {\em spatially separated} charge
distributions, see Fig.~\ref{fig:cxexplain}.  In this way strong
interactions are avoided, there is only the electromagnetic
interaction between the ions and this simple reaction mechanism is not
spoiled by nuclear effects. It is useful to take the photon propagator
in coordinate space $\vec r$ and only Fourier transformed with respect
to $t$, i.e. we consider $D_{\mu \nu}(\vec r,\omega)$; in this
representation the condition of {\em no nuclear contact} ($r<r_p$) can
be implemented easily.

The condition $r<r_p$ is used in the semiclassical approach here,
where the projectile on the orbit $\vec{r}_p(t)$ does not overlap with
the target.  In a distorted wave Born approximation (DWBA) with, e.g.,
only Coulomb waves (CWBA) the {\em no nuclear contact} condition
$r<r_p$ is also useful: for low collision energies, where the
classical turning point is outside the target nucleus, the Coulomb
wave function is very small inside the target nucleus and its
contribution can be neglected.

As we will see below, the information about the hadronic system to be
studied can be expressed in terms of electromagnetic matrix-elements
{\em at the photon point.}

The interaction of two currents is a standard problem in
electrodynamics and QED. The electromagnetic field is described by the
vector potential $A_\mu$ and its interaction as $M_{int}=\int d^4x \: j \cdot
A$.  We define the scalar product of two four-vectors as
$a\cdot b=g^{\mu \nu} a_\mu b_\nu = a_\mu a^\mu = a_0b_0-\vec{a} \cdot
\vec{b}$, where we use the metric $g=\mbox{diagonal}(1,-1,-1,-1)$
throughout this article. In our case the vector potential $A_{\mu}$ is
generated by the moving charge distribution $J_{\mu}$.  It is given by
\begin{equation}
A_{\mu}(x)=\int d^4x' D_{\mu \nu}(x,x') J_{\nu}(x').
\end{equation}
The Fourier transform of the current distribution is defined as
\begin{eqnarray}
j_\mu(x) &=& \frac{1}{(2\pi)^4} \int d^4k \exp(i k \cdot x) \hat j_\mu (k)\\
\hat j_\mu(k) &=& \int d^4x \exp(- i k \cdot x) j_\mu(x),
\end{eqnarray}
with the notation $d^4k=dk_0 d^3k$ and $k=(k_0=\omega,\vec k)$. 
We have current conservation, which is expressed as 
\begin{equation}
\partial_\mu j^\mu 
= \partial j_0 / \partial x_0 - \vec \nabla \cdot \vec j = 0,
\end{equation}
or $k \cdot \hat j =0$ in momentum space.

The electromagnetic interaction between two current distributions
$j_1$ and $j_2$ can be written as
\begin{equation}
M_{int} = \int \!\! \int d^4x d^4x' j_{1\mu}(x) D_{\mu\nu}(x-x') j_{2\nu}(x').
\end{equation}
Alternatively it is given in terms of the Fourier transform of the propagator 
\begin{eqnarray}
D_{\mu\nu}(k) &=&
 \int d^4x \exp(-ikx) D_{\mu\nu}(x)\\
\end{eqnarray}
and the currents (see above) as
\begin{eqnarray}
M_{int} &=& \frac{1}{(2\pi)^4}
\int d^4k \; \hat j_{1\mu}(-k) D_{\mu\nu}(k) \hat j_{2\nu}(k).
\label{eq:hintk}
\end{eqnarray}

In the Born approximation which is often used, e.g., in lepton
scattering, see Fig.~\ref{fig:eeprime}, Eq.~(\ref{eq:hintk})
simplifies as we can use plane waves for initial and final states.  We
consider the reaction $1+2 \rightarrow 1' +2'$, with the four-momenta
$p_1+p_2= p_1' +p_2'$. The momentum transfer $q$ is given by
$q=p_1-p_1' =-p_2+p_2'$. The transition currents can be written as
$j_{1\mu}(x)=a_{\mu} \exp(i(p_1-p_1')x)$ and $j_{2\mu}=b_{\mu}
\exp(i(p_2-p_2')x')$ where $a_{\mu}$ and $b_{\mu}$ are independent of
$x$, $x'$.  With this we can write $M_{int}$ as
\begin{equation}
M_{int}=(2\pi)^4 \delta(p_1+p_2-p_1'-p_2' ) a_{\mu}  D_{\mu \nu}(q) b_{\nu}.
\end{equation}

The fundamental property of the electromagnetic interaction is {\em
gauge invariance:} We can add to any vector potential $A_{\mu}$ a
gradient term of the form $\partial_{\mu} \chi(x)$. Due to current
conservation $\partial^{\mu} j_{\mu} =0$ this leaves the
electromagnetic interaction $M_{int}$ invariant.  Various gauges exist
also for $D_{\mu \nu}$, corresponding to various gauges of
$A_{\mu}$. E.g., \cite{LandauLQED} gives a very instructive discussion
on the application of different gauges, see also \cite{Lee81}
p.~109ff.  For interesting historical remarks (e.g., that the Lorentz
condition is originally due to Lorenz) see \cite{JacksonO01}.  A
rather general covariant form of the propagator
is given by the {\em linear lambda gauge}
\begin{equation}
D_{\mu\nu}(k) = \frac{4\pi}{k^2+i\epsilon} 
\left( g_{\mu\nu} + \lambda \frac{k_\mu k_\nu}{k^2} \right),
\end{equation}
where  $\lambda$ is an  arbitrary parameter. 

In the Coulomb gauge (characterized by $\mbox{div} \vec A=0$) one has
explicitly \cite{LandauLQED}
\begin{eqnarray}
D_{00} &=& - \frac{4\pi}{{\vec k}^2} \nonumber \\
D_{0i} &=& D_{i0} = 0 \nonumber\\
D_{il} &=& - \frac{4\pi}{k^2} \left( \delta_{il} 
- \frac{k_i k_l}{{\vec k}^2} \right),
\end{eqnarray}
(with $k^2=\omega^2-{\vec k}^2$ and $\omega=k_0$) where $i,l=1,2,3$
denote the spatial indices. (We leave out the $+i\epsilon$ for
simplicity.)  In this gauge we have
\begin{equation}
a_{\mu} D_{\mu\nu}(k) b_{\nu} = 4\pi \left(
\frac{- a_0 b_0}{{\vec k}^2} - \frac{\vec a \cdot \vec b}{k^2} + 
\frac{(\vec k \cdot \vec a)(\vec k \cdot \vec b)}{k^2 {\vec k}^2}
\right).
\end{equation}
Using current conservation we can rewrite this as
\begin{eqnarray}
&=& 4\pi \left[ - \frac{\vec a \cdot \vec b}{k^2} - 
\frac{ a_0 b_0}{{\vec k}^2} 
\left(1-\frac{k_0^2}{k^2}\right) \right]
\nonumber\\
&=& \frac{4 \pi}{k^2}\left( a_0 b_0 - \vec a \cdot \vec b \right) 
= \frac{4 \pi}{k^2} a \cdot b.
\end{eqnarray}
Due to gauge invariance one obtains the same result in the Coulomb
gauge, as well as in the Feynman gauge ($\lambda=0$), Lorentz gauge
($\lambda=-1$) or in any other gauge, as it should be.

In the general case, which is relevant for electromagnetic excitation
with nuclei, the transition currents are not associated with a
definite momentum transfer $q$.  There is an integration over the
momentum transfer $k$, see Eq.~(\ref{eq:hintk}) above.  In the static
limit (no retardation, corresponding to $k_0=\omega=0$) one gets
\begin{eqnarray}
M_{int} &=& \frac{1}{(2\pi)^4} \int d^4k \left( \frac{\hat j_{1,0}(-k) \hat j_{2,0}(k) 
- \vec{\hat j_1}(-k) \cdot \vec{\hat j_2}(k)}{{\vec k}^2} \right) \nonumber\\
&=&
\int dt\int\!\! \int d^3r_1 d^3r_2 \frac{\rho_1(r_1) \rho_2(r_2) 
- \vec j_1(r_1) \cdot \vec j_2(r_2) /c^2}{\left| \vec r_1 - \vec r_2\right|} 
\nonumber\\
&=& 
\int dt \; W(1,2),
\end{eqnarray}
which corresponds to Eq.~(\ref{eq:theory:W12}) in the main text, the
starting point for the nonrelativistic Coulomb excitation in
\cite{AlderW75}, see also p.84 of \cite{EisenbergG88}.

Now we want to indicate how to express the interaction in terms of the
electromagnetic multipole matrix-elements of the charge-current
distribution. (The details can be found in the textbooks, see, e.g.,
\cite{EisenbergG88}.)  This is possible for non-overlapping charge
distributions.  We only deal with the monopole-multipole part of the
interaction, see Fig.~\ref{fig:cxexplain}. Then we can assume that the
charge $Z_p$ is a point charge. The static case is well known, the
relativistic case can be handled as well.

In \cite{EisenbergG88} the Lorentz condition is used for the
propagator.  See especially Sec.~6.1 Eq.~(8) and~(22) and Sec.~6.2 in
that reference.  We do not give here all the details of the multipole
expansion for interacting charges and currents, see, e.g.,
\cite{EisenbergG88,JacksonED,AlderW66}. Let us emphasize the main
points related to the question of {\em penetrating} (like in the plane
wave case) versus spatially separated charge distributions.  The
propagator (``Green's function'') is expanded into multipoles
in analogy to  Eq.~(\ref{eq:elmulti}) (see also, e.g.,
\cite{EisenbergG88}, Eq.~(23))
\footnote{We follow here the convention of \cite{AbramowitzS64} in the
definition of the spherical Bessel function, which is
$h^{(1)}_L(z)=j_L(z)+i y_L(z)$, where $j_L$ and $y_L$ are the (regular)
spherical Bessel and the (irregular) Neumann function. 
This is different than the one used
in \cite{Messiah85}.}
\begin{equation}
\frac{\exp( i\omega|\vec{r}-\vec{r_p}|)}{|\vec{r}-\vec{r_p}|}
= 4\pi i \omega \sum_{LM} j_L(\omega r_<)
h^{(1)}_L(\omega r_>) Y^*_{LM}(\hat r_<) Y_{LM}(\hat r_>),
\label{eq:proprr}
\end{equation}
where $r_<$ and $r_>$ are the smaller and larger of $r_p$ and r
respectively.  In the case of spatially separated charge distributions
we always have $r_<=r$ and $r_>=r_p$.

This is a generalization of the multipole expansion of the static
Coulomb interaction
\footnote{For $L=0$ we have a term proportional to $1/r_>$. This term
gives a contribution to nuclear excitation only if $r>r_p$. Monopole
transitions can occur, e.g., in the case of internal conversion, due
to the penetration of the electron charge cloud into the nucleus, see,
e.g., \cite{AchieserB62}.  In this case we have $r>r_e$, where $r_e$
and $r$ denote the electronic and nuclear variables,
respectively. Usually, the electron cloud is outside of the nucleus,
i.e. $r<r_e$ and the probability of internal conversion is
proportional to the $B(\pi,\lambda)$-value of the corresponding
$\gamma$-transition.  This is in close analogy to the Coulomb
excitation discussed here.},
\begin{equation}
\frac{1}{|\vec{r}-\vec{r_p}|}=4\pi \sum_{LM} 
\frac{1}{2L+1} \frac{r_<^L}{r_>^{L+1}} Y^*_{LM}( \hat r_<)Y_{LM}(\hat r_>),
\end{equation}
which can be obtained from Eq.~(\ref{eq:proprr}) by performing 
the limit $\omega\rightarrow 0$ and using the expansion 
\begin{equation}
j_L(x)=\frac{x^L}{(2L+1)!!},\quad\mbox{and}\quad
h^{(1)}_L(x)=-i \frac{(2L-1)!!}{x^{L+1}}
\label{eq:lwl_jh}
\end{equation}
for small values of the argument $x=\omega r$. 

On the other hand one can Fourier transform the (retarded) interaction
into momentum space by means of the Bethe integral
\begin{equation}
\frac{\exp( i\omega|\vec{r}-\vec{r_p}|)}{|\vec{r}-\vec{r_p}|}
=\frac{1}{2\pi^2} \int d^3k 
\frac{\exp( i\vec{k}\cdot (\vec{r}-\vec{r_p}))}{{\vec k}^2-\omega^2 - i \epsilon},
\end{equation}
with $\epsilon$ being an infinitesimally small positive number.
Using the plane-wave expansion of $\exp(i \vec k \cdot \vec r)$ into
multipoles, integrating over the angular part of $\vec k$, and making
use of the orthogonality of the spherical harmonics we obtain the
formula
\begin{equation}
\int_0^\infty dk k^2 \frac{j_L(kr) j_L(kr_p)}{k^2-\omega^2 - i \epsilon}
=\frac{i \pi \omega}{2} j_L(\omega r_<) h^{(1)}_L(\omega r_>).
\label{eq:app:int1}
\end{equation}
In the static limit $\omega=0$ one obtains
\begin{equation}
\int_0^\infty dk j_L(kr)j_L(kr_p)=\frac{\pi }{2(2L+1)}
\frac{r_<^L}{r_>^{L+1}}.
\label{eq:app:int2}
\end{equation}

In \cite{AlderBHM56} the Coulomb gauge is used to derive the full
expression for first order Coulomb excitation, see their Sec.~II~B.1.
The result in this approach is the same as in \cite{EisenbergG88},
where the Lorentz gauge is used; this must be so due to gauge
invariance. It is important to note that the electromagnetic
matrix-elements {\em at the photon point} (see Eqs.~(\ref{eq:appME})
and~(\ref{eq:appMM}) below) enter in the expression for the Coulomb
excitation amplitude.  In the approach using the Coulomb gauge one
also needs the equation (Eq.~(2.B.11) in \cite{AlderBHM56})
\begin{equation}
\int_0^\infty dk \frac{j_L(kr) j_L(k r_p)}{k^2-\omega^2 - i \epsilon}= 
\frac{i \pi}{2\omega} j_L(\omega r_<) h^{(1)}_L(\omega r_>)
- \frac{\pi }{2(2L+1)\omega^2}
\frac{r_<^L}{r_>^{L+1}},
\label{eq:app:int3}
\end{equation}
which can either be found in \cite{Watson44} or follows from the
identity
\begin{equation}
\frac{k^2}{\omega^2(k^2-\omega^2)}-\frac{1}{\omega^2}=
\frac{1}{k^2-\omega^2},
\end{equation}
together with the two expressions Eqs.~(\ref{eq:app:int1}) and
Eqs.~(\ref{eq:app:int2}) above.

These expressions are now used in the calculation of the
matrixelements for electromagnetic excitation, either in a
semiclassical or in a quantum-mechanical framework (for all the
necessary details on the multipole expansion of the electromagnetic
field and on vector spherical harmonics see, e.g.,
\cite{AlderBHM56,EisenbergG88,BlattW79}).  A good discussion of vector
spherical harmonics and electromagnetic interactions is also given in
\cite{BohrM69}.  One clearly sees how the separation into the $r$ and
$r_p$ parts comes about when we can use the no penetration condition
$r_< = r$ and $r_> = r_p$. The expressions factorize naturally into an
orbital part (associated with $h_L^{(1)}(\omega r_p$)) and an
electromagnetic matrixelement (associated with the $j_l(\omega r)$).

The Coulomb excitation amplitude is expressed in terms of the
electromagnetic multipole moments, which are defined as (see, e.g.,
\cite{AlderBHM56})
\begin{eqnarray}
M(E\lambda\mu,q) &=& 
\frac{(2l+1)!!}{q^{l+1} c (l+1)}
\int d^3r 
\vec j \cdot \vec \nabla \times \vec L \left\lgroup j_l(qr) Y_{lm}(\hat r)
\right\rgroup
\label{eq:appME}\\
M(M\lambda\mu,q) &=& 
-i \frac{(2l+1)!!}{q^l c (l+1)} \int d^3r \vec j(\vec r) \cdot \vec L
\left\lgroup j_l(qr) Y_{lm}(\hat r)\right\rgroup,
\label{eq:appMM}
\end{eqnarray}
with $\vec L = -i \vec r \times \vec \nabla$ (see Eqs.~(II.1.3),
(II.1.4) of \cite{AlderW75}).

One can see that the integration over the momenta $k$ (see
Eqs.~(\ref{eq:app:int1}) and~(\ref{eq:app:int2})) conspires in such a
way that only the electromagnetic matrixelements at the photon point
$k=\omega$ (or the long-wave-length expression, respectively) appear.

An even more intriguing version of an addition theorem to expand the
electromagnetic field into multipoles and into $r_<$ and $r_>$ parts
is used by Winther and Alder \cite{WintherA79} to handle the
semiclassic relativistic straight line case.  In this case the
electromagnetic potential is described by the Li\'enard-Wiechert
potential
\begin{equation}
\phi(\vec {r'},t)=
\frac{Z_p e \gamma}{\sqrt{(b-x')^2+y'^2+\gamma^2(z'-vt)^2}},\quad
\vec A(\vec {r'},t) = \frac{\vec v}{c} \phi(\vec {r'},t),
\label{eq:lienardwiechert}
\end{equation}
which can be Fourier transformed with respect to $t$ as
\begin{equation}
\phi(\vec {r'},\omega)=
\frac{2 Z_p e}{v} \exp\left(i \frac{\omega}{v} z'\right) K_0\left( 
\frac{\omega}{v\gamma} \sqrt{(b-x')^2+y'^2}\right).
\end{equation}
One expands this expression into multipole components
\begin{equation}
\phi(\vec {r'},\omega) = \sum_{\lambda\mu} W_{\lambda\mu}(r',\omega)
Y_{\lambda\mu}^*(\hat {r'}).
\label{eq:relatmultipole}
\end{equation}
By the use of the Graf addition theorem (see Eq.~(9.1.79) of
\cite{AbramowitzS64}) an analytic expression for $W_{\lambda\mu}$ is
obtained.  Again a complete separation of the excitation probability
in electromagnetic multipole moments and quantities which describe the
projectile motion is reached, see Eq.~(\ref{eq:afirelat}) of
Sec.~\ref{sec:theory}.

This is to be contrasted to the plane wave case (e.g.  commonly used
in inelastic electron scattering).  In this case there is a definite
(space-like) momentum transfer $q=p_1-p_1'$, see
Fig.~\ref{fig:eeprime}.  By varying $q$ one can probe the structure of
an object in a way not possible with real photons. An important
example is deep inelastic electron scattering.

For small values of $-q^2$ we can think of the exchanged photon as
quasireal.  For small energy loss and small angle scattering we have
the kinematical relation $-q^2=q_\perp^2+(\omega/(\gamma v))^2$.  As
the main contribution to the total cross section comes from small
values of $-q^2$ (where the photon propagator becomes very large),
this cross section is dominated by values of $-q^2\approx
(\omega/(\gamma v))^2$, which, especially for large values of
$\gamma$, can be quite small and therefore $q^2\approx 0$ ({\em
quasireal}) photons dominate in this case as well.

This kind of equivalent photon approximation is different from the one
discussed mainly in this review: here we have used the semiclassical
approximation, as it is appropriate for the nucleus-nucleus
collisions. This leads to impact parameter dependent equivalent photon
spectra. These spectra can be integrated over the impact parameters
$b>R_{min}$.  Still, the plane-wave and the semiclassical approach
have some qualitative features in common, like the logarithmic rise of
the photon number with the Lorentz factor $\gamma$.  As the name
implies,
the EPA is an {\em approximation}, with certain ranges of validity,
different for the semiclassical and plane wave approaches, see, e.g.,
\cite{BaurB97} and \cite{BeneshHF96}.  In \cite{RubehnMT97}
experimental data for electromagnetic excitation with heavy ions were
analysed using the EPA spectrum of \cite{BeneshHF96}.  In contrast to
semiclassical calculations, systematically lower cross sections are
obtained that cannot reproduce the experimental results.

How the EPA works and where its limits are, is explicitly studied, e.g.,
in the case of $\bar H^0$ production,
\cite{MungerBS94,MeierHHT98,BertulaniB98}; we can only refer to these
papers here. The PWBA in the Coulomb gauge is also studied very lucidly in a
classic paper by Fano \cite{Fano63}; a more recent work is
\cite{VoitkivU01}.


\begin{thebibliography}{100}

\bibitem{AlderW75}
K.~Alder and A.~Winther,
\newblock {\em Electromagnetic excitation},
\newblock North-Holland, Amsterdam, 1975.

\bibitem{AlderBHM56}
{K. Alder, A. Bohr, T. Huus, B. Mottelson, and A. Winther},
\newblock {\em Rev. Mod. Phys.} 28 (1956) 432.

\bibitem{CDR}
H.~H. Gutbrod et~al., editors,
\newblock {\em An International Accelerator Facility for Beams of Ions and
  antiprotons},
\newblock Gesellschaft f\"ur Schwerionenforschung, Darmstadt, 2001.

\bibitem{riawhitepaper}
{National Superconducting Cyclotron Laboratory},
\newblock Scientific Opportunities with Fast Fragmentation Beams from RIA,
\newblock available at http://www.nscl.msu.edu/research/ria/whitepaper.pdf,
  2000.

\bibitem{riaweb}
{National Superconducting Cylcotron Laboratory},
\newblock Workshops and White Papers,
\newblock see webpage at
  http://www.nscl.msu.edu/future/ria/process/whitepapers/.

\bibitem{ribfweb}
RIKEN RI Beam Factory,
\newblock see webpage at http://www.rarf.riken.go.jp/ribf/.

\bibitem{BertulaniB88}
C.~A. Bertulani and G.~Baur,
\newblock {\em Phys. Rep.} 163 (1988) 299.

\bibitem{DreitleinP62}
J.~Dreitlein and H.~Primakoff,
\newblock {\em Phys. Rev.} 125 (1962) 1671.

\bibitem{PomeranchuckS61}
I.~Y. Pomeranchuck and I.~M. Shmushkevich,
\newblock {\em Nucl. Phys.} 23 (1961) 452.

\bibitem{Emling94}
H.~Emling,
\newblock {\em Prog. Part. Nucl. Phys.} 33 (1994) 729.

\bibitem{AumannBE98}
T.~Aumann, P.~F. Bortignon, and H.~Emling,
\newblock {\em Annu. Rev. Nucl. Part. Sci.} 48 (1998) 351.

\bibitem{BaurB86}
G.~Baur and C.~Bertulani,
\newblock {\em Phys. Lett.~B} 174 (1986) 23.

\bibitem{TypelB94a}
S.~Typel and G.~Baur,
\newblock {\em Nucl. Phys.~A} 573 (1994) 486.

\bibitem{TypelB94b}
S.~Typel and G.~Baur,
\newblock {\em Phys. Rev.~C} 50 (1994) 2104.

\bibitem{TypelWB97}
S.~Typel, H.~H. Wolter, and G.~Baur,
\newblock {\em Nucl. Phys.~A} 613 (1997) 147.

\bibitem{BaurBR86}
G.~Baur, C.~A. Bertulani, and H.~Rebel,
\newblock {\em Nucl. Phys.~A} 458 (1986) 188.

\bibitem{BaurR94}
G.~Baur and H.~Rebel,
\newblock {\em J. Phys.~G} 20 (1994) 1.

\bibitem{BaurR96}
G.~Baur and H.~Rebel,
\newblock {\em Ann. Rev. Nucl. Part. Sci.} 46 (1996) 321.

\bibitem{eurograd}
G.~Baur,
\newblock Photon-Hadron, Photon-Photon Interactions and Nuclear Astrophysics at
  Heavy Ion Accelerators,
\newblock lecture held at the ``Europ{\"a}isches Graduiertenkolleg
  Basel-T{\"u}bingen'', April 8--12, 2002, see webpage at
  http://www.physik.unibas.ch/eurograd/Vorlesung/Baur/.

\bibitem{Messiah85}
A.~Messiah,
\newblock {\em Quantenmechanik}, volume~2,
\newblock Walter de Gruyter, Berlin, New York, 1985.

\bibitem{EisenbergG88}
J.~M. Eisenberg and W.~Greiner,
\newblock {\em Nuclear Theory}, volume 2: Excitation Mechanisms of the Nucleus,
\newblock North-Holland, Amsterdam, Oxford, New York, Tokyo, third revised
  edition edition, 1988.

\bibitem{EsbensenB02}
H.~Esbensen and G.~F. Bertsch,
\newblock {\em Nucl. Phys.~A} 706 (2002) 477.

\bibitem{EsbensenB02c}
H.~Esbensen and C.~A. Bertulani,
\newblock {\em Phys. Rev.~C} 65 (2002) 024605.

\bibitem{WintherA79}
A.~Winther and K.~Alder,
\newblock {\em Nucl. Phys.~A} 319 (1979) 518.

\bibitem{Fermi24}
E.~Fermi,
\newblock {\em Z. Phys.} 29 (1924) 315.

\bibitem{Fermi25}
E.~Fermi,
\newblock {\em Nuovo Cimento} 2 (1925) 143.

\bibitem{FermiGW02}
E.~Fermi,
\newblock in {\em Proceedings of the workshop on ``Electromagnetic Probes of
  Fundamental Physics'', Erice, Italy, Oct. 16--21, 2001}, edited by
  W.~Marciano and S.~White, p. 243, Singapore, 2003, World Scientific,
\newblock translated from the Italian by M. Gallinaro and S. White, available
  as e-print hep-th/0205086.

\bibitem{Weizsaecker34}
C.~F. Weizs{\"a}cker,
\newblock {\em Z. Phys.} 88 (1934) 612.

\bibitem{Williams34}
E.~J. Williams,
\newblock {\em Phys. Rev.} 45 (1934) 729.

\bibitem{JacksonED}
J.~D. Jackson,
\newblock {\em Classical Electrodynamics},
\newblock John Wiley, New York, 1975.

\bibitem{AguiarAB90}
C.~E. Aguiar, A.~N.~F. Aleixo, and C.~A. Bertulani,
\newblock {\em Phys. Rev.~C} 42 (1990) 2180.

\bibitem{AleixoB89}
A.~N.~F. Aleixo and C.~A. Bertulani,
\newblock {\em Nucl. Phys.~A} 505 (1989) 448.

\bibitem{CharagiG90}
S.~K. Charagi and S.~K. Gupta,
\newblock {\em Phys. Rev.~C} 41 (1990) 1610.

\bibitem{BertulaniCG02}
C.~A. Bertulani, C.~M. Campbell, and T.~Glasmacher,
\newblock {\em Computer Phys. Comm.} 152 (2003) 317,
\newblock e-print nucl-th/0207035.

\bibitem{BertulaniSMD03}
C.~A. Bertulani et~al.,
\newblock Intermediate energy Coulomb excitation as a probe of nuclear
  structure at radioactive beam facilities,
\newblock e-print nucl-th/0305001, 2003.

\bibitem{Bohr48}
N.~Bohr,
\newblock {\em Mat. Fys. Medd. Dan. Vid. Selsk.} 18 (1948) No. 8.

\bibitem{MuendelB96}
A.~M{\"u}ndel and G.~Baur,
\newblock {\em Nucl. Phys.~A} 609 (1996) 254.

\bibitem{BertulaniN93}
C.~A. Bertulani and A.~M. Nathan,
\newblock {\em Nucl. Phys.~A} 554 (1993) 158.

\bibitem{BetheJ68}
H.~A. Bethe and R.~W. Jackiw,
\newblock {\em Intermediate Quantum Mechanics},
\newblock Benjamin, New York, 1968.

\bibitem{Baur91c}
G.~Baur,
\newblock {\em Nucl. Phys.~A} 531 (1991) 685.

\bibitem{AsteBHT03}
A.~Aste et~al.,
\newblock An eikonal approach to Coulomb corrections in quasielastic electron
  scattering on heavy nuclei,
\newblock (in preparation), 2003.

\bibitem{BaurHAT03}
G.~Baur et~al.,
\newblock Multiphoton Exchange Processes in Ultra Peripheral Relativistic Heavy
  Ion Collisions,
\newblock submitted to Nucl. Phys. A, 2003.

\bibitem{LevyS69}
M.~Levy and J.~Sucher,
\newblock {\em Phys. Rev.} 186 (1969) 1656.

\bibitem{HayotI72}
F.~Hayot and C.~Itzykson,
\newblock {\em Phys. Lett.~B} 39 (1972) 521.

\bibitem{BraunMunzinger85}
P.~Braun-Munzinger,
\newblock Proposal 814 submitted to the AGS Program Committee, SUNY at Stony
  Brook, accepted 1985 (unpublished).

\bibitem{Merzbacher70}
E.~Merzbacher,
\newblock {\em Quantum Mechanics},
\newblock Wiley, New York, 2nd edition, 1970.

\bibitem{Glauber63}
R.~J. Glauber,
\newblock {\em Phys. Rev.} 131 (1963) 2766.

\bibitem{KlauderS85}
J.~R. Klauder and B.-S. Skagerstam,
\newblock {\em Coherent states},
\newblock World Scientific, Singapore, 1985.

\bibitem{BertulaniB88c}
C.~A. Bertulani and G.~Baur,
\newblock {\em Nucl. Phys.~A} 482 (1988) 313c.

\bibitem{Boretzki96}
K.~Boretzki et~al.,
\newblock {\em Phys. Lett.~B} 384 (1996) 30.

\bibitem{Ritman93}
J.~Ritman et~al.,
\newblock {\em Phys. Rev. Lett.} 70 (1993) 533.

\bibitem{Schmidt93}
R.~Schmidt et~al.,
\newblock {\em Phys. Rev. Lett.} 70 (1993) 1767.

\bibitem{Bertulani02}
C.~A. Bertulani,
\newblock in {\em Proceedings of the workshop on ``Electromagnetic Probes of
  Fundamental Physics'', Erice, Italy, Oct. 16--21, 2001}, edited by
  W.~Marciano and S.~White, p. 203, Singapore, 2003, World Scientific,
\newblock available as e-print nucl-th/0201060.

\bibitem{BaurBD00}
G.~Baur, C.~A. Bertulani, and D.~Dolci,
\newblock {\em Eur. Phys. J. A} 7 (2000) 55.

\bibitem{GuW01}
J.~Z. Gu and H.~A. Weidenm{\"u}ller,
\newblock {\em Nucl. Phys.~A} 690 (2001) 382.

\bibitem{Carlson99a}
B.~V. Carlson et~al.,
\newblock {\em Ann. Phys.} 276 (1999) 111.

\bibitem{Carlson99b}
B.~V. Carlson et~al.,
\newblock {\em Phys. Rev.~C} 60 (1999) 014604.

\bibitem{BaurB86e}
G.~Baur and C.~A. Bertulani,
\newblock in {\em Proc. Int. School of Heavy Ion Physics, Oct. 12--22, 1986},
  edited by R.~A. Broglia and G.~F. Bertsch, p. 343, Plenum Press, 1986.

\bibitem{ChomazF95}
P.~Chomaz and N.~Frascaria,
\newblock {\em Phys. Rep.} 252 (1995) 275.

\bibitem{MordechaiM91}
S.~Mordechai and C.~F. Moore,
\newblock {\em Nature} 352 (1991) 393.

\bibitem{MelezhikB99}
V.~S. Melezhik and D.~Baye,
\newblock {\em Phys. Rev.~C} 59 (1999) 3232.

\bibitem{EsbensenBB95}
H.~Esbensen, G.~F. Bertsch, and C.~A. Bertulani,
\newblock {\em Nucl. Phys.~A} 581 (1995) 107.

\bibitem{Utsunomia99}
H.~Utsunomia et~al.,
\newblock {\em Nucl. Phys.~A} 654 (1999) 928c.

\bibitem{TypelW99}
S.~Typel and H.~H. Wolter,
\newblock {\em Z. Naturforsch.} 54a (1999) 63.

\bibitem{TypelB01}
S.~Typel and G.~Baur,
\newblock {\em Phys. Rev.~C} 64 (2001) 024601.

\bibitem{Nakamura99}
T.~Nakamura,
\newblock {\em Phys. Rev. Lett.} 84 (1999) 1112.

\bibitem{Nakamura94}
T.~Nakamura,
\newblock {\em Phys. Lett.~B} 331 (1994) 296.

\bibitem{Tostevin99}
J.~A. Tostevin,
\newblock in {\em 2nd International Conference on Fission and Neutron Rich
  Nuclei, St. Andrews, Scotland, June 28 -- July 2 1999}, edited by J.~H.
  Hamilton et~al., Singapore, 2000, World Scientific.

\bibitem{TostevinRJ98}
J.~A. Tostevin, S.~Rugmai, and R.~C. Johnson,
\newblock {\em Phys. Rev.~C} 57 (1998) 3225.

\bibitem{MortimerTT02}
J.~Mortimer, I.~J. Thompson, and J.~A. Tostevin,
\newblock {\em Phys. Rev.~C} 65 (2002) 064619.

\bibitem{MatsumotoKOI03}
T.~Matsumoto et~al.,
\newblock New treatment of breakup continuum in the method of continuum
  discretized coupled channels,
\newblock submitted to Phys. Rev. C, available as e-print nucl-th/0302034,
  2003.

\bibitem{Marta02}
H.~D. Marta et~al.,
\newblock {\em Phys. Rev.~C} 66 (2002) 024605.

\bibitem{TokimotoEA01}
Y.~Tokimoto et~al.,
\newblock {\em Phys. Rev.~C} 63 (2001) 035801.

\bibitem{EsbensenB96}
H.~Esbensen and G.~Bertsch,
\newblock {\em Nucl. Phys.~A} 600 (1996) 37.

\bibitem{EsbensenB95}
H.~Esbensen and G.~F. Bertsch,
\newblock {\em Phys. Lett.~B} 359 (1995) 13.

\bibitem{Austern70}
N.~Austern,
\newblock {\em Direct Reaction Theory},
\newblock Wiley, New York, 1970.

\bibitem{BaurT72}
G.~Baur and D.~Trautmann,
\newblock {\em Phys. Lett.~B} 42 (1972) 31.

\bibitem{Jarczyk72}
L.~Jarczyk et~al.,
\newblock {\em Phys. Lett.} 39B (1972) 191.

\bibitem{BaurHTT01}
G.~Baur et~al.,
\newblock {\em Prog. Part. Nucl. Phys.} 46 (2001) 99.

\bibitem{RybickiA71}
F.~Ribycki and N.~Austern,
\newblock {\em Phys. Rev.~C} 6 (1971) 1525.

\bibitem{AltIM03}
E.~O. Alt, B.~F. Igarziev, and A.~M. Mukhamedzhanov,
\newblock {\em Phys. Rev. Lett.} 90 (2003) 122701.

\bibitem{BaurT72b}
G.~Baur and D.~Trautmann,
\newblock {\em Nucl. Phys.~A} 191 (1972) 321.

\bibitem{Galonsky94}
A.~Galonsky et~al.,
\newblock {$n-n$} Correlations with exotic nuclei,
\newblock Preprint MSUCL-951, 1994,
\newblock available from www.nscl.msu.edu.

\bibitem{BaurPT74}
G.~Baur, M.~Pauli, and D.~Trautmann,
\newblock {\em Nucl. Phys.~A} 224 (1974) 477.

\bibitem{BaurBK95}
G.~Baur, C.~A. Bertulani, and D.~M. Kalassa,
\newblock {\em Nucl. Phys.~A} 550 (1995) 107.

\bibitem{BaurT76}
G.~Baur and D.~Trautmann,
\newblock {\em Phys. Rep.} 25C (1976) 293.

\bibitem{BaurRT84}
G.~Baur et~al.,
\newblock {\em Phys. Rep.} 111 (1984) 333.

\bibitem{ShyamBB92}
R.~Shyam, P.~Banerjee, and G.~Baur,
\newblock {\em Nucl. Phys.~A} 540 (1992) 341.

\bibitem{BaurHT01}
G.~Baur, K.~Hencken, and D.~Trautmann,
\newblock in {\em Proc. of ENAM 2001, H\"ameenlinna, Finland, July 2001}, p.
  161, Heidelberg, 2002, Springer.

\bibitem{LandauLQED}
L.~D. Landau and E.~M. Lifschitz,
\newblock {\em Quantenelektrodynamik}, volume~IV of {\em Lehrbuch der
  theoretischen Physik},
\newblock Akademie Verlag, Berlin, 1986.

\bibitem{Nordsieck54}
A.~Nordsieck,
\newblock {\em Phys. Rev.} 93 (1954) 785.

\bibitem{AbramowitzS64}
M.~Abramowitz and I.~A. Stegun,
\newblock {\em Handbook of Mathematical Functions},
\newblock National Bureau of Standars, Washington, DC, 1964.

\bibitem{TrautmannHB03}
D.~Trautmann, K.~Hencken, and G.~Baur,
\newblock A realistic solvable model for the Coulomb breakup of Neutron Halo
  Nuclei,
\newblock to be published, 2003.

\bibitem{BanerjeeBHST02}
P.~Banerjee et~al.,
\newblock {\em Phys. Rev.~C} 65 (2002) 064602.

\bibitem{BonaccorsoBB03}
A.~Bonaccorso, D.~M. Brink, and C.~A. Bertulani,
\newblock Proton vs. neutron halo breakup,
\newblock Pisa preprint IFUP-TH 11/2003, available as e-print nucl-th/0302001,
  2003.

\bibitem{Bonaccorso99}
A.~Bonaccorso,
\newblock {\em Phys. Rev.~C} 60 (1999) 054604.

\bibitem{HenckenBE96}
K.~Hencken, G.~Bertsch, and H.~Esbensen,
\newblock {\em Phys. Rev.~C} 54 (1996) 3043.

\bibitem{Schiller05}
F.~Schiller,
\newblock {\em Wilhelm Tell},
\newblock J. B. Cotta'sche Buchhandlung, T{\"u}bingen, 1804.

\bibitem{Frahn85}
W.~E. Frahn,
\newblock {\em Diffractive Processes in Nuclear Physics},
\newblock Clarendon Press, Oxford, 1985.

\bibitem{intweb}
``Reaction Theory for Nuclei Far From Stability'',
\newblock INT Workshop 02-26W, Seattle, September 16 - 20, 2002, see webpage at
  http://int.phys.washington.edu/$\sim$int\_talk/WorkShops/int\_02\_26W/.

\bibitem{Serber47}
R.~Serber,
\newblock {\em Phys. Rev.} 72 (1947) 1008.

\bibitem{Glauber55}
R.~J. Glauber,
\newblock {\em Phys. Rev.} 99 (1955) 1515.

\bibitem{Sitenko90}
A.~G. Sitenko,
\newblock {\em Theory of Nuclear Reactions},
\newblock World Scientific, Singapore, 1990.

\bibitem{Hansen96}
P.~G. Hansen,
\newblock {\em Phys. Rev. Lett.} 77 (1996) 1016.

\bibitem{HuefnerN81}
J.~H{\"u}fner and M.~C. Nemes,
\newblock {\em Phys. Rev.~C} 23 (1981) 2538.

\bibitem{HansenS01}
P.~G. Hansen and B.~M. Sherrill,
\newblock {\em Nucl. Phys.~A} 693 (2001) 133.

\bibitem{Tanihata96}
I.~Tanihata,
\newblock {\em J. Phys.~G} 22 (1996) 157.

\bibitem{HansenJJ95}
P.~G. Hansen, A.~S. Jensen, and B.~Jonson,
\newblock {\em Annu. Rev. Nucl. Sci.} 45 (1995) 591.

\bibitem{Bonaccorso03a}
A.~Bonaccorso,
\newblock Reaction Mechanisms with Exotic Nuclei,
\newblock Pisa preprint IFUP-TH 7/03, available as e-print nucl-th/0301030,
  2003.

\bibitem{Bonaccorso03b}
A.~Bonaccorso,
\newblock Theoretical developments for low energy experiments with radioactive
  beams,
\newblock Pisa preprint IFUP-TH 6/03, available as e-print nucl-th/0301031,
  2003.

\bibitem{BertulaniBH91}
C.~A. Bertulani, G.~Baur, and M.~S. Hussein,
\newblock {\em Nucl. Phys.~A} 526 (1991) 751.

\bibitem{PerreyP76}
C.~M. Perrey and F.~G. Perrey,
\newblock {\em At. Data and Nucl. Data Tables} 17 (1976) 1.

\bibitem{VarnerTAL91}
R.~L. Varner et~al.,
\newblock {\em Phys. Rep.} 201 (1991) 57.

\bibitem{KoningD03}
A.~J. Koning and J.~P. Delaroche,
\newblock {\em Nucl. Phys.~A} 713 (2003) 231.

\bibitem{Kobos82}
A.~M. Kobos et~al.,
\newblock {\em Nucl. Phys.~A} 384 (1982) 65.

\bibitem{Brandan88}
M.~E. Brandan and G.~R. Satchler,
\newblock {\em Nucl. Phys.~A} 487 (1988) 477.

\bibitem{BertschBML77}
G.~Bertsch et~al.,
\newblock {\em Nucl. Phys.~A} 284 (1977) 399.

\bibitem{KhoaSO95}
D.~T. Khoa, R.~Satchler, and W.~{van Oertzen},
\newblock {\em Phys. Rev.~C} 51 (1995) 2069.

\bibitem{HusseinRB91}
M.~S. Hussein, R.~Rego, and C.~A. Bertulani,
\newblock {\em Phys. Rep.} 201 (1991) 279.

\bibitem{Ray79}
L.~Ray,
\newblock {\em Phys. Rev.~C} 20 (1979) 1857.

\bibitem{Glauber59}
R.~J. Glauber,
\newblock in {\em Lectures in Theoretical Physics}, edited by W.~E. Brittin and
  L.~C. Dunham, volume~1, p. 315, Interscience, New York, 1959.

\bibitem{JoachainQ74}
C.~J. Joachain and C.~Quigg,
\newblock {\em Rev. Mod. Phys.} 46 (1974) 279.

\bibitem{EsbensenB99}
H.~Esbensen and G.~F. Bertsch,
\newblock {\em Phys. Rev.~C} 59 (1999) 3240.

\bibitem{SakuragiYK86}
Y.~Sakuragi, M.~Yahiro, and M.~Kamimura,
\newblock {\em Prog. Theor. Phys. Suppl.} 89 (1986) 136.

\bibitem{MoroCNT02}
A.~M. Moro et~al.,
\newblock {\em Phys. Rev.~C} 66 (2002) 024612.

\bibitem{YahiroNIK82}
M.~Yahiro et~al.,
\newblock {\em Prog. Theor. Phys.} 67 (1982) 1464.

\bibitem{YahiroIKK86}
M.~Yahiro et~al.,
\newblock {\em Prog. Theor. Phys. Suppl.} 89 (1986) 32.

\bibitem{AusternIKK87}
N.~Austern et~al.,
\newblock {\em Phys. Rep.} 154 (1987) 125.

\bibitem{VargaPSW02}
K.~Varga et~al.,
\newblock {\em Phys. Rev.~C} 66 (2002) 034611.

\bibitem{BertulaniM92}
C.~A. Bertulani and K.~W. McVoy,
\newblock {\em Phys. Rev.~C} 46 (1992) 2638.

\bibitem{BanerjeeS93}
P.~Banerjee and R.~Shyam,
\newblock {\em Phys. Lett.~B} 349 (1993) 421.

\bibitem{SagawaT94}
H.~Sagawa and N.~Takigawa,
\newblock {\em Phys. Rev.~C} 50 (1994) 985.

\bibitem{BarrancoVB96}
F.~Barranco, E.~Vigezzi, and R.~A. Broglia,
\newblock {\em Z. Phys. A} 356 (1996) 45.

\bibitem{OgawaYS92}
Y.~Ogawa, K.~Yabana, and Y.~Suzuki,
\newblock {\em Nucl. Phys.~A} 543 (1992) 722.

\bibitem{OgawaSY94}
Y.~Ogawa, Y.~Suzuki, and K.~Yabana,
\newblock {\em Nucl. Phys.~A} 571 (1994) 784.

\bibitem{MargueronBB02}
J.~Margueron, A.~Bonaccorso, and D.~M. Brink,
\newblock {\em Nucl. Phys.~A} 703 (2002) 105.

\bibitem{ChatterjeeS02}
R.~Chatterjee and R.~Shyam,
\newblock {\em Phys. Rev.~C} 66 (2002) 061601(R).

\bibitem{EsbensenB02a}
H.~Esbensen and G.~F. Bertsch,
\newblock {\em Nucl. Phys.~A} 706 (2002) 383.

\bibitem{EsbensenB02b}
H.~Esbensen and G.~F. Bertsch,
\newblock {\em Phys. Rev.~C} 66 (2002) 044609.

\bibitem{AkhiezerS57}
A.~I. Akhiezer and A.~G. Sitenko,
\newblock {\em Phys. Rev.} 106 (1957) 1236.

\bibitem{MargueronBB03}
J.~Margueron, A.~Bonaccorsa, and D.~M. Brink,
\newblock {\em Nucl. Phys.~A} 720 (2003) 337.

\bibitem{AlaviHarati02}
A.~Alavi-Harati et~al.,
\newblock {\em Phys. Rev. Lett.} 89 (2002) 072001.

\bibitem{CareyCC90}
D.~Carey et~al.,
\newblock {\em Phys. Rev. Lett.} 64 (1990) 357.

\bibitem{Schmidt00}
K.-H. Schmidt et~al.,
\newblock {\em Nucl. Phys.~A} 665 (2000) 221.

\bibitem{Heinz03}
A.~Heinz et~al.,
\newblock {\em Nucl. Phys.~A} 713 (2003) 3.

\bibitem{Abreu99}
M.~C. Abreu et~al.,
\newblock {\em Phys. Rev.~C} 59 (1999) 876.

\bibitem{BaurB89}
G.~Baur and C.~A. Bertulani,
\newblock {\em Nucl. Phys.~A} 505 (1989) 835.

\bibitem{BaurHT98}
G.~Baur, K.~Hencken, and D.~Trautmann,
\newblock {\em Topical Review, J. Phys. G} 24 (1998) 1657.

\bibitem{BaurHT02}
G.~Baur et~al.,
\newblock {\em Phys. Rep.} 364 (2002) 359.

\bibitem{ChiuEA01}
M.~Chiu et~al.,
\newblock {\em Phys. Rev. Lett.} 89 (2002) 012302.

\bibitem{Klein01}
S.~R. Klein,
\newblock {\em Nucl. Instrum. Methods} A59 (2001) 51.

\bibitem{BaltzCW98}
A.~Baltz, C.~Chasman, and S.~N. White,
\newblock {\em Nucl. Instrum. Methods} 417 (1998) 1.

\bibitem{HenckenW02}
K.~Hencken and S.~White,
\newblock {\em Cern Courier} 42 (2002) 15.

\bibitem{KleinN99}
S.~Klein and J.~Nystrand,
\newblock {\em Phys. Rev.~C} 60 (1999) 014903.

\bibitem{Adler02}
C.~Adler et~al.,
\newblock {\em Phys. Rev. Lett.} 89 (2002) 272303.

\bibitem{KraussGS97}
F.~Krauss, M.~Greiner, and G.~Soff,
\newblock {\em Prog. Part. Nucl. Phys.} 39 (1997) 503.

\bibitem{BaurBC02}
G.~Baur et~al.,
\newblock in {\em Proceedings of the workshop on ``Electromagnetic Probes of
  Fundamental Physics'', Erice, Italy, Oct. 16--21, 2001}, edited by
  W.~Marciano and S.~White, p. 235, Singapore, 2003, World Scientific,
\newblock available as e-print hep-ex/0201034.

\bibitem{Glasmacher98}
T.~Glasmacher,
\newblock {\em Annu. Rev. Nucl. Part. Sci.} 48 (1998) 1.

\bibitem{Glasmacher01}
T.~Glasmacher,
\newblock {\em Nucl. Phys.~A} 693 (2001) 90.

\bibitem{Motobayashi95}
T.~Motobayashi et~al.,
\newblock {\em Phys. Lett.~B} 346 (1995) 9.

\bibitem{Scheit96}
H.~Scheit et~al.,
\newblock {\em Phys. Rev. Lett.} 77 (1996) 3967.

\bibitem{Anne95}
R.~Anne et~al.,
\newblock {\em Z. Phys. A} 352 (1995) 397.

\bibitem{Nakamura97}
T.~Nakamura et~al.,
\newblock {\em Phys. Lett.~B} 394 (1997) 11.

\bibitem{Fauerbach97}
M.~Fauerbach et~al.,
\newblock {\em Phys. Rev.~C} 56 (1997) R1.

\bibitem{BertulaniCH95}
C.~A. Bertulani, L.~F. Canto, and M.~S. Hussein,
\newblock {\em Phys. Lett.~B} 353 (1995) 413.

\bibitem{TypelB95}
S.~Typel and G.~Baur,
\newblock {\em Phys. Lett.~B} 356 (1995) 186.

\bibitem{Iwasaki01}
H.~Iwasaki,
\newblock {\em Phys. Lett.~B} 522 (2001) 227.

\bibitem{Glasmacherweb}
Homepage of Thomas Glasmacher,
\newblock see webpage at www.nscl.msu.edu/$\sim$glasmacher.

\bibitem{Cottle02}
P.~D. Cottle et~al.,
\newblock {\em Phys. Rev. Lett.} 88 (2002) 172502.

\bibitem{UchiyamaM85}
T.~Uchiyama and H.~Morinaga,
\newblock {\em Z. Phys. A} 320 (1985) 273.

\bibitem{BlattW79}
J.~M. Blatt and V.~F. Weisskopf,
\newblock {\em Theoretical nuclear physics},
\newblock Springer, New York, 1979.

\bibitem{HansenJ87}
P.~G. Hansen and B.~Jonson,
\newblock {\em Europhys. Lett.} 4 (1987) 409.

\bibitem{Ikeda92}
K.~Ikeda,
\newblock {\em Nucl. Phys.~A} 538 (1992) 355c.

\bibitem{Bertsch98}
G.~F. Bertsch,
\newblock in {\em Trends in nuclear physics, 100 years later}, edited by
  H.~Nifenecker et~al., volume Session LXVI of {\em Les Houches}, p. 123,
  Elsevier, Amsterdam, 1998.

\bibitem{AlhassidBG82}
Y.~Alhassid, M.~Gai, and G.~Bertsch,
\newblock {\em Phys. Rev. Lett.} 49 (1982) 1482.

\bibitem{BertulaniHK02}
C.~A. Bertulani, H.-W. Hammer, and U.~van Kolck,
\newblock {\em Nucl. Phys.~A} 712 (2002) 37.

\bibitem{KalassaB94}
D.~M. Kalassa and G.~Baur,
\newblock The effective range approach to the electromagnetic dissociation of
  loosely bound nuclei,
\newblock in {\em Proc. of the Int. Conference on ``Physics with GeV-Particle
  Beams''}, edited by H. Machner and K. Sistemich, p.468, Singapore, 1994,
  Forschungszentrum J{\"u}lich, World Scientific.

\bibitem{KalassaB96}
D.~M. Kalassa and G.~Baur,
\newblock {\em J. Phys.~G} 22 (1996) 115.

\bibitem{MengoniOI95}
A.~Mengoni et~al.,
\newblock {\em Phys. Rev.~C} 52 (1995) R2334.

\bibitem{OtsukaIFN94}
T.~Otsuka et~al.,
\newblock {\em Phys. Rev.~C} 49 (1994) R2289.

\bibitem{Aumann01}
T.~Aumann et~al.,
\newblock {\em Nucl. Phys.~A} 687 (2001) 103c.

\bibitem{Horvath02}
A.~Horvath et~al.,
\newblock {\em Astrophys. J.} 570 (2002) 926.

\bibitem{BohmW70}
G.~Bohm and F.~Wysotzki,
\newblock {\em Nucl. Phys.~B} 15 (1970) 628.

\bibitem{Juric73}
M.~Juric et~al.,
\newblock {\em Nucl. Phys.~B} 52 (1973) 1.

\bibitem{Lyuboshits90}
V.~L. Lyuboshits,
\newblock {\em Sov. J. Nucl. Phys.} 51 (1990) 648.

\bibitem{DanilinTVZ98}
B.~V. Danilin et~al.,
\newblock {\em Nucl. Phys.~A} 632 (1998) 383.

\bibitem{DanilinVEH97}
B.~V. Danilin et~al.,
\newblock {\em Phys. Rev.~C} 55 (1997) R577.

\bibitem{Aumann99}
T.~Aumann et~al.,
\newblock {\em Phys. Rev.~C} 59 (1999) 1252.

\bibitem{Leistenschneider01}
A.~Leistenschneider et~al.,
\newblock {\em Phys. Rev. Lett.} 86 (2001) 5442.

\bibitem{SagawaS99}
H.~Sagawa and T.~Suzuki,
\newblock {\em Phys. Rev.~C} 59 (1000) 3116.

\bibitem{Kobayashi89}
T.~Kobayashi et~al.,
\newblock {\em Phys. Lett.~B} 232 (1989) 51.

\bibitem{Shimoura95}
S.~Shimoura et~al.,
\newblock {\em Phys. Lett.~B} 348 (1995) 29.

\bibitem{Zinser97}
M.~Zinser et~al.,
\newblock {\em Nucl. Phys.~A} 619 (1997) 151.

\bibitem{EsbensenBI93}
H.~Esbensen, G.~F. Bertsch, and K.~Ieki,
\newblock {\em Phys. Rev.~C} 48 (1993) 326.

\bibitem{EsbensenB92}
H.~Esbensen and G.~Bertsch,
\newblock {\em Phys. Rev.~C} 46 (1992) 1552.

\bibitem{IekiG96}
K.~Ieki et~al.,
\newblock {\em Phys. Rev.~C} 54 (1996) 1589.

\bibitem{BertschHE98}
G.~F. Bertsch, K.~Hencken, and H.~Esbensen,
\newblock {\em Phys. Rev.~C} 77 (1998) 1366.

\bibitem{Meister02}
M.~Meister et~al.,
\newblock {\em Nucl. Phys.~A} 700 (2002) 3.

\bibitem{WangGK02}
J.~Wang et~al.,
\newblock {\em Phys. Rev.~C} 65 (2002) 034036.

\bibitem{Marquez01}
F.~M. Marquez et~al.,
\newblock {\em Phys. Rev.~C} 64 (2001) 061301.

\bibitem{MinemuraMSM02b}
T.~Minemura et~al.,
\newblock {\em RIKEN Accel. Prog. Rep.} 35 (2002) 59.

\bibitem{hirscheggweb}
``Nuclear Structure and Dynamics at the Limits'',
\newblock Hirschegg, Kleinwalsertal, January 12--18, 2003 see webpage at
  http://theory.gsi.de/hirschegg/.

\bibitem{UtsunomiyaYA01}
H.~Utsunomiya et~al.,
\newblock {\em Phys. Rev.~C} 63 (2001) 018801.

\bibitem{Austin02}
S.~Austin,
\newblock in {\em EMI 2001 Int. Symp. Electromagnetic Interactions in Nuclear
  and Hadron Physics}, Singapore, 2002, World Scientific,
\newblock available as e-print nucl-th/0201010.

\bibitem{KienerGR91}
J.~Kiener et~al.,
\newblock {\em Phys. Rev.~C} 44 (1991) 2195.

\bibitem{NollettLS97}
K.~M. Nollett, M.~Lemoine, and D.~N. Schramm,
\newblock {\em Phys. Rev.~C} 56 (1997) 1144.

\bibitem{HesselbarthK91}
J.~Hesselbarth and T.~K. Kn{\"o}pfle,
\newblock {\em Phys. Rev. Lett.} 67 (1991) 2773.

\bibitem{Robertson81}
R.~G.~H. Robertson et~al.,
\newblock {\em Phys. Rev. Lett.} 47 (1981) 1867.

\bibitem{Cecil96}
F.~E. Cecil et~al.,
\newblock {\em Phys. Rev.~C} 53 (1996) 1967.

\bibitem{Mohr94}
P.~Mohr et~al.,
\newblock {\em Phys. Rev.~C} 50 (1994) 1543.

\bibitem{Shotter84}
A.~Shotter et~al.,
\newblock {\em Phys. Rev. Lett.} 53 (1984) 1539.

\bibitem{Gazes92}
S.~B. Gazes et~al.,
\newblock {\em Phys. Rev. Lett.} 68 (1992) 150.

\bibitem{Zecher98}
P.~D. Zecher et~al.,
\newblock {\em Phys. Rev.~C} 57 (1998) 959.

\bibitem{RosswogFT01}
S.~K. Rosswog, C.~Freiburghaus, and F.-K. Thielemann,
\newblock {\em Nucl. Phys.~A} 688 (2001) 344.

\bibitem{MaoC91}
Z.~Q. Mao and A.~E. Champagne,
\newblock {\em Nucl. Phys.~A} 522 (1991) 568.

\bibitem{MalaneyF89}
R.~A. Malaney and W.~A. Fowler,
\newblock {\em Astrophys. J.} 345 (1989) L5.

\bibitem{Rauscher94}
T.~Rauscher et~al.,
\newblock {\em Astrophys. J.} 429 (1994) 499.

\bibitem{Descouvement93}
P.~Descouvement,
\newblock {\em Astrophys. J. Lett.} 405 (1993) 518.

\bibitem{KobayashiIH02}
H.~Kobayashi et~al.,
\newblock {\em Phys. Rev.~C} 67 (2003) 015806.

\bibitem{WiescherGT90}
M.~Wiescher, J.~G{\"o}rres, and F.~Thielemann,
\newblock {\em Astrophys. J.} 363 (1990) 340.

\bibitem{Descouvement00}
P.~Descouvement,
\newblock {\em Nucl. Phys.~A} 675 (2000) 559.

\bibitem{Beer92}
H.~Beer et~al.,
\newblock {\em Astrophys. J.} 387 (1992) 258.

\bibitem{Kiener01}
J.~Kiener,
\newblock in {\em NATO Advanced Study Institute: Nuclei Far from Stability and
  Astrophysics}, edited by D.~N. Poenaru, H.~Rebel, and J.~Wentz, p. 271, New
  York, 2001, Plenum.

\bibitem{Tatischeff95}
V.~Tatischeff et~al.,
\newblock {\em Phys. Rev.~C} 51 (1995) 2789.

\bibitem{BaurW89}
G.~Baur and M.~Weber,
\newblock {\em Nucl. Phys.~A} 504 (1989) 352.

\bibitem{Fleurot02}
F.~Fleurot,
\newblock {\em {$^{16}$O} Coulomb dissociation: a means to determine {$^{12}$C
  + $\alpha$} fusion rate in stars},
\newblock PhD thesis, Rijksuniversiteit Groningen, 2002,
\newblock (unpublished), available from \\
  http://www.ub.rug.nl/eldoc/dis/science/f.fleurot/.

\bibitem{Motobayashi91}
T.~Motobayashi et~al.,
\newblock {\em Phys. Lett.~B} 264 (1991) 259.

\bibitem{Kiener93}
J.~Kiener et~al.,
\newblock {\em Nucl. Phys.~A} 552 (1993) 63.

\bibitem{Decrock91}
P.~Decrock et~al.,
\newblock {\em Phys. Rev. Lett.} 67 (1991) 808.

\bibitem{SerataMSA02}
M.~Serata et~al.,
\newblock {\em RIKEN Accel. Prog. Rep.} 35 (2002) 62.

\bibitem{Levebvre95}
A.~Levebvre et~al.,
\newblock {\em Nucl. Phys.~A} 595 (1995) 69.

\bibitem{WiescherT89}
M.~Wiescher and F.~Thielemann,
\newblock {\em Astrophys. J.} 343 (1989) 352.

\bibitem{Motobayashi02}
T.~Motobayashi et~al.,
\newblock {\em Eur. Phys. J. A} 13 (2002) 207.

\bibitem{MinemuraMSM02a}
T.~Minemura et~al.,
\newblock {\em RIKEN Accel. Prog. Rep.} 35 (2002) 58.

\bibitem{GomiMYK02}
T.~Gomi et~al.,
\newblock {\em RIKEN Accel. Prog. Rep.} 35 (2002) 69.

\bibitem{Rolfs73}
C.~Rolfs,
\newblock {\em Nucl. Phys.~A} 217 (1973) 29.

\bibitem{Morlock97}
R.~Morlock et~al.,
\newblock {\em Phys. Rev. Lett.} 79 (1997) 3837.

\bibitem{Liang00}
J.~F. Liang et~al.,
\newblock {\em Phys. Lett.~B} 491 (2000) 23.

\bibitem{BertulaniD02}
C.~A. Bertulani and P.~Danielewicz,
\newblock {\em Nucl. Phys.~A} 717 (2003) 199.

\bibitem{nobelprice02}
{The Royal Swedish Academy of Sciences},
\newblock The Nobel Prize in Physics 2002,
\newblock see webpage at http://www.nobel.se/physics/laureates/2002/.

\bibitem{Ahmad01}
Q.~Ahmad et~al.,
\newblock {\em Phys. Rev. Lett.} 87 (2001) 071301.

\bibitem{Ahmad02a}
Q.~Ahmad et~al.,
\newblock {\em Phys. Rev. Lett.} 89 (2002) 011301.

\bibitem{Ahmad02b}
Q.~Ahmad et~al.,
\newblock {\em Phys. Rev. Lett.} 89 (2002) 011302.

\bibitem{Eguchi02}
K.~Eguchi et~al.,
\newblock {\em Phys. Rev. Lett.} 90 (2003) 021802.

\bibitem{BahcallPB01}
J.~N. Bahcall, M.~H. Pinsonneault, and S.~Basu,
\newblock {\em Astrophys. J.} 555 (2001) 990.

\bibitem{BahcallBP98}
J.~N. Bahcall, S.~Basu, and M.~H. Pinsonneault,
\newblock {\em Phys. Lett.~B} 433 (1998) 1.

\bibitem{BargerMW02}
V.~Barger, D.~Marfatia, and K.~Whisnant,
\newblock {\em Phys. Rev. Lett.} 88 (2002) 011302.

\bibitem{bahcallweb}
Homepage of John N Bahcall,
\newblock see webpage at http:/www.sns.ias.edu/$\sim$jnb/.

\bibitem{JunghansMS02}
A.~R. Junghans et~al.,
\newblock {\em Phys. Rev. Lett.} 88 (2002) 0411011.

\bibitem{Hammache98}
F.~Hammache et~al.,
\newblock {\em Phys. Rev. Lett.} 80 (1998) 928.

\bibitem{Strieder01}
F.~Strieder et~al.,
\newblock {\em Nucl. Phys.~A} 696 (2001) 219.

\bibitem{BabyBG02}
L.~T. Baby et~al.,
\newblock {\em Phys. Rev. Lett.} 90 (2003) 022501.

\bibitem{Terrasi01}
F.~Terrasi et~al.,
\newblock {\em Nucl. Phys.~A} 688 (2001) 539.

\bibitem{Taube94}
G.~Taube,
\newblock {\em Science} 266 (1994) 1157.

\bibitem{Iwasa99}
N.~Iwasa et~al.,
\newblock {\em Phys. Rev. Lett.} 83 (1999) 2910.

\bibitem{Kelley96}
J.~H. Kelley et~al.,
\newblock {\em Phys. Rev. Lett.} 77 (1996) 5020.

\bibitem{Davids01}
B.~Davids et~al.,
\newblock {\em Phys. Rev.~C} 63 (2001) 065806.

\bibitem{Motobayashi94}
T.~Motobayashi et~al.,
\newblock {\em Phys. Rev. Lett.} 73 (1994) 2680.

\bibitem{Schwarzenberg96}
J.~{von Schwarzenberg} et~al.,
\newblock {\em Phys. Rev.~C} 53 (1996) R2598.

\bibitem{Kikuchi97}
T.~Kikuchi et~al.,
\newblock {\em Phys. Lett.~B} 391 (1997) 261.

\bibitem{Iwasa96}
N.~Iwasa et~al.,
\newblock {\em J. of Phys. Soc. of Japan} 65 (1996) 1256.

\bibitem{Kikuchi98}
T.~Kikuchi et~al.,
\newblock {\em Eur. Phys. J. A} 3 (1998) 213.

\bibitem{DavidsAA98}
B.~Davids et~al.,
\newblock {\em Phys. Rev. Lett.} 81 (1998) 2209.

\bibitem{DavidsEA01}
B.~Davids et~al.,
\newblock {\em Phys. Rev. Lett.} 86 (2001) 2750.

\bibitem{Adelberger98}
E.~G. Adelberger et~al.,
\newblock {\em Rev. Mod. Phys.} 70 (1998) 1265.

\bibitem{Schuemann03}
F.~Sch\"umann et~al.,
\newblock {\em Phys. Rev. Lett.} 90 (2003) 232501.

\bibitem{DescouvementB94}
P.~Descouvement and D.~Baye,
\newblock {\em Nucl. Phys.~A} 567 (1994) 341.

\bibitem{DavidsT03}
B.~Davids and S.~Typel,
\newblock Electromagnetic dissociation of $^8B$ and the astrophysical $S$
  factor for $^7Be(p,\gamma)^8B$,
\newblock submitted to Phys. Rev. C, available as e-print nucl-th/0304054,
  2003.

\bibitem{Gai03}
M.~Gai,
\newblock Solar Fusion and The Coulomb Dissociation of 8B; What Have We Learned
  and Where Do We Go From Here?,
\newblock Paper for the 19th Conf. on Nuclear Dynamics, Breckenridge, Colorado,
  available as e-print nucl-th/0303009, 2003.

\bibitem{townmeeting99}
Opportunities in Nuclear Astrophysics,
\newblock Conclusions of a Town Meeting held at the University of Notre Dame
  7--8, June 1999.

\bibitem{Nupecc00}
Radioactive Nuclear Beam Facilities,
\newblock NuPECC Report, 2000.

\bibitem{KaeppelerWT98}
F.~K{\"a}ppeler, M.~Wiescher, and F.~Thielemann,
\newblock {\em Annu. Rev. Nucl. Part. Sci.} 48 (1998) 175.

\bibitem{RolfsR88}
C.~E. Rolfs and W.~S. Rodney,
\newblock {\em Cauldrons in the Cosmos},
\newblock University of Chicago Press, Chicago, 1988.

\bibitem{CowanTT91}
J.~J. Cowan, F.-K. Thielemann, and J.~W. Truran,
\newblock {\em Phys. Rep.} 208 (1991) 267.

\bibitem{tommyweb}
Homepage of Thomas Rauscher,
\newblock see webpage at http://quasar.physik.unibas.ch/$\sim$tommy/.

\bibitem{Rauscher98}
T.~Rauscher et~al.,
\newblock {\em Phys. Rev.~C} 57 (1998) 2031.

\bibitem{Goriely98}
S.~Goriely,
\newblock {\em Phys. Lett.~B} 436 (1998) 10.

\bibitem{Gai97}
M.~Gai,
\newblock ISOL workshop, Columbus/Ohio, July 30 -- August 1, 1997, 1997.

\bibitem{Hoyle54}
F.~Hoyle,
\newblock {\em Astrophys. J.} Suppl. 1 (1954) 121.

\bibitem{GorresHTW95}
J.~G{\"o}rres et~al.,
\newblock {\em Phys. Rev.~C} 52 (1995) 2231.

\bibitem{EfrosBH96}
V.~D. Efros et~al.,
\newblock {\em Z. Phys. A} 355 (1996) 101.

\bibitem{AjzenbergSelove88}
F.~Ajzenberg-Selove,
\newblock {\em Nucl. Phys.~A} 490 (1988) 1.

\bibitem{KuechlerRW87}
G.~Kuechler, A.~Richter, and W.~{von Witsch},
\newblock {\em Z. Phys. A} 326 (1987) 447.

\bibitem{UtsunomiyaGa01}
H.~Utsunomia et~al.,
\newblock in {\em EMI 2001 Int. Symp. Electromagnetic Interactions in Nuclear
  and Hadron Physics}, Singapore, 2001, World Scientific.

\bibitem{Schatz98}
H.~Schatz et~al.,
\newblock {\em Phys. Rep.} 294 (1998) 167.

\bibitem{WiescherS01}
M.~Wiescher and H.~Schatz,
\newblock {\em Nucl. Phys.~A} 693 (2001) 269.

\bibitem{Baraffe97}
I.~Baraffe et~al.,
\newblock Nuclear and Particle Astrophysics, F.-K. Thielemann (convener),
\newblock in NuPECC Report: {\it Nuclear Physics in Europe: Highlights and
  Opportunities}, 1997.

\bibitem{GorresWT95}
J.~G{\"o}rres, M.~Wiescher, and T.-K. Thielemann,
\newblock {\em Phys. Rev.~C} 51 (1995) 392.

\bibitem{Lee81}
T.~D. Lee,
\newblock {\em Particle Physics and Introduction to Field Theory},
\newblock Harwood academic publishers, London, 1981.

\bibitem{JacksonO01}
J.~D. Jackson and L.~B. Okun,
\newblock {\em Rev. Mod. Phys.} 73 (2001) 663.

\bibitem{AlderW66}
K.~Alder and A.~Winther, editors,
\newblock {\em Coulomb Excitation},
\newblock Perspectives in Physics, Academic Press, New York, London, 1966.

\bibitem{AchieserB62}
A.~I. Achieser and W.~B. Berestezki,
\newblock {\em Quanten-Elektrodynamik},
\newblock Teubner-Verlag, Leipzig, 1962.

\bibitem{Watson44}
G.~N. Watson,
\newblock {\em Theory of Bessel Functions},
\newblock Cambridge University Press, New York, 1944.

\bibitem{BohrM69}
A.~Bohr and B.~Mottelson,
\newblock {\em Nuclear Structure}, volume I: Single Particle Motion,
\newblock Benjamin, New York, 1969.

\bibitem{BaurB97}
G.~Baur and C.~Bertulani,
\newblock {\em Phys. Rev.~C} 56 (1997) 581.

\bibitem{BeneshHF96}
C.~J. Benesh, A.~C. Hayes, and J.~L. Friar,
\newblock {\em Phys. Rev.~C} 54 (1996) 1404.

\bibitem{RubehnMT97}
T.~Rubehn, W.~F.~J. M{\"u}ller, and W.~Trautmann,
\newblock {\em Phys. Rev.~C} 56 (1997) 1165.

\bibitem{MungerBS94}
C.~T. Munger, S.~J. Brodsky, and I.~Schmidt,
\newblock {\em Phys. Rev.~D} 49 (1994) 3228.

\bibitem{MeierHHT98}
H.~Meier et~al.,
\newblock {\em Eur. Phys. J. C} 5 (1998) 287.

\bibitem{BertulaniB98}
C.~A. Bertulani and G.~Baur,
\newblock {\em Phys. Rev.~D} 58 (1998) 034005.

\bibitem{Fano63}
U.~Fano,
\newblock {\em Annu. Rev. Nucl. Sci.} 13 (1963) 1.

\bibitem{VoitkivU01}
A.~B. Voitkiv and J.~Ullrich,
\newblock {\em J. Phys.~B} 34 (2001) 4513.

\end{thebibliography}

\end{document}